%% file: Main.tex
\documentclass{jfm}

\usepackage[]{amsmath}
\usepackage[dvipsnames]{xcolor}
\usepackage{wrapfig} 

\usepackage{graphicx}
\usepackage[position=t,singlelinecheck=off]{subfig}
\usepackage{caption}
\usepackage{floatrow}

\usepackage{upgreek}
\usepackage{overpic}
\usepackage[]{multirow}
\usepackage[]{url}
\usepackage[]{nicefrac}
\usepackage{bm}
\usepackage{wasysym} 
\usepackage{enumerate}

\begin{document}

\newtheorem{lemma}{Lemma}
\newtheorem{corollary}{Corollary}

\shorttitle{Explorative gradient method for active drag reduction} 
\shortauthor{Y.~Li et al.} 
\title[Explorative gradient methods]{Explorative gradient method 
for active drag reduction of the fluidic pinball and slanted Ahmed body}
\author{Yiqing Li\aff{1,2,3}, Wenshi Cui\aff{2,3}, Qing Jia\aff{2,3}, Qiliang Li\aff{2,3}, Zhigang Yang\aff{2,3,4}, Marek Morzy\'nski\aff{5} and Bernd R. Noack\aff{1,6}\corresp{\email{zhigangyang@tongji.edu.cn, bernd.noack@hit.edu.cn}}}
\affiliation{
\aff{1} Center for Turbulence Control,
Harbin Institute of Technology, Shenzhen, 
Room 312, Building C, University Town, Xili, Shenzhen 518058, 
People’s Republic of China 
\aff{2}  Shanghai Automotive Wind Tunnel Center, Tongji University,
4800 CaoAn Road, Jiading District, Shanghai 201804 
\aff{3}  Shanghai Key Lab of Vehicle Aerodynamics and Vehicle Thermal Management Systems, Shanghai 201804, China
\aff{4}  Beijing Aeronautical Science \& Technology Research Institute, Beijing, 102211, China
\aff{5}  Pozna\'n University of Technology, Chair of Virtual Engineering, 
Jana Pawla II 24, PL 60-965 Pozna\'n, Poland 
\aff{6}  Hermann-F\"ottinger-Institut, Technische Universit\"at Berlin,
  M\"uller-Breslau-Stra{\ss}e 8, D-10623 Berlin, Germany
  }

\pubyear{???}
\volume{???}
\pagerange{???--???}
\date{?; revised ?; accepted ?. - To be entered by editorial office}
\setcounter{page}{1}

\maketitle




\input{S0.tex}
\input{S1.tex}
\input{S2.tex}
\input{S3.tex}
\input{S4.tex}
\input{S5.tex}
\input{S6.tex}
\input{Acknowledgements}


\begin{appendix}
\input{SA.tex}

\end{appendix}

\bibliographystyle{jfm}
\bibliography{Main_Anne,Main_Bernd,Main_RANS}
\end{document}

%% file: S0.tex

\begin{abstract}

\begin{wrapfigure}[22]{r}[0pt]{0.55\textwidth}
    \vspace*{-8mm}
\begin{flushright}
    \includegraphics[width = \textwidth]{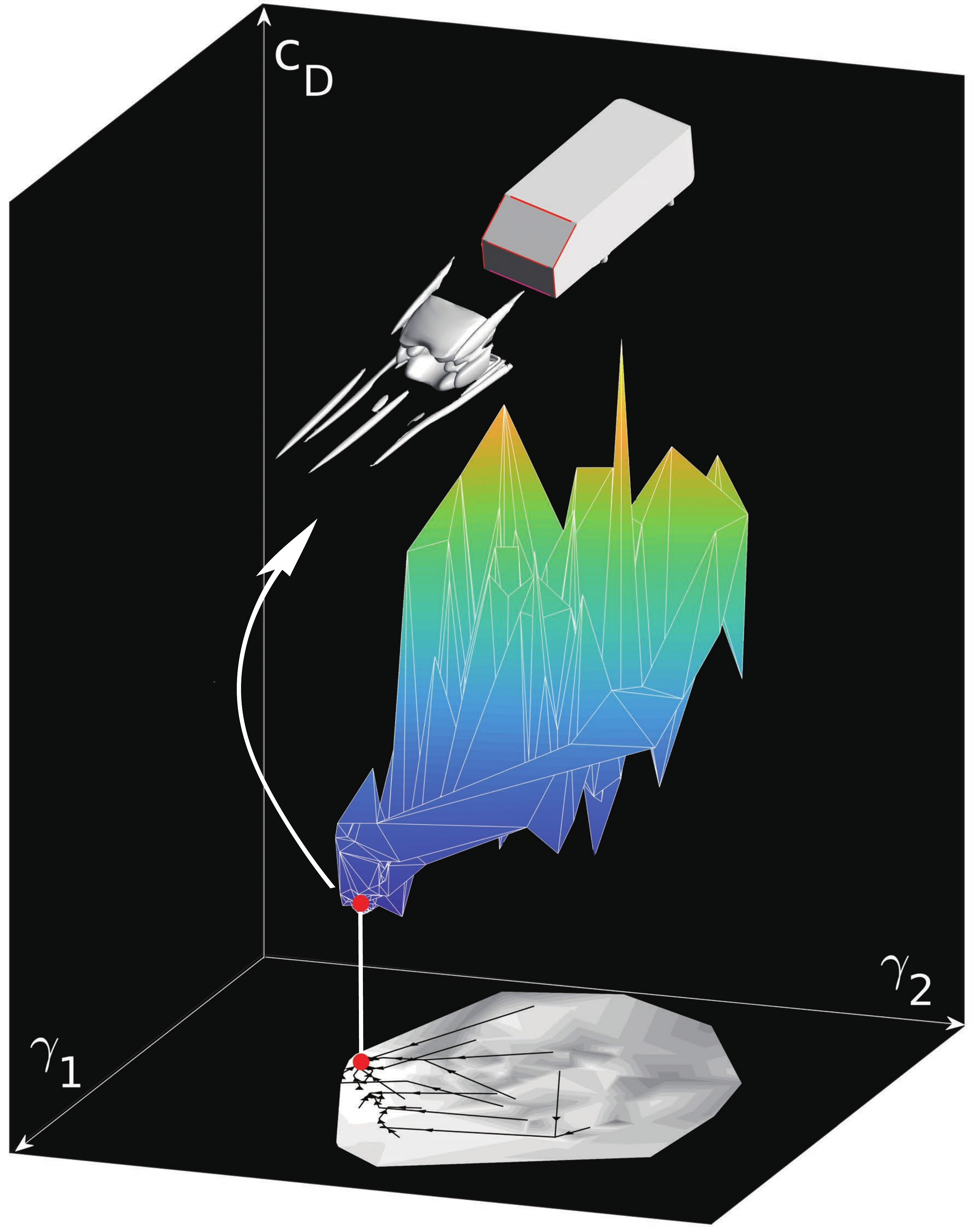}
\end{flushright}
\end{wrapfigure}    


We address a challenge of active flow control: 
the optimization of many actuation parameters 
guaranteeing fast convergence
and avoiding suboptimal local minima.
This challenge is addressed by a new optimizer, called explorative gradient method (EGM).
EGM alternatively performs one exploitive downhill simplex step 
and an explorative Latin hypercube sampling iteration.
Thus, the convergence rate of a gradient based method is guaranteed
while, at the same time, better minima are explored.
For an analytical multi-modal test function,
EGM is shown to significantly outperform the downhill simplex method,
the random restart variant, Latin hypercube sampling,
Monte Carlo iterations and the genetic algorithm.

EGM is applied to minimize the net drag power 
of the two-dimensional fluidic pinball benchmark
with three cylinder rotations as actuation parameters.
The net drag power is reduced by 40$\>$\% employing 
direct numerical simulations at a Reynolds number of $100$
based on the cylinder diameter.
This optimal actuation leads to 98$\>$\% drag reduction 
employing Coanda forcing for boat tailing and partial stabilization of vortex shedding.
The price is an actuation energy corresponding to 58\% of the unforced parasitic drag power. 

EGM is also used to minimize drag of the $35^\circ$ slanted Ahmed body
employing distributed steady blowing with 10 inputs.
17\% drag reduction are achieved using Reynolds-Averaged Navier-Stokes simulations (RANS)
at the  Reynolds number $Re_H=1.9 \times 10^5$ based on the height of the Ahmed body.
The wake is controlled with seven local jet slot actuators at all trailing edges.
Symmetric operation corresponds to five independent actuator groups at top, midle, bottom, top sides and bottom sides.
Each slot actuator produces a uniform jet with the velocity and angle as free parameters,
yielding  10 actuation parameters as free inputs. 
The optimal actuation emulates boat tailing 
by inward-directed  blowing with velocities which are  comparable to the oncoming velocity.
We expect that EGM will be employed as efficient optimizer in many future active flow control plants. 

\end{abstract}

%% file: S1.tex
\section{Introduction}
\label{ToC:Introduction}

In this study, we propose an optimizer for active flow control 
focusing on multi-actuator bluff-body drag reduction.
This optimizer combines the convergence rate of a gradient-based method
with an explorative method for identifying the global minimum. 
Actuators and sensors become increasingly cheaper, powerful and reliable.
This trend makes active flow control of increasing interest to industry.
In addition, distributed actuation 
can give rise performance benefits over single actuator solutions.
Here, we focus on the simple case of open-loop control 
with steady or periodic operation of multiple actuators.

Even for this simple case, the optimization of actuation
constitutes an algorithmic challenge.
Often the budget for optimization is limited to $O(100)$ high fidelity simulations,
like direct numerical simulations (DNS) or large-eddy simulations (LES)
or $O(100)$ water tunnel experiments,
or $O(1000)$ Reynolds Averaged Navier-Stokes (RANS) simulations, 
or a similar amount of wind-tunnel experiments.
Moreover, the optimization may need to be performed for multiple operating conditions.

Evidently, efficient optimizers are of large practical importance.
Gradient-based optimizers, like the downhill simplex method 
have the advantage of rapid convergence against a cost minimum,
but this minimum may easily be a suboptimal local one,
particularly for high-dimensional search spaces.
Random restart variants have a larger probability of finding the global minimum
but come with a dramatic increase of testing.
In contrast to gradient-based approaches,
Latin hypercube sampling performs an ideal exploration 
by guaranteeing a close geometric coverage of the search space---obviously with a poor associated convergence rate and the price of extensive evaluations of unpromising territories.
Monte Carlo sampling has similar advantages and disadvantages.
Genetic algorithms elegantly combine exploration with mutation
and exploitation with crossover operations.
These are routinely used optimizers and the focus of our study.

Myriad of  other optimizers have been invented for different niche applications.
Deterministic gradient-based optimizers may be augmented by estimators for the gradient.
These estimators  become particularly challenging for sparse data.
This challenge is addressed by 
stochastic gradient methods which aim at navigating through a high-dimensional search space
with insufficient derivative information.
Many biologically inspired optimization methods, like 
ant colony and participle swarm optimization 
also aim at balancing exploitation and exploration,
like the genetic algorithm.
A new avenue is opened by including the learning 
of the response model from actuation to cost function
 during the optimization process
and using this model for identifying promising actuation parameters.
Another new path is ridgeline inter- or extrapolation \citep{Fernex2020prf},
exploiting the topology of the control landscape.
In this study, these extensions are not included in the comparative analysis,
as the additional complexity of these methods 
with many additional tuning parameters can hardly be objectively performed.

Our first flow control benchmark is the fluidic pinball \citep{Ishar2019jfm,Deng2020jfm}.
This two-dimensional flow around three equal, parallel, equidistantly placed cylinders
can be changed by the three rotation velocities of the cylinders.
The dynamics is rich in nonlinear behaviour, 
yet geometrically simple and physically interpretable.
With suitable rotation of the cylinders many literature-known wake stabilizing
and drag-reducing mechanisms can be realized:
(1) Coanda actuation \citep{Geropp1995patent,Geropp2000ef},  
(2) circulation control (Magnus effect), 
(3) base bleed \citep{Wood1964jras}, 
(4) high-frequency forcing \citep{Thiria2006jfm}, 
(5) low-frequency forcing \citep{Glezer2005aiaaj} 
and (6) phasor control \citep{Protas2004pf}.
In this study, constant rotations are optimized for net drag power reduction
accounting for the actuation energy. 
This search space implies the first three mechanisms.
The fluidic pinball study will foreshadow key results of the Ahmed body.
This includes the drag reducing actuation mechanism 
and the visualization tools for high-dimensional search spaces.

The main application focus of this study 
is on active drag reduction behind a generic car model 
using Reynolds-averaged Navier-Stokes (RANS) simulations.
Aerodynamic drag is a major contribution of traffic-related costs,
from airborne to ground and marine traffic.
A small drag reduction would have a dramatic economic effect 
considering that transportation accounts 
for approximately $20\%$ of global energy consumption
\citep{Gadelhak2006book,Kim2011ptrsa}.  
While the drag of airplanes and ships is largely caused by skin-friction,
the resistance of cars and trucks is mainly caused by pressure or bluff-body drag.
\citet{Hucho2011book} defines bodies with a pressure drag 
exceeding the skin-friction contribution as bluff and as streamlined otherwise.

The pressure drag of cars and trucks 
originates from the excess pressure 
at the front scaling with the dynamic pressure
and a low-pressure region at the rear side of lower but negative magnitude.
The reduction of the pressure contribution from the front side 
often requires significant changes of the aerodynamic design.
Few active control solutions for the front drag reduction 
have been suggested \citep{Minelli2020jfm}.
In contrast, 
the contribution at the rearward side can significantly be changed 
with passive or active means.
Drag reductions of 10\% to 20\% are common,
\citep{Pfeiffer2014aiaa} have even achieved 25\% drag reduction with active blowing.
For a car at a speed of 120 km/h, 
this would reduce consumption by about 1.8 liter per 100 km.
The economic impact of drag reduction 
is significant for trucking fleets 
with a profit margin of only 2-3\%.
Two thirds of the operating costs are from fuel consumption.
Hence, a 5\% reduction of fuel costs from aerodynamic drag 
corresponds to over 100\% increase of the profit margin.

The car and truck design is largely determined 
by practical and aesthetic considerations.
In this study, we focus on  drag reduction 
by active means at the rearward side.
Intriguingly most drag reductions of bluff body fall 
in the categories of Kirchhoff solution and aerodynamic boat tailing.
The first strategy may be idealized by the Kirchhoff solution,
i.e.\ potential flow around the car with infinitely thin shear-layers from the rearward separation lines,
separating the oncoming flow and a dead-water region.
The low-pressure region due to curved shear-layers 
is replaced by an elongated, ideally infinitely long wake
with small, ideally vanishing curvature of the shear-layer.
Thus, the pressure of the dead water region 
is elevated to the outer pressure, i.e., the wake does not contribute to the drag.
This wake elongation is achieved by reducing entrainment through the shear-layer,
e.g. by phasor-control control mitigating vortex shedding \citep{Pastoor2008jfm} or
by energetization of the shear-layer with high-frequency actuation \citep{Barros2016jfm}.
Wake disrupters also decrease drag, 
yet by energetizing the shear layer
\citep{Park2006jfm} or delaying separation \citep{Aider2010ef}.
Arguably, the drag of the Kirchhoff solution can be considered as achievable limit
with small actuation energy.

The second strategy targets drag reduction 
by aerodynamic boat tailing.
\citet{Geropp1995patent,Geropp2000ef} 
have pioneered this approach by Coanda blowing.
Here, the shear-layer originating at the bluff body 
is vectored inward and gives thus rise to a more streamlined wake shape.
\citet{Barros2016jfm} has achieved 20\% drag reduction
of a square-back Ahmed body with high-frequency Coanda blowing 
in a high-Reynolds-number experiment.
A similar drag reduction was achieved with steady blowing
but at higher $C_\mu$ values.

This study focuses on drag reduction of the low-drag Ahmed body 
with rear slant angle of 35 degrees.
This Ahmed body idealizes the shape of many cars.
\citet{Bideaux2011ija,Gillieron2013ef} have achieved 20\% drag reduction
for this configuration in an experiment.
High-frequency blowing was applied orthogonal to the upper corner
of the slanted rear surface.
Intriguingly, the maximum drag reduction
was achieved in a narrow range of frequencies and actuation velocities
and its effect rapidly deteriorated for slightly changed parameters.
In addition, the actuation is neither Coanda blowing 
nor an ideal candidate for shear-layer energization.

The literature on active drag reduction of the Ahmed body
indicates that small changes of actuation 
can significantly change its effectiveness.
Actuators have been applied with beneficial effects 
at all rearward edges \citep{Barros2016jfm}, 
thus further complicating the optimization task.
A systematic optimization of the actuation at all edges,
including amplitudes and angles of blowing,
is beyond reach of current experiments. 
In this study, a systematic RANS optimization 
is performed in a rich parametric space
comprising the angles and amplitudes of steady blowing
of five actuator groups: one on the top, middle and bottom edge
and two symmetric actuators at the corners of the slanted and vertical surface.
High-frequency forcing is not considered,
as the RANS tends to be overly dissipative to the actuation response.

The manuscript is organized as follows. 
The employed optimization algorithms are introduced in \S~\ref{ToC:Method} 
and compared in \S~\ref{ToC:Method Comparison}.
\S~\ref{ToC:Fluidic Pinball} optimizes the net drag power for the fluidic pinball, 
which features 2-dimensional flow controlled in a three-dimensional actuation space based on DNS.
A simulation-based optimization of actuation for the three-dimensional low-drag Ahmed body  is given in \S~\ref{ToC:Ahmed}. 
Here, up to 10 actuation commands controlling the velocity and direction 
of five rearward slot actuator groups are optimized.
Our results are summarized in \S~\ref{ToC:Conclusions}.

%% file: S2.tex
\section{Optimization algorithms}
\label{ToC:Method}
In this section, 
the employed optimization algorithms
for the actuation parameters are described.
Let $J ( \bm{b})$ be the cost function---here the drag coefficient---depending 
on $N$ actuation parameters $\bm{b}=(b_1, \ldots, b_N)$ in the domain $\Omega$,
\begin{equation}
\bm{b}=[b_1,\ldots,b_N]^{\rm T} \in \Omega \subset {\cal R}^N,
\end{equation}
there the superscript `$\rm T$' denotes the transpose.
Permissible values of each parameter define an interval, $b_i \in \left [ b_{i,\rm min}, b_{i, \rm max} \right ]$, $i=1,\ldots,N$.
In other words, optimization is performed in rectangular search space,
\begin{equation}
\Omega = \left [ b_{1,\rm min}, b_{1, \rm max} \right ] \times \ldots \times 
            \left [ b_{N,\rm min}, b_{N, \rm max} \right ].
\label{Eqn:Domain}
\end{equation}
The optimization goal is to find the global minimum of $J$ in $\Omega$,
\begin{equation}
\bm{b}^{\star} = \arg\min\limits_{b \in \Omega} J(\bm{b}).
\label{Eqn:OptimizationProblem}
\end{equation}

Several common optimization methods are investigated.
Benchmark is the \emph{Downhill Simplex Method} (DSM) \citep[see, e.g.,][]{Press2007book}
as robust data-driven representative for gradient-based method (\S~\ref{ToC:Method:Simplex}).
This algorithm exploits gradient information 
from neighboring points to descent to a local minimum. 
Depending on the initial condition, this search may yield any local minimum.
In the \emph{random restart simplex (RRS) method},
the chance for finding a global minimum is increased 
by multiple runs with random initial conditions.
The geometric coverage of the search space is the focus 
of  \emph{Latin Hypercube Sampling (LHS)}  \citep[see, again,][]{Press2007book},
which optimally explores the whole domain $\Omega$
independently of the cost values, 
i.e., ignores any gradient information.
Evidently, LHS has the larger chance of getting close to the global minimum
while the simplex algorithm is more efficient in descending to a  minimum,
potentially a suboptimal one.
\emph{Monte Carlo Sampling} (MCS) \citep[see, again,][]{Press2007book}, 
is a simpler and more common exploration strategy
by taking random values for each argument,
again, ignoring any cost value information.
\emph{Genetic Algorithms} (GA) start with an MCS
in the first generation but then employ genetic operations 
to combine explorative and exploitive features in the following generations 
\citep[see, e.g.,][]{Wahde2008book}.

Sections \ref{ToC:Method:LHS}--\ref{ToC:Method:GA}
outline the non-gradient based explorative methods
from the most explorative Latin hypercube sampling, 
to Monte Carlo sampling and the partially exploitive genetic algorithm.
Sections \ref{ToC:Method:Simplex} \& \ref{ToC:Method:RRS}
recapitulate the downhill simplex method and its random restart variant.
These are commonly used methods for data-driven optimization 
with unknown analytical cost function.

\begin{figure}
	\centering
        \includegraphics[width=0.8\textwidth]{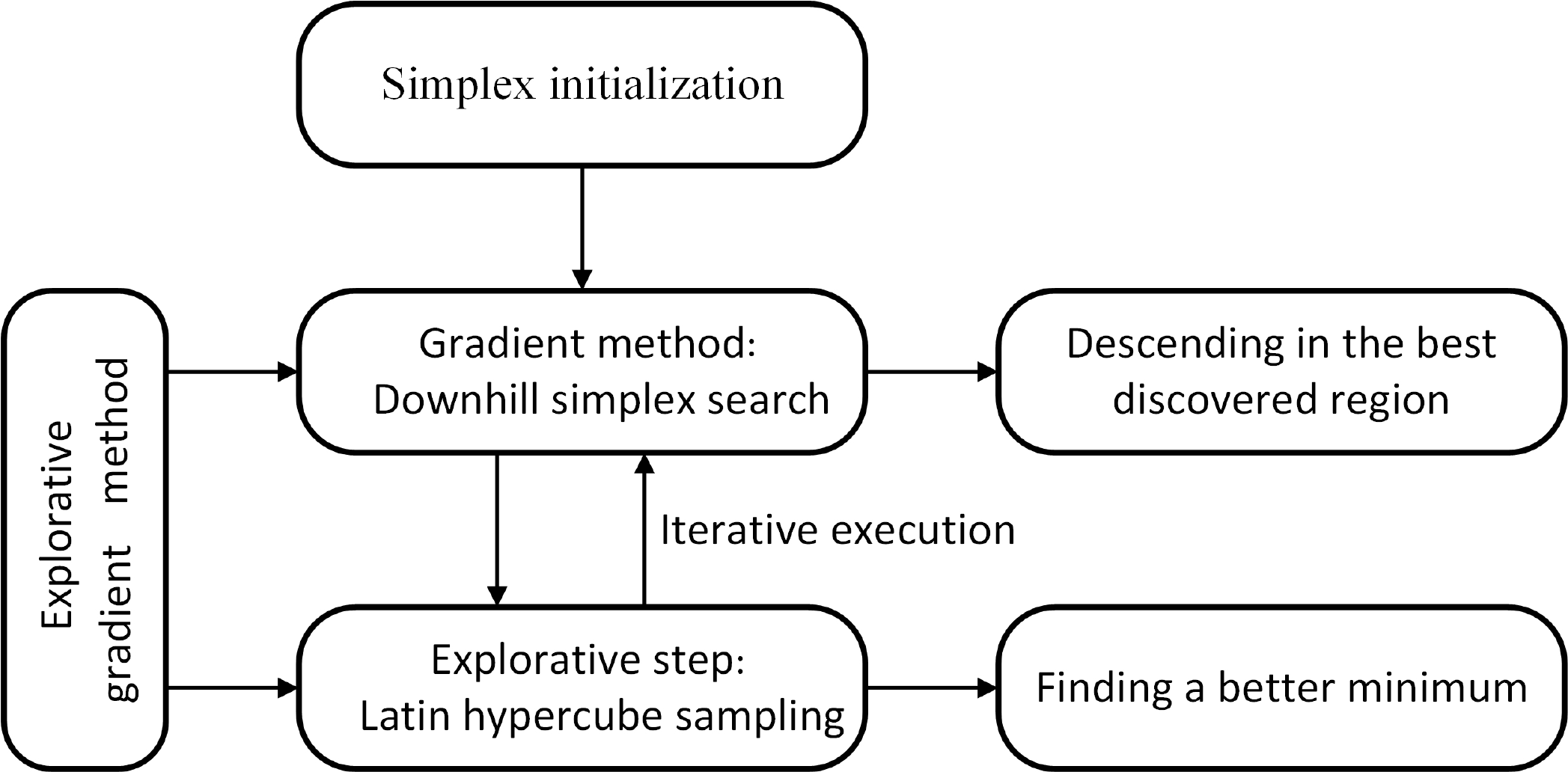}
	\caption{Sketch of the explorative gradient method. For details, see text.}
	\label{Fig:Method:EGM}
\end{figure}

In section \ref{ToC:Method:EGM},
we combine the advantages of the simplex method in exploiting a local minimum
and of the LHS in exploring the global one 
in a new \emph{explorative gradient method}
by an alternative execution (see figure \ref{Fig:Method:EGM}).
\S~\ref{ToC:Method:Accelerators} 
discusses auxiliary accelerators
which are specific to the performed computational fluid dynamics optimization.

\subsection{Latin hypercube sampling---Deterministic exploration}
\label{ToC:Method:LHS}
While our  DSM benchmark exploits neighborhood information to slide down to a local minimum,
Latin Hypercube Sampling (LHS) \citep{McKay1979t} aims to explore 
the parameter space irrespective of the cost values.
We employ a space-filling variant which effectively covers the whole permissible domain of parameters. 
This explorative strategy  (`\texttt{maximin}' criterion in Mathematica) 
minimizes the maximum minimal distance between the points:
$$ \left \{ \bm{b}_m \right \}_{m=1}^M 
   = \arg\max\limits_{\bm{b}_m \in \Omega}  
              \min\limits_{i =1,\ldots,M-1, \atop  j=i+1,\ldots, M} 
              \left \Vert \bm{b}_i -\bm{b}_j \right \Vert. $$
In other words, 
there is no other sampling of $M$ parameters
with a larger minimum distance.
$M$ can be any positive integral number.

For better comparison with the simplex algorithm,
we employ an iterative variant.
Note that once  $M$ sample points are created they cannot be augmented anymore, 
for instance when learning by LHS was not satisfactory.
We create a large number of LHS candidates $\bm{b}^\star_j$, $j=1,\ldots, M^\star$ 
for a dense coverage of the parameter space $\Omega$ at the beginning, typically $M^\star = 10^6$.
As first sample $\bm{b}_1$, the center of the initial simplex is taken.
The second parameter is taken from $\bm{b}^\star_j$, $j=1,\ldots,M^\star$
maximizing the distance to $\bm{b}_1$,
$$ \bm{b}_2 = \hbox{argmax}_{j=1,\ldots,M^\star} \Vert \bm{b}^\star_j - \bm{b}_1 \Vert. $$
The third parameter $\bm{b}_3$ is taken from the same set so that the minimal distance to $\bm{b}_1$ and $\bm{b}_2$ is maximized
and so on.
This procedure allows to recursively refine sample points
and to start with an initial set of parameters.

\subsection{Monte Carlo Sampling---Stochastic exploration}
\label{ToC:Method:MCS}
The employed space-filling variant of LHS requires the solution of an optimization problem
guaranteeing a uniform geometric coverage of the domain.
In high-dimensional domains, this coverage many not be achievable. 
A much easier and far more commonly used exploration strategy is Monte Carlo Sampling (MCS).
Here, the $m$th sample $\bm{b}_m = \left [ b_{1,m}, \ldots, b_{N,m}  \right]^{\rm T}$ 
is given by 
\begin{equation}
b_{i,m} = b_{i,\rm min} + \zeta_{i,m} \left ( b_{i, \rm max} - b_{i, \rm min} \right),
\label{Eqn:MCS}
\end{equation}
$\zeta_{i,m} \in [0,1]$ are random numbers with uniform probability distribution in the unit domain.
The relative performance between LHS and MCS is a debated topic.
We will wait for the results for an analytical problem in section \ref{ToC:Method Comparison}.

\subsection{Genetic algorithm---Biologically inspired exploration and exploitation}
\label{ToC:Method:GA}
The Genetic Algorithm (GA) mimics natural selection process.
We refer to \citet{Wahde2008book} as excellent reference.
In the following, the method is briefly outlined to highlight the specific version and the chosen parameters.

Any parameter vector $  \bm{b} = [b_1, b_2, \ldots, b_N]^{\rm T} \in \Omega \subset {\cal R}^N$
comprises the real values $b_i$, also called \emph{alleles}.
This real value is encoded as a binary number and called \emph{gene}. 
The \emph{chromosome} comprises alle genes and 
represents the parameter vector  \citep{wright1991genetic}.


The genetic algorithm evolves one generation of $I$ parameters, also called \emph{individuals},
into a new generation with the same number of parameters using biological inspired genetic operations.
The first generation is based on MCS, i.e., represents completely random genoms.
The individuals $J_i^1$, $i=1,\ldots,I$ are evaluated and sorted by their costs
$$ J_1^1 \le J_2^1 \le \ldots \le J_I^1.$$

The next generation is computed with \emph{elitism}  and two genetic operations.
Elitism copies the $N_e$ best performing individuals in the new generation.
$P_e = N_e/I$ denotes the relative quota.
The two genetic operations include
\emph{mutation}, which randomly changes parts of the genom, and
\emph{crossover}, which randomly exchanging parts of the genoms of two individuals.
Mutation serves explorative purposes and crossover has the tendency to breed better individuals.
In an outer loop,
the genetic operations are randomly chosen with probabilities $P_m$ and $P_c$  
for  mutation and crossover,  respectively.
Note that $P_e+P_m+P_c=1$ by design.

In the inner loop, i.e., after the genetic operation is determined,
individuals from the current generation are chosen.
Higher performing individuals have higher probability to be chosen.
Following the genetic algorithm matlab routine, 
this probability is proportional to the inverse square-root of its relative rank $p \propto 1/\sqrt{i}$.  

The genetic algorithm terminates according to a predetermined stop criterion,
here a maximum number of generations $L$ or corresponding number of evaluations $M=IL$.
For reasons of comparison, we renumber the individuals in the order of their evaluation,
i.e., $m \in \{1,\ldots, I\}$ belongs to the first generation,
$m \in \{I+1, \ldots, 2I \}$ to the second generation, etc.

The chosen parameters are the default values of matlab,
e.g. $P_r=0.05$ $P_c=0.8$, $P_m=0.15$, $N_e = 3$.
Further details are provided in the appendix \ref{ToC:GA:Matlab}.

\subsection{Downhill simplex search---A robust gradient method}
\label{ToC:Method:Simplex}
The Downhill Simplex Method (DSM) by \citet{Nelder1965jc} 
is a very simple, robust and widely used gradient method.
This method does not require any gradient information 
and is well suited for expensive function evaluations, 
like the considered RANS simulation for the drag coefficients,
and for experimental optimizations with inevitable noise.
A price is a slow convergence for the minimization of smooth functions
as compared to algorithms which can exploit gradient and curvature information.

We briefly outline the employed downhill simplex algorithm,
as there are many variants.
First, $N+1$ vertices $\bm{b}_m$, $m=1,\ldots,N+1$ in $\Omega$ 
are initialized as detailed in the respective sections.
Commonly, $\bm{b}_{N+1}$ is placed somewhere in the middle of the domain
and the other vertices explore steps in all directions,
$\bm{b}_m = \bm{b}_{N+1} + h \> \bm{e}_{m}$, $m=1,\ldots,N$.
Here,  $\bm{e}_i = [\delta_{i1}, \ldots, \delta_{iN}]^{\rm T}$ 
is a unit vector in $i$-th direction
and $h$ is a step size which is small compared to the domain.
Evidently, all vertices must remain in the domain $\bm{b}_m \in \Omega$.

The goal of the simplex transformation iteration 
is to replace the worst argument $\bm{b}_h$ of the considered simplex
by a new better one $\bm{b}_{N+2}$.
This is archived in following steps:
\begin{description}
\item[1) Ordering: ]
Without loss of generality, we assume that the vertices are 
sorted in terms of the cost values 
$J_m = J(\bm{b}_m)$: $J_1 \le J_2 \le \ldots \le J_{N+1}$.
\item[2) Centroid: ] In the second step,
the centroid of the best side opposite to the worst vertex $\bm{b}_{N+1}$ is computed:
$$ 
\bm{c} =\frac{1}{N} \sum\limits_{m=1}^{N} \bm{b}_m.
$$
\item[3) Reflection: ] 
Reflect the worst simplex $\bm{b}_{N+1}$ at the best side,
$$ \bm{b}_r = \bm{c} +  \left( \bm{c} - \bm{b}_{N+1} \right) $$
and compute the new cost $J_r=J(\bm{b}_{r})$. 
Take $\bm{b}_{r}$ as new vertex, 
if $J_1 \le J_r \le J_{N}$.
$\bm{b}_m$, $m=1\ldots,N$ and $\bm{b}_r$ 
define the new simplex for the next iteration.
Renumber the indices to the $1 \ldots N+1$ range.
Now, the cost is better than the second worst value $J_N$,
but not as good as the best one $J_1$.
Start a new iteration with step 1.
\item[4) Expansion: ] 
If $J_{r} < J_1$, 
expand in this direction further by a factor $2$,
$$ 
\bm{b}_{e}  = \bm{c} +  2  \left(\bm{b}_{N+1} - \bm{c} \right) . 
$$
Take the best vertex of $\bm{b}_r$ and $\bm{b}_e$  
as  $\bm{b}_{N+1}$ replacement
and start a new iteration.
\item[5) Single contraction: ] 
At this stage,  $J_{r} \ge J_{N}$.
Contract the worst vertex half-way towards centroid,
$$ \bm{b}_c = \bm{c} + \frac{1}{2} \left( \bm{c} - \bm{b}_{N+1} \right). $$ 
Take $\bm{b}_c$ as new vertex ($\bm{b}_{N+1}$ replacement), 
if it is better than the worst one, i.e., $J_c \le J_{N+1}$.
In this case, start the next iteration.
\item[6) Shrink / multiple contraction: ] 
At this stage,  none of the above operations was successful.
Shrink the whole simplex by a factor $1/2$ towards the best vertex,
i.e., replace all vertices by
$$ 
\bm{b_m} \mapsto \bm{b}_1 + \frac{1}{2} \> \left( \bm{b}_m-\bm{b}_1\right), m=2,\ldots,N+1.  
$$
This shrinked simplex represents the one for the next iteration.
It should be noted that this shrinking operation is the last resort
as it is very expensive with $N$ function evaluations.
The rationale behind this shrinking is that a smaller simplex
may better follow local gradients.
\end{description}

\subsection{Random restart simplex method---Preparing for multiple minima}
\label{ToC:Method:RRS}

The downhill simplex method of the previous section may be equipped with a random restart initialization \citep{Humphrey2000jc}.
As random initial condition, we chose a Monte Carlo sample as main vertex of the simplex
and explore all coordinate directions by a positive shift of 10\% of the domain size.
It is secured that all vertices are inside the domain $\Omega$.
These initial simplexes attribute the same probability to the whole search space.
The chosen small edge length makes a locally smooth behaviour probable---in absence of any other information.
The downhill search is stopped after a fixed number of evaluations.
We chose 50 evaluations as safe upper bound for convergence.
It should be noted that the number of simplex iterations is noticeably smaller,
as one iteration implies one to $N+2$ evaluations.

Evidently, the random restart algorithm may be improved 
by appreciating the many recommendations of literature, 
e.g., avoiding  closeness to explored parameters. 
We trade these improvements in all optimization strategies
for simplicity of the algorithms.

\subsection{Explorative gradient method--Combining exploration and gradient method}
\label{ToC:Method:EGM}
In this section, 
we combine the advantages of the exploitive DSM
and the explorative LHS in a single algorithm.
\begin{description}	
\item[Step 0---Initialize. ] First, $\bm{b}_{m}$, $m=1,\ldots,M+1$ 
are initialized for the DSM.
\item[Step 1---Downhill simplex. ] 
Perform one simplex iteration (\S\ \ref{ToC:Method:Simplex}) 
with the best $M+1$ parameters discovered so far.

\item[Step 2---LHS. ] Compute the cost $J$ of a new LHS parameter $\bm{b}$.
As described above, 
we take a parameter from a precomputed list which is the furthest away 
from all hitherto employed parameters.
\item[Step 3---Loop. ] Continue with Step 1 until a convergence criterion is met.
\end{description}
Sometimes, the simplex may degenerate to one with small volume,
for instance, when it crawls through a narrow valley.
In this case, the vertices lie in a subspace
and valuable gradient information is lost.
This degeneration is diagnosed and cured after step 1 as follows.
Let $\bm{b}_c$ be the geometric center of the simplex.
Compute the distance $D$ between each vertex and their geometric center point. 
If the minimum ${D}_{\rm min}$ is smaller than half the maximum distance ${D}_{\rm max}/2$, 
the simplex is deemed degenerated.
This degeneration is removed as follows.
Draw a sphere with ${D}_{max}$ around the simplex center.
This sphere contains all vertices by construction.
Obtain $1000$ random points in this sphere. 
Replace the vertex with the highest cost $J_{max}$
with one of these point to create simplex with the largest volume.
Of course, the cost of this changed point needs to be evaluated.

The algorithm is intuitively appealing.
If the LHS discovers a parameter with a cost $J$ in the top $M+1$ values,
this parameter is included in the new simplex 
and corresponding iteration may slide down in another better minimum.
It should be noted that LHS exploration 
does not come with the toll of 
having to evaluate the cost at $N+1$ vertices
and subsequent iterations.
The downside of a single evaluation is that 
we miss potentially important gradient information
pointing to an unexplored much better minimum.
Relative to random-restart gradients searches requiring a many evaluations for a converging iteration, 
LHS exploration becomes increasingly better in rougher landscapes,
i.e., more complex multi-modal behaviour.

\subsection{Computational accelerators}
\label{ToC:Method:Accelerators}
The RANS based optimization may be accelerated by enablers 
which are specific to the chosen flow control problem.
The computation time for each RANS simulation is based 
on the choice of the initial condition,
as it affects the convergence time for the steady solution.
The first simulation of an optimization starts with the unforced flow as initial condition.
The next iterations exploit 
that the  averaged velocity field $\overline{\bm{u}}(\bm{x})$
is a function of the actuation parameter $\bm{b}$.
The initial condition of the $m$th simulation 
is obtained with the 1-nearest-neighbour approach:
The velocity field associated with the closest hitherto computed actuation vector
is taken as initial condition for the RANS simulation.
This simple choice of initial condition saves 
about 60$\>$\% CPU time in reduced convergence time.

Another 30$\>$\% reduction of the CPU time
is achieved by avoiding RANS computations with very similar actuations.
This is achieved by a quantization of the $\bm{b}$ vector: 
The actuation velocities are quantized with respect to integral $\rm m/s$ values.
This corresponds to increments of $U_{\infty}/30$  with  $U_{\infty}=30 \> \> \rm m/s$.
All actuation vectors are rounded with respect to this quantization.
If the optimization algorithm yields a rounded actuation vector 
which has already been investigated, 
the drag is taken from the corresponding simulation
and no new RANS simulation is performed.
Similarly, the angles are discretized into integral degrees.

%% file: S3.tex
\section{Comparative optimization study}
\label{ToC:Method Comparison}

In this section, the six optimization methods of \S~\ref{ToC:Method}
are compared for an analytical function with 4 local minima.
\S~\ref{ToC:ToyModel} describes this function.
In \S~\ref{ToC:ToyParameters}, the optimization methods with corresponding parameters are discussed.
\S~\ref{ToC:ToyVisualization} shows the tested individuals.
The learning rates are detailed in \S~\ref{ToC:ToyLearning}.
Finally, the results are summarized (\S~\ref{ToC:ToyDiscussion}).

\subsection{Analytical function}
\label{ToC:ToyModel}
The considered analytical cost function
\begin{equation} 
\begin{split}
J(b_1, b_2) = 1 & \>\>-\>\>             e^{ -2 (b_1-1)^2 - 2 (b_2-1) ^2 }
                  - \frac{1}{2} e^{ -2 (b_1+1)^2 - 2 (b_2-1) ^2 }\\
                & - \frac{1}{3} e^{ -2 (b_1-1)^2 - 2 (b_2+1) ^2 }
                 - \frac{1}{4} e^{ -2 (b_1+1)^2 - 2 (b_2+1) ^2 }        
\end{split}                                   
\label{Eqn:Cost:Toy}
\end{equation}
is characterized 
by a global minimum near $[1,1]^{\rm T}$ and 
three local minimum separately near $[1,-1]^{\rm T}$, $[-1,1]^{\rm T}$, and $[-1,-1]^{\rm T}$.
The cost reaches a plateau  $J=1$ far away from the origin.
The investigated parameter domain is $\Omega = [-3,3] \times [-3,3]$.

\subsection{Optimization methods and their parameters}
\label{ToC:ToyParameters}

Latin hypercube sampling (LHS) is performed as described in \S~\ref{ToC:Method:LHS}.
We take  $M^\star = 10^3$ random points for the optimization of the coverage.
The Monte Carlo sampling (MCS) is uniformly distributed 
over the parameter domain $\Omega$.

The most important parameters of genetic algorithms (GA) are summarized 
from appendix \ref{ToC:GA:Matlab}:
The generation size  is $I = 50$ and the iterations 
are terminated with generation $L = 20$.
The crossover and mutation probabilities 
are $P_c = 0.80$ and $P_m = 0.15$, respectively.
The number of elite individuals $N_e=3$ correspond to the probability 
complementary probability $P_e = 5\>\%$.

The downhill simplex method follows exactly the description of \S~\ref{ToC:Method:Simplex}
with an expansion rate of $2$, single contraction rate of $1/2$ and a shrink rate of $1/2$.
The random restart variant (RRS, \S~\ref{ToC:Method:RRS}) has an evaluation limit of 50 for 20 random restarts.
The step size for each initial simplex is  $h=0.35$.
The explorative gradient method (EGM) builds on the LHS and downhill simplex iterations discussed above.

\subsection{Tested individuals in the parameter space}
\label{ToC:ToyVisualization}

\begin{figure}[htb]
	\centering
	\subfloat{
		\includegraphics[width=0.4\textwidth]{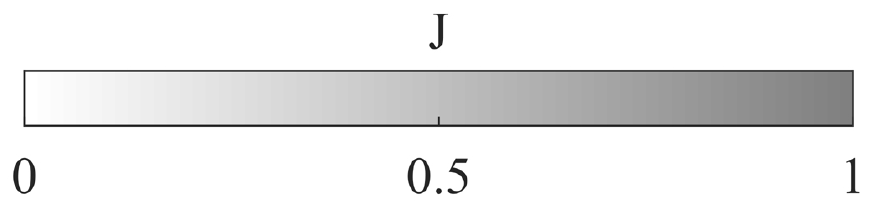}}
  
	\setcounter{subfigure}{0}
		\subfloat[]{
		\label{Fig:ToySystem:tested individuals:a}
		\includegraphics[width=0.31\textwidth]{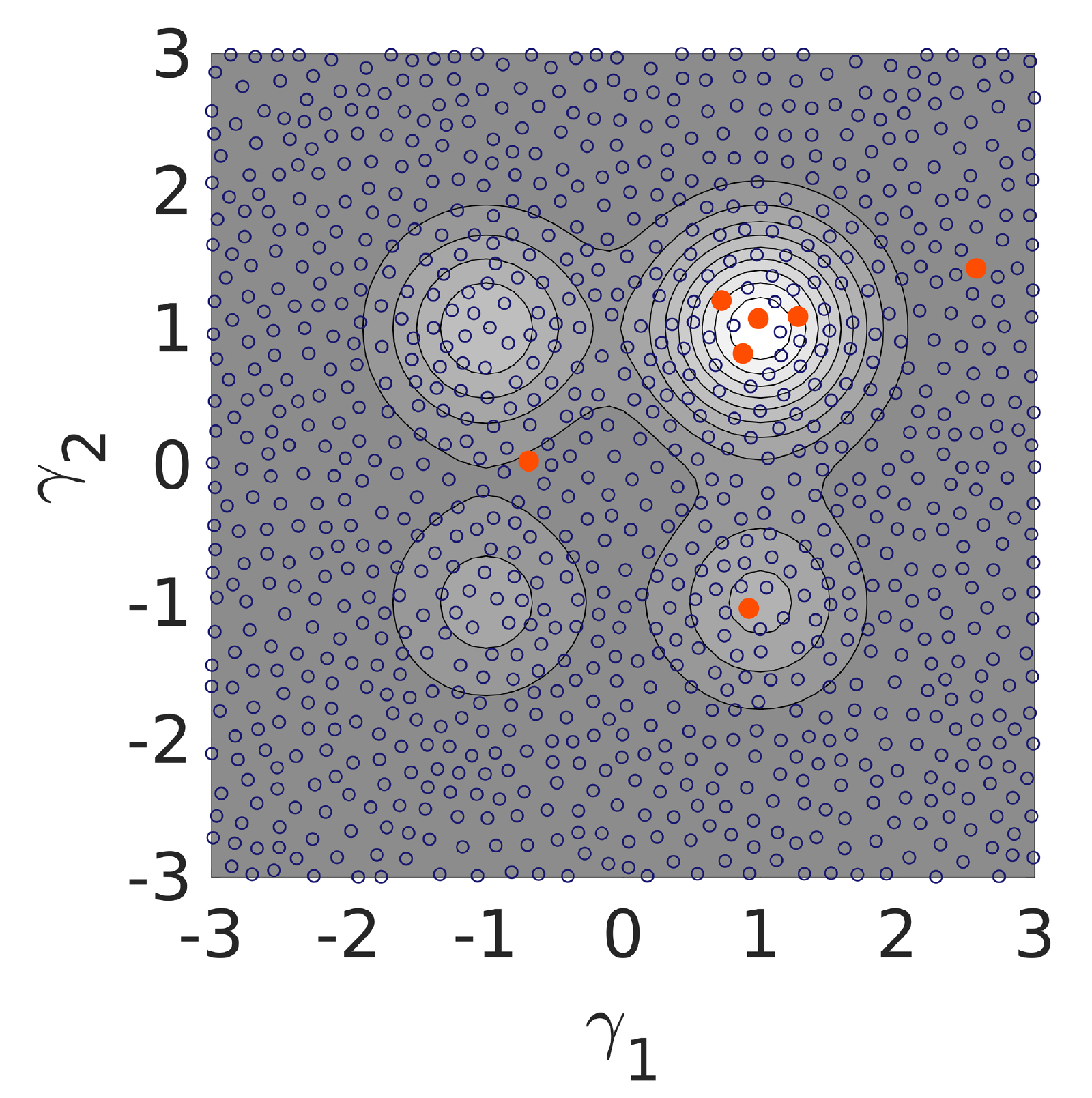}}
	\hfil
	\subfloat[]{
		\label{Fig:ToySystem:tested individuals:b} 
		\includegraphics[width=0.31\textwidth]{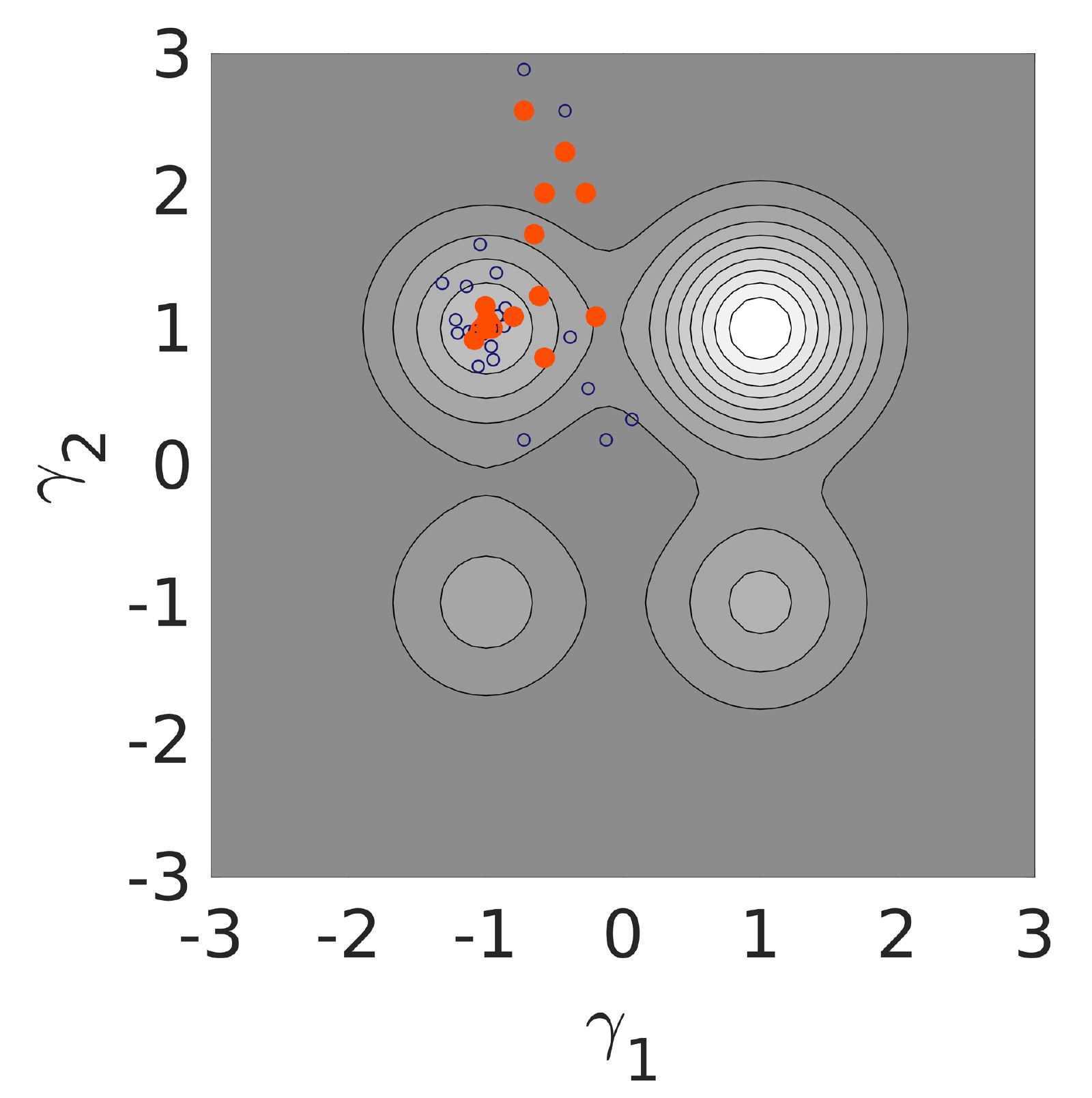}}
  
	\subfloat[]{
		\label{Fig:ToySystem:tested individuals:c} 
		\includegraphics[width=0.31\textwidth]{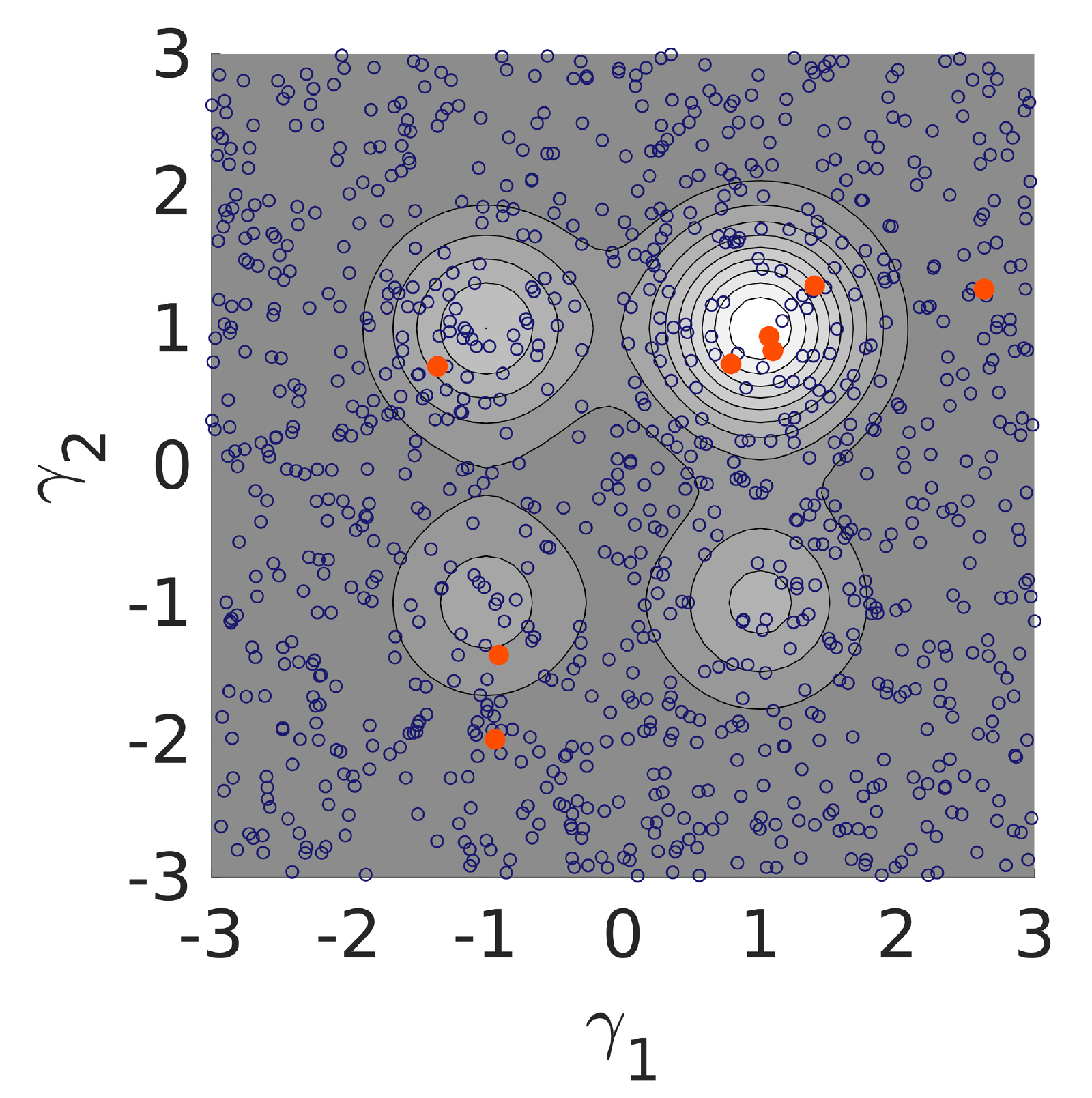}}
	\hfil
	\subfloat[]{
		\label{Fig:ToySystem:tested individuals:d}  
		\includegraphics[width=0.31\textwidth]{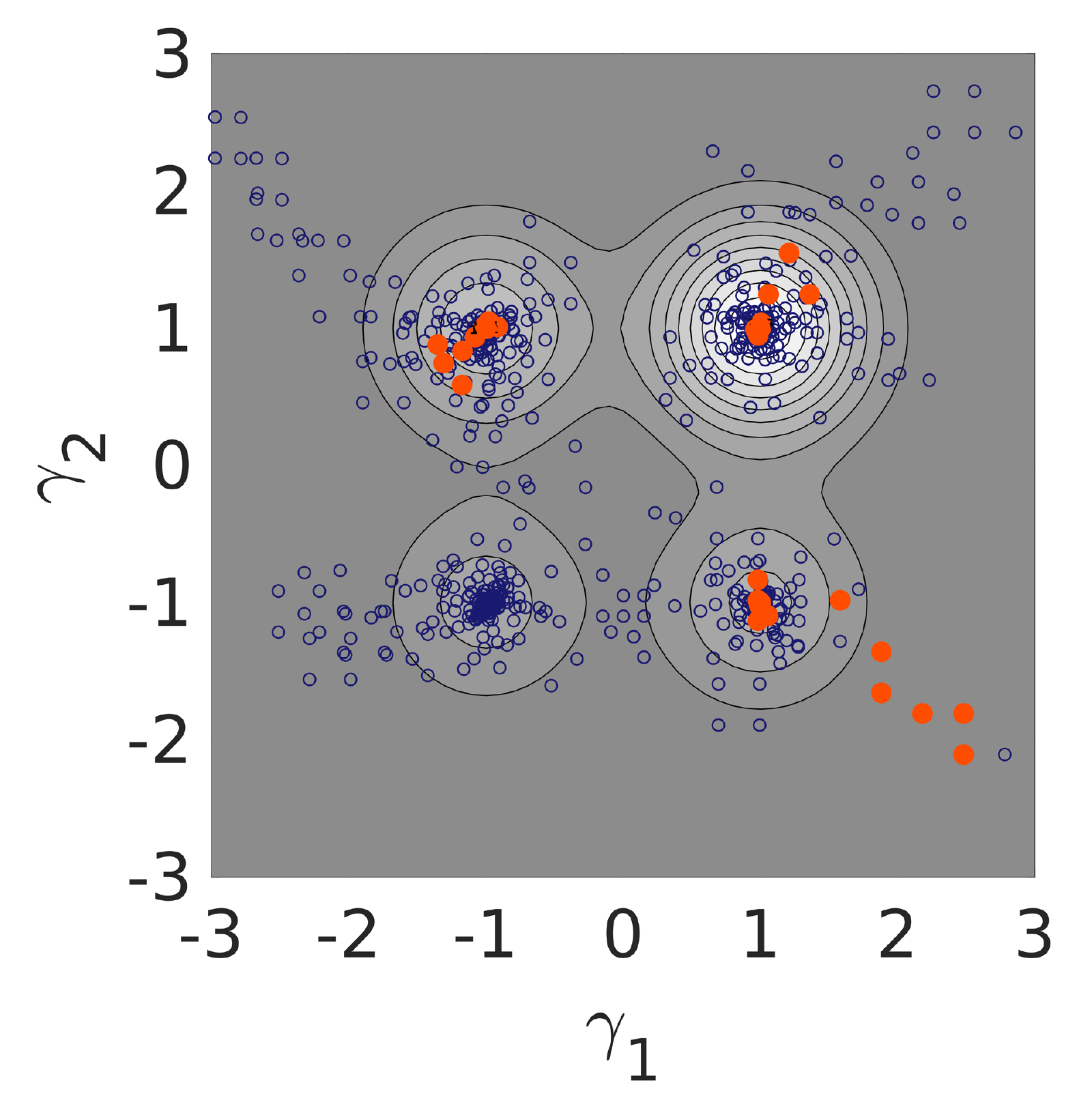}}

	\subfloat[]{
		\label{Fig:ToySystem:tested individuals:e} 
		\includegraphics[width=0.31\textwidth]{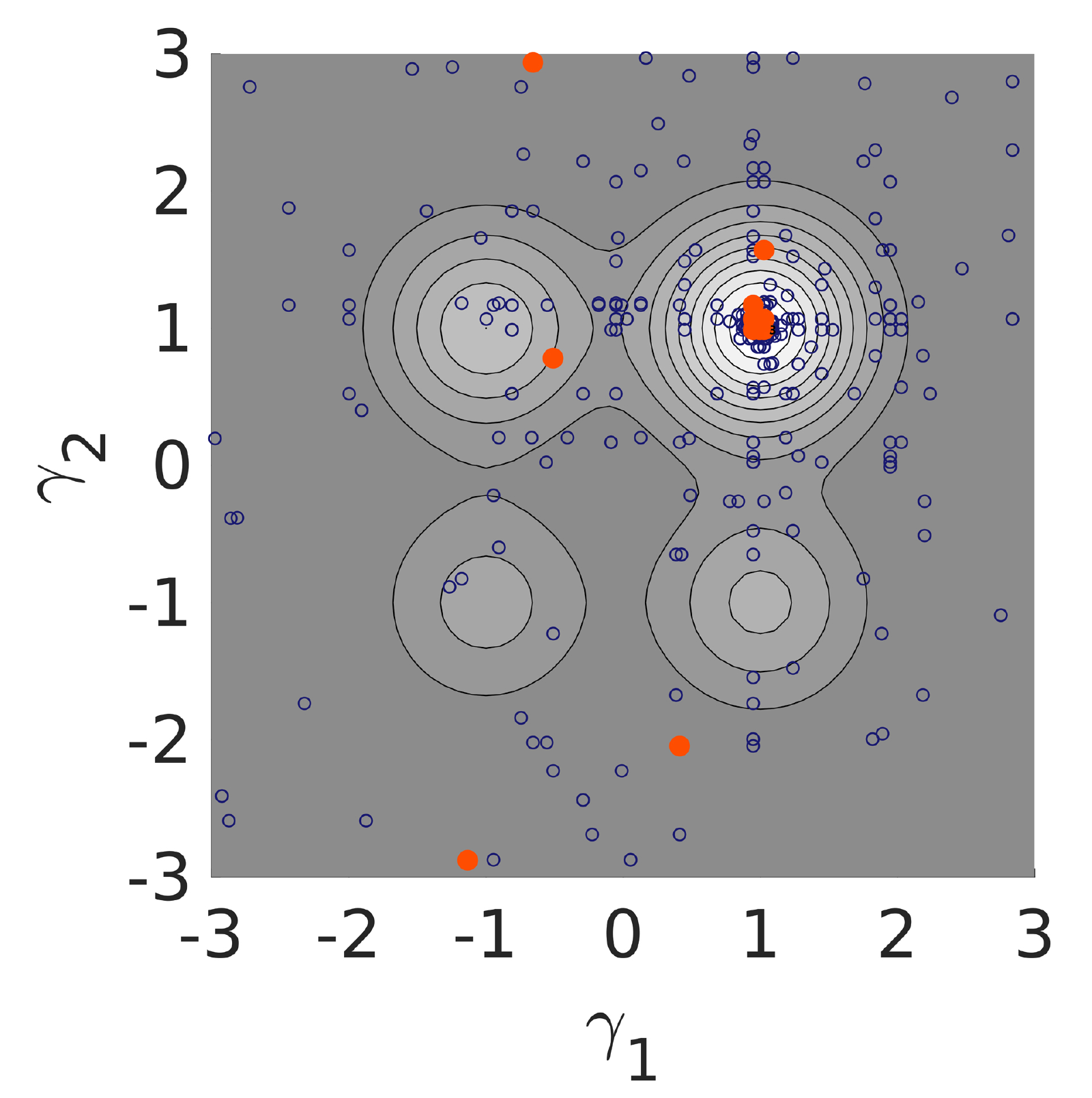}}
	\hfil
	\subfloat[]{
		\label{Fig:ToySystem:tested individuals:f} 
		\includegraphics[width=0.31\textwidth]{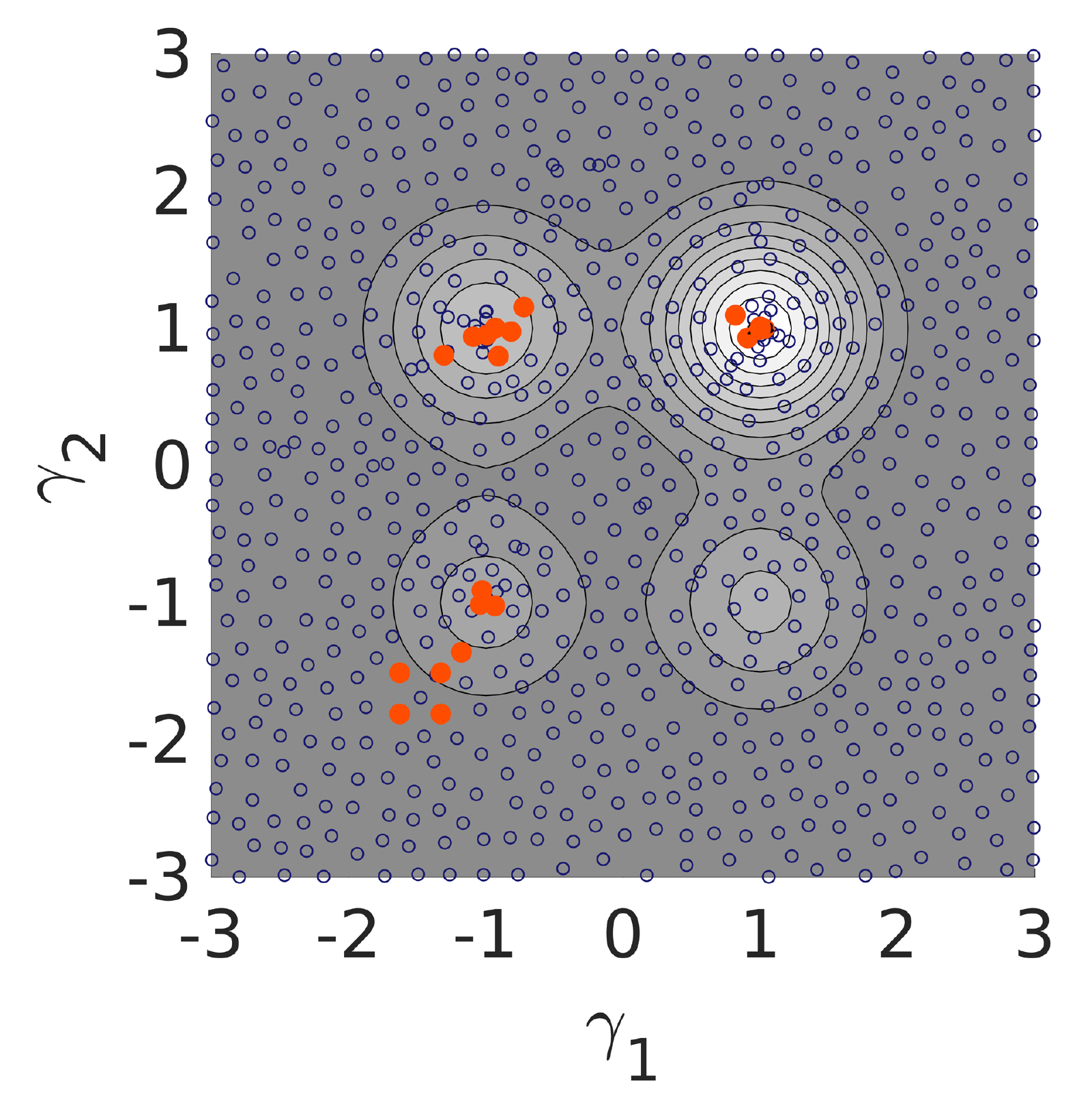}}	  
	\caption{Comparison of all optimizers for the analytical function \eqref{Eqn:Cost:Toy}.
                 Tested individuals of \textit{a}) LHS, \textit{b}) DSM, 
			\textit{c}) MCS, \textit{d}) RRS, \textit{e}) GA 
			and \textit{f}) EGM from a typical optimization with 1000 individuals.
			The red solid circles mark new local minima during the iteration 
			while the blue open circles represent suboptimal tested parameters.}
	\label{Fig:ToySystem:tested individuals}
\end{figure}
Figure \ref{Fig:ToySystem:tested individuals} illustrates the iteration
of all six algorithms in the parameter space.
LHS shows a uniform coverage of the domain.
In contrast, MCS leads to local `lumping' of close individuals, i.e., indications of redundant testing,
and local untested regions, both undesirable features. 
Thus, LHS is clearly seen to perform better than MCS.
The genetic algorithm is seen to sparsely test the plateau
while densely populating the best minima.
This is clearly a desirable feature over LHS and MCS.

The standard downhill simplex method converges to a local minimum
in this realization, while the random restart variant  (RRS) finds all minima,
including the global optimal one.
Clearly, the random restart initialization 
is a security policy against sliding into a suboptimal minimum.
The proposed new explorative gradient method (EGM)
finds all four minima and converges against the global one.
By construction, the exploration is less dense as LHS.
The 1000 iterations comprise about 250 LHS steps
and about 250 downhill simplex iterations with an average of 3 evaluations for each.

Arguably, the EGM is seen to be superior to all downhill simplex variant 
with more dense exploration and convergence to the global optimum.
EGM also performs better than LHS, MCS, and genetic algorithm,
as it invests in a more dense coverage of the parameter domain
while about $75\>\%$ of the evaluations serve the convergence.

The conclusions are practically independent 
of the chosen realization of the optimization algorithm,
except that the downhill simplex method slides 
into the global minimum in about $27\>\%$ of the cases.

We note that some of our conclusions are tied 
to the low dimension of the parameter space.
In, cubical domain of 10 dimensions,
the first $2^{10}=1024$ LHS individuals 
would populate the corners before the interior is explored.
A geometric coverage of higher dimensions is incompatible 
with a  budget of 1000 evaluations.

\subsection{The learning curve}
\label{ToC:ToyLearning}

\begin{figure}[htb]
	\centering
  
	\subfloat[]{
		\label{Fig:ToySystem:learning curve:a}
		\includegraphics[width=0.45\textwidth]{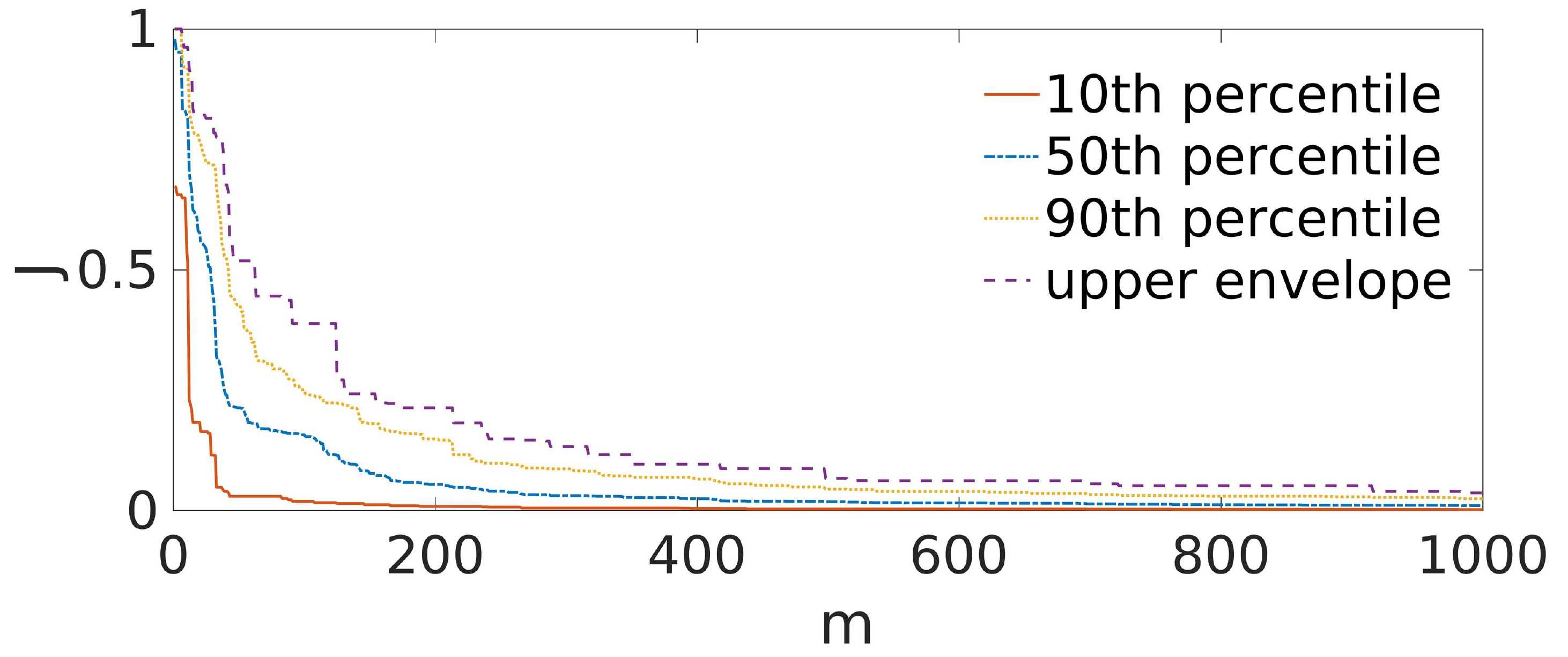}}
	\hfil
	\subfloat[]{
		\label{Fig:ToySystem:learning curve:b} 
		\includegraphics[width=0.45\textwidth]{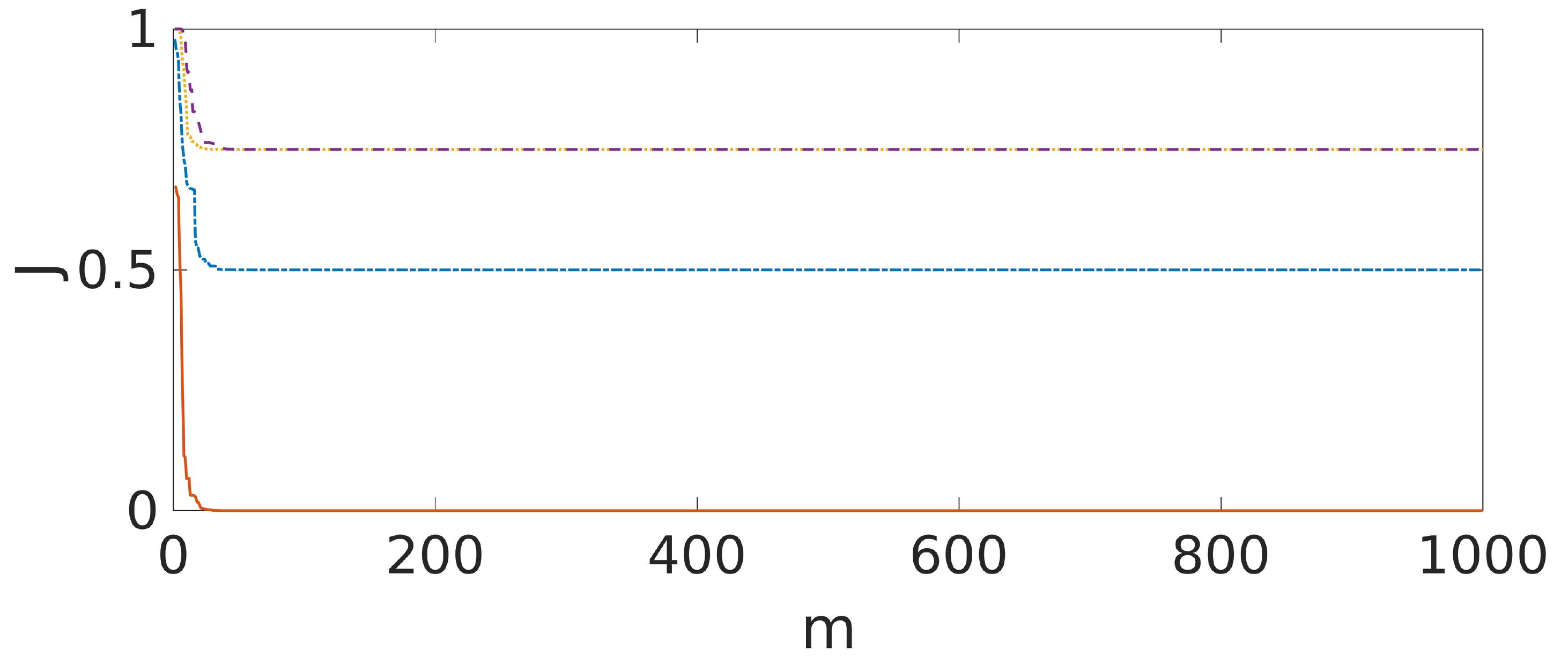}}
  
	\subfloat[]{
		\label{Fig:ToySystem:learning curve:c} 
		\includegraphics[width=0.45\textwidth]{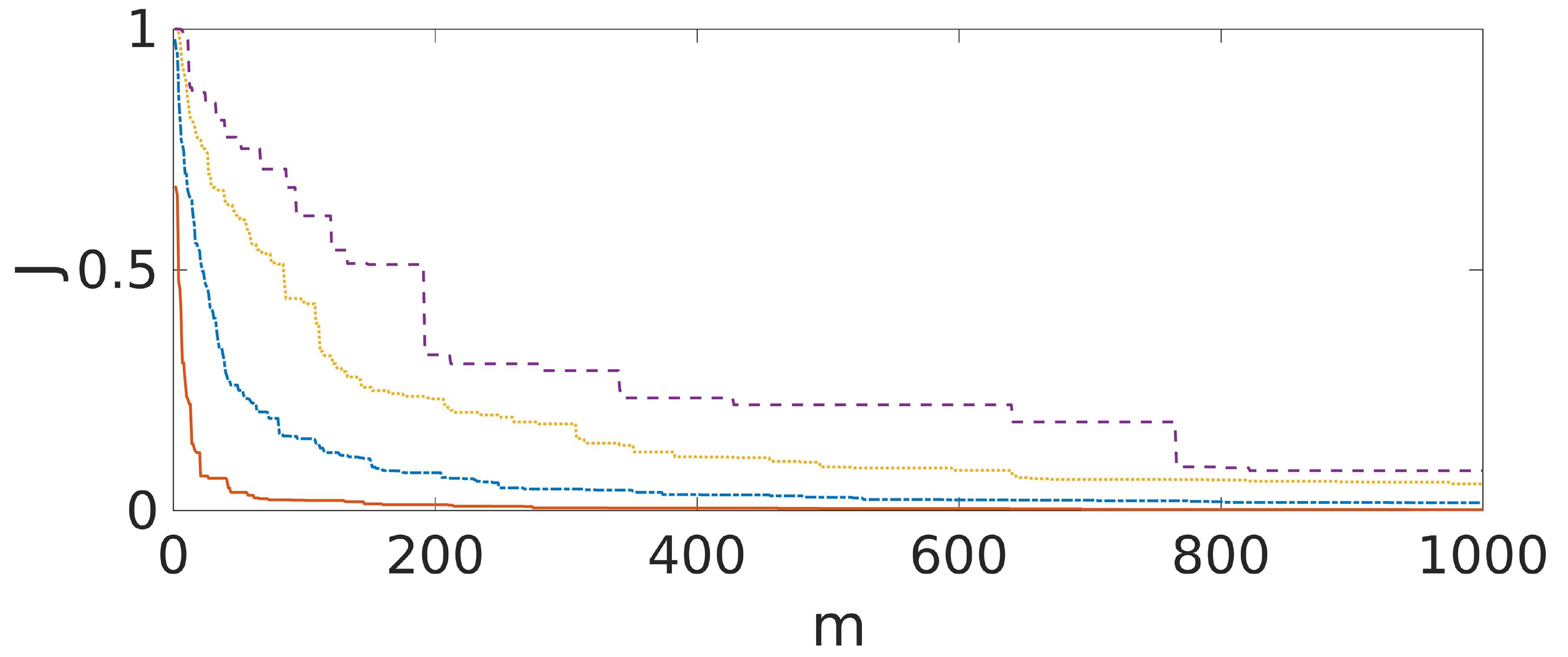}}
	\hfil
	\subfloat[]{
		\label{Fig:ToySystem:learning curve:d}  
		\includegraphics[width=0.45\textwidth]{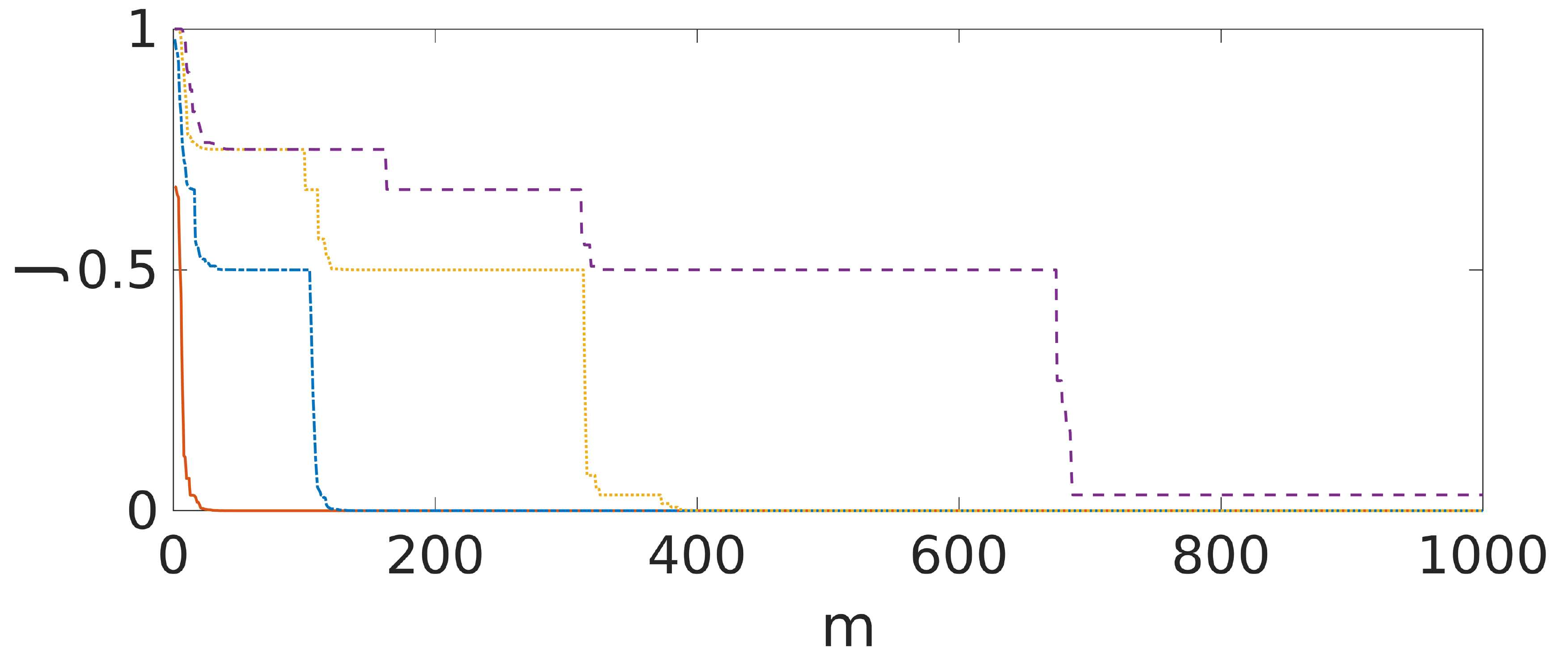}}

	\subfloat[]{
		\label{Fig:ToySystem:learning curve:e} 
		\includegraphics[width=0.45\textwidth]{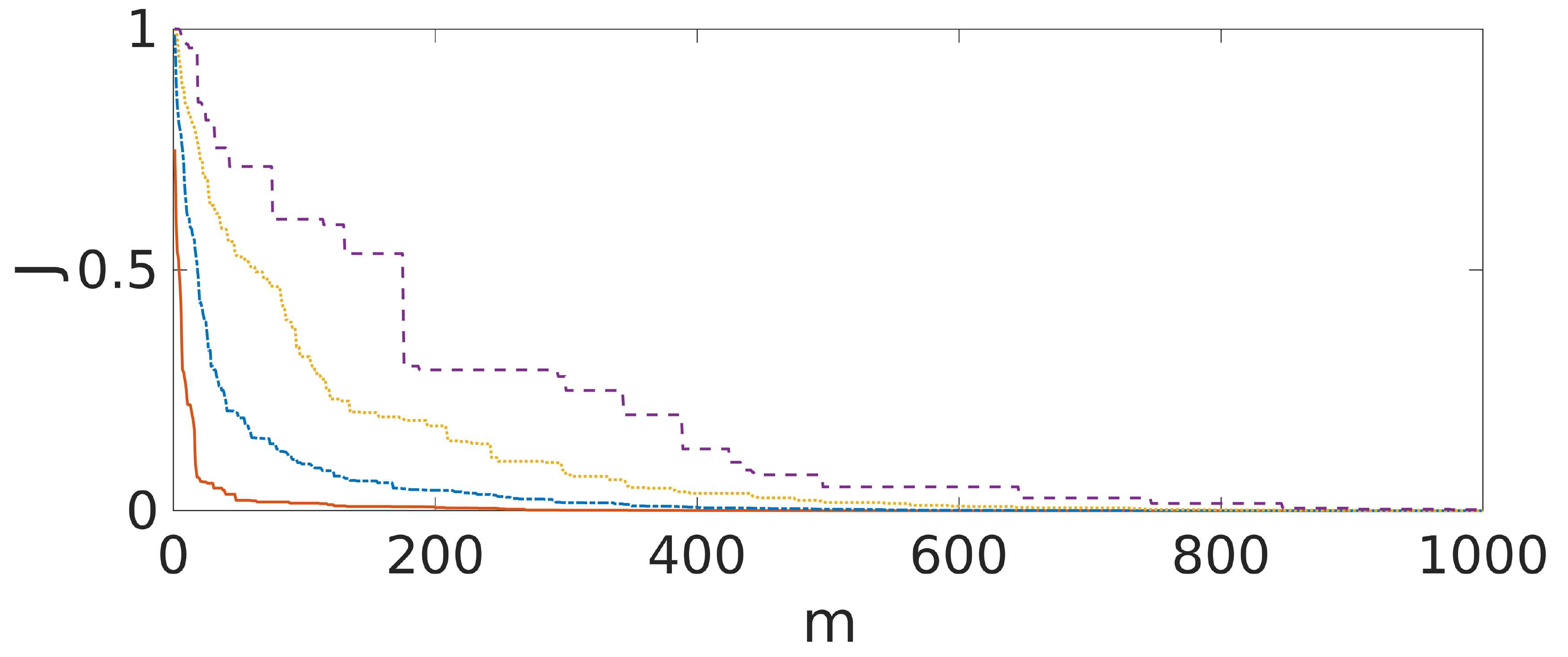}}
	\hfil
	\subfloat[]{
		\label{Fig:ToySystem:learning curve:f} 
		\includegraphics[width=0.45\textwidth]{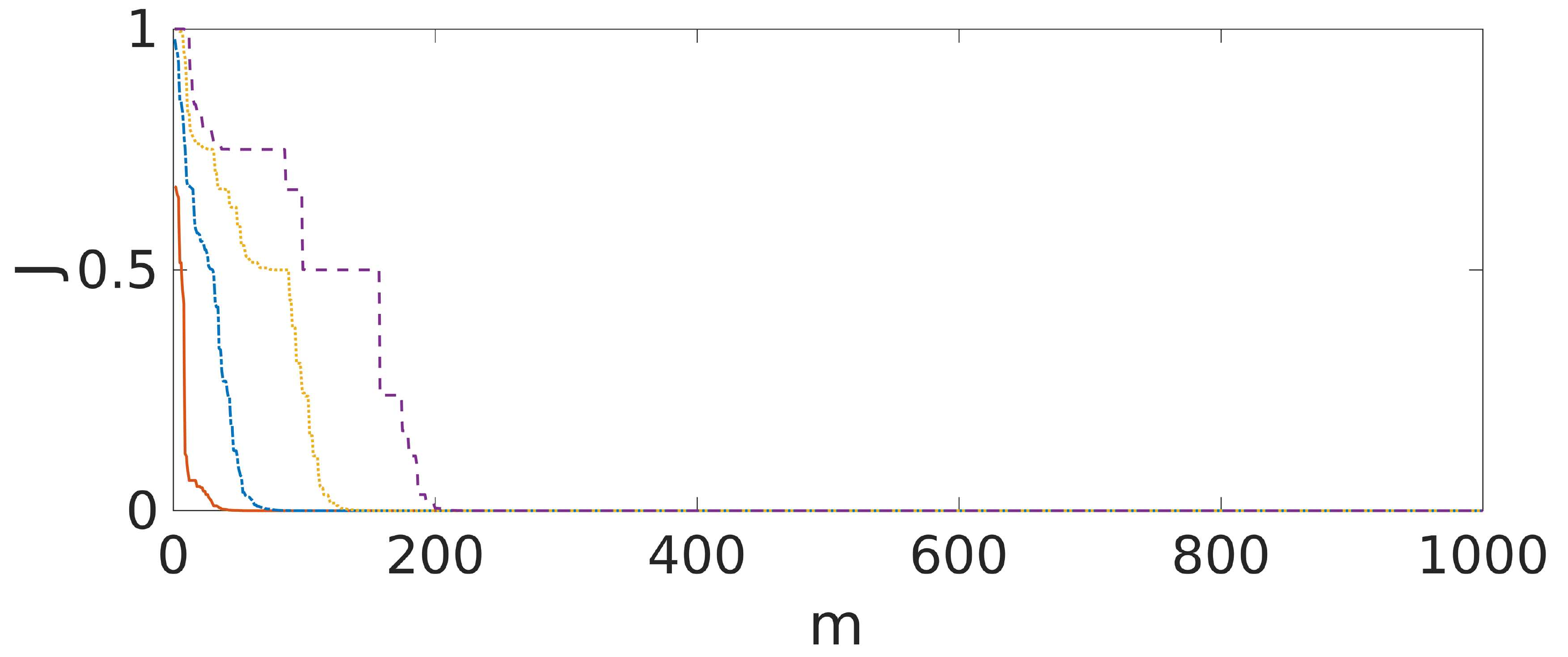}}	  
	\caption{Comparison of all optimizers for the analytical function \eqref{Eqn:Cost:Toy}.
	         Learning curve of (\textit{a}) LHS, (\textit{b}) Simplex, 
			(\textit{c}) Monte Carlo, (\textit{d}) RRS, (\textit{e}) GA
			and (\textit{f}) EGM in 100 runs. $10th$, $50th$, and $90th$ percentile 
			indicates the J value below which 10, 40 and 90 percentage of runs 
			at current evaluation falls.}
	\label{Fig:ToySystem:learning curve}
\end{figure}
In figure \ref{Fig:ToySystem:learning curve}, 
we investigate the learning curve of each algorithm
for 100 realizations with randomly chosen initial conditions.
The learning curve shows the best cost value found with $m$ evaluations.
In this statistical analysis, 
 the $10$, $50$ and $90\>\%$ percentiles of the learning curves are displayed.
The $10\>\%$ percentile at $m$ evaluations implies
that $10\>\%$ of the realizations yield better and $90\>\%$ yield worse cost values. 
The $50\>\%$ and $90\>\%$ percentiles are defined analogously.

The gradient-free algorithms (LHS, MC, GA) in the left column show smooth learning curves.
All iterations eventually converge against the global optimum as seen from the upper envelope.
The $10\>\%$ and $50\>\%$ percentile curves are comparable.
Focusing on the bad case ($90\>\%$ percentile) and worst case performance (upper envelope),
LHS is seen to beat both MCS and GA.
MCS has the worst outliers, because it neither exploits the cost function, like the GA,
nor comes with advantage of guaranteed good geometric coverage, like LHS.

The gradient-based algorithms reveal other  features.
The downhill simplex method can arrive at the global optimum 
much faster than any of the gradient-free algorithms.
But is has also a $73\>\%$ probability terminating in one of the suboptimal local minima.
The random restart version mitigates this risk to practically zero.
In RRS, $50\>\%$ of the runs reach the minimum before 300 evaluations.

The learning curve of all gradient-based algorithms have jumps.
Once the initial condition is in the attractive basin of one minimum,
the convergence to that minimum is very fast, leading to a step decline of the learning curve.
We notice that the worst case scenario does not exactly converge to zero.
The reason is the degeneration of the simplex to points 
on a line which does not go through the global minimum.
Only the EGM takes care of this degeneration
with a geometric correction after step 1 as described in \S~\ref{ToC:Method:EGM}. 
Expectedly, EGM also outperforms all other optimizers with respect to $50\>\%$ percentile, $90\>\%$ percentile and the worst case scenario.
The global minimum is consistently found in less than 200 evaluations.
It is much more efficient to invest 50 points in LHS exploration than to iterate into a suboptimal minimum.
The price to be paid with EGM is that the best case performance is mitigated by gradient descent 
which is distracted by roughly $25\>\%$ LHS iterations as insurance policy.

\subsection{Discussion}
\label{ToC:ToyDiscussion}

\begin{table}
	\centering
	\def~{\hphantom{0}}
	\setlength{\tabcolsep}{5mm}	
		\begin{tabular}{c|cccc|c}
			\multirow{2}*{Method} & \multicolumn{4}{c|}{Evaluation}  & \multirow{2}*{Failure rate}\\
			~			& $20^{th}$ & $100^{th}$ & $500^{th}$ & $1000^{th}$ \\[3pt]
			\hline
			LHS         & 0.5163   & 0.1456   & 0.0218   & 0.0129   & 0.55\\
			MCS          & 0.4810   & 0.1863   & 0.0441   & 0.0221   & 0.61\\
			GA          & 0.4269   & 0.1441   & 0.0065   & 0.0002   & 0\\
			DSM         & 0.4893   & 0.4675   & 0.4673   & 0.4673   & 0.73\\
			RRS         & 0.4893   & 0.3208   & 0.0211   & 0.0003   & 0.01\\
			EGM         & 0.5121   & 0.0621   & 0.0000   & 0.0000   & 0
		\end{tabular}
	\caption{Comparison of all optimizers for the analytical function \eqref{Eqn:Cost:Toy}.
	         Average cost of different algorithms during $m=20$, $100$, $500$ and $1000$ evaluations in 100 runs.}
	\label{Tab:ToySystem:averge performance}
\end{table}
The relative strengths and weaknesses 
of the different optimizers are summarized 
in table \ref{Tab:ToySystem:averge performance} for the average performance
after $m=20$, $100$, $500$ and $1000$ evaluations.
The averaging is performed over the costs of all 100 realizations after $m$ evaluations.
The iteration is considered as failed if the value is $1\>\%$, i.e., $0.01$ above the global minimum.

First, we observe that the downhill simplex algorithm
has the worst failure rate with $73\>\%$, followed by $61\>\%$ of MCS and $55\>\%$ of LHS.
The failures of the simple simplex method are more severe
as the converged parameters significantly depart from the global minimum in $73\>\%$ of the runs.
In case of LHS and MCS, the failure is only the result of pure convergence against the right global  minimum.

Second, after 20 evaluations, the average cost of all algorithms is close to $0.50$, i.e., very similar.

After 100 evaluations, algorithms with explorative steps, i.e., LHS, MCS, GA and EGM have 
a distinct advantage over the downhill simplex method and even over the random restart version.
About 4 restarts are necessary to avoid the convergence to a suboptimal minimum in $99\>\%$ of the cases.
EGM is already  better than the other algorithms by a large factor.

After 500 evaluations, 
EGM corroborates its distinct superiority over the other algorithms,
followed by the RRS and GA.
Intriguingly, GA with its exploitive crossover operation
performs better than all other optimizers after 500 evaluations,
except for  EGM.
LHS and MCS keep a significant error, 
lacking gradient-based optimization.

Summarizing,  algorithms combining exploration and exploitation,
i.e., EGM, GA and RRS, perform
better than purely explorative or purely exploitive  algorithms (LHS, MCS and Simplex).
For the `pure' algorithms, LHS has the fastest decrease of cost function 
while simplex has the fastest convergence.
EGM turns out to be the best combined algorithm 
by making a balance of exploration and exploit 
from LHS and simplex respectively.
This superiority is already apparent after 100 evaluations.

We note that the conclusions have been drawn 
for a single analytical example for an optimization in a low-dimensional parameter space with few minima. 
From many randomly created analytical functions, 
we observe that EGM tends to outperform other optimizers 
in the case of few smooth minima and for low-dimensional search spaces.
Yet, in higher-dimensional search spaces, 
LHS becomes increasingly inefficient and MCS may turn out to perform better.
The number of minima also has an impact on the performance.
For a single minimum with parabolic growth, the DSM can be expected to outperform the other algorithms.
In case of many local shallow minima, the advantage of gradient-based approaches will become smaller
and exploration will correspondingly increase in importance.

%% file: S4.tex
\section{Drag optimization of fluidic pinball with three actuators}
\label{ToC:Fluidic Pinball}
As first flow control example, 
the explorative gradient method is applied 
to the two-dimensional fluidic pinball \citep{Deng2020jfm,Cornejo2019pamm},
the wake behind a cluster of three rotating cylinders.
In \S~\ref{ToC:fluidic pinball:Configuration},
 the benchmark problem is described: Minimize the net drag power with the cylinder rotations as input parameters.
In \S~\ref{ToC:fluidic pinball:EGM optimization},
the explorative gradient method yields a surprising non-symmetric result, 
consistent with other fluidic pinball simulations \citep{Cornejo2019pamm} 
and experiments \citep{Raibaudo2020pf}.
The learning process of  the Downhill Simplex Method (DSM) 
and Latin Hypercube Sampling (LHS) are investigated 
in \S~\ref{ToC:fluidic pinball:gradient optimization} and \ref{ToC:fluidic pinball:explorative optimization},

\subsection{Configuration}
\label{ToC:fluidic pinball:Configuration}
The fluidic pinball is a benchmark configuration for wake control
which is geometrically simple yet  rich in nonlinear dynamics behaviours.
This two-dimensional configuration consists 
of a cluster of three equal, parallel and equidistantly spaced cylinders
pointing in opposite to  uniform flow.
The wake can be controlled by the cylinder rotation.
The fluidic pinball comprises most known wake stabilization mechanisms,
like phasor control, circulation control, Coanda forcing, base bleed
as well as high and low-frequency forcing.
In this study, we focus on steady open-loop forcing minimizing the drag power
corrected by actuation energy.

The viscous incompressible two-dimensional flow
has uniform oncoming flow with  speed $U_\infty$
and a fluid with constant density $\rho$ and kinematic viscosity $\nu$.
The three equal circular cylinders have radius $R$ and their centers 
form an equilateral triangle with sidelength $3R$ pointing upstream.
Thus, the transverse dimension of the cluster reads $L = 5R$.

\begin{figure}[htb]
    \centering
	\subfloat[]{
		\label{Fig:fluidic pinball:configuration:a}
		\includegraphics[width=0.45\textwidth]{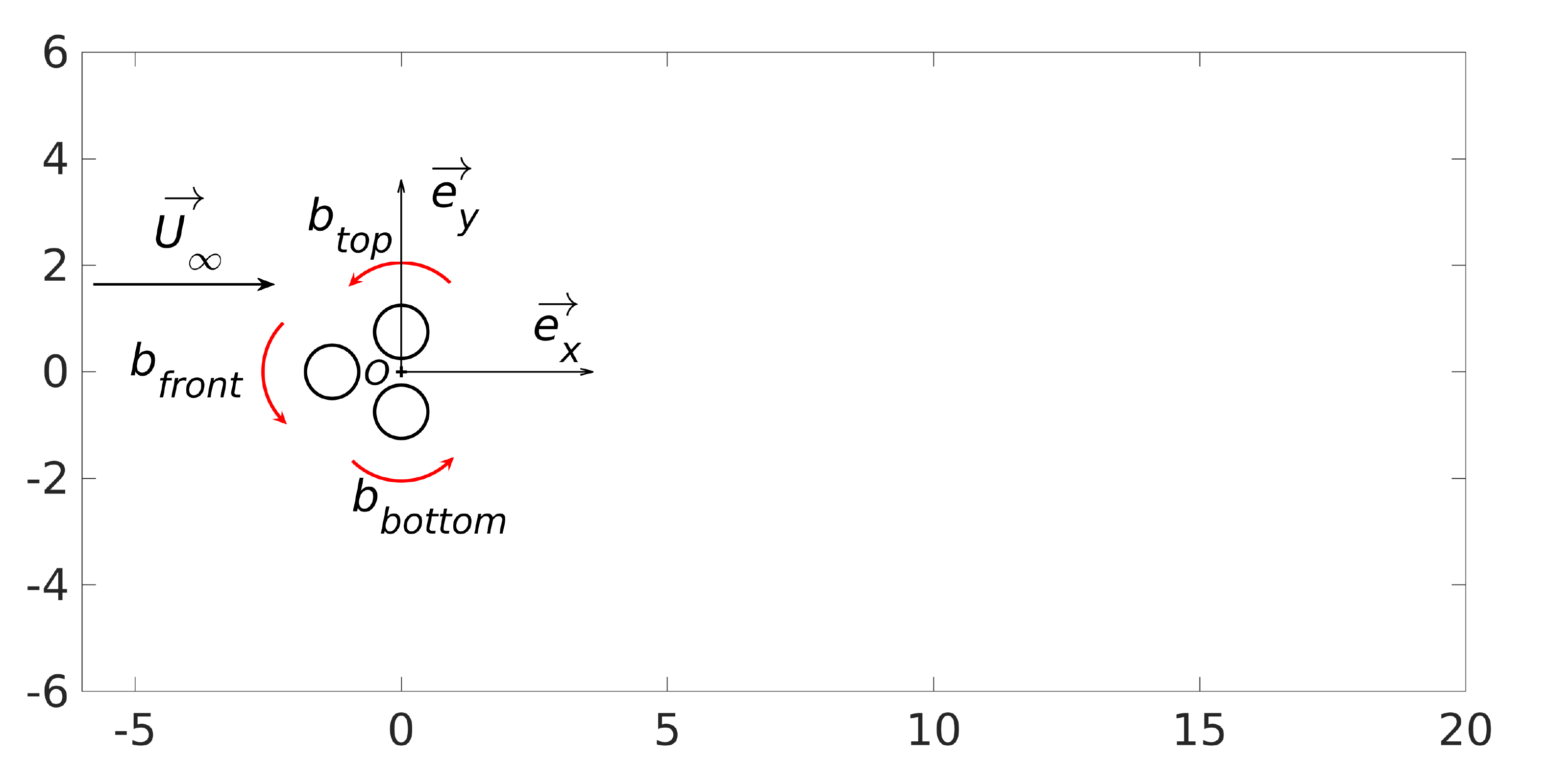}}		
	\hfil
	\subfloat[]{
		\label{Fig:fluidic pinball:configuration:b}
		\includegraphics[width=0.45\textwidth]{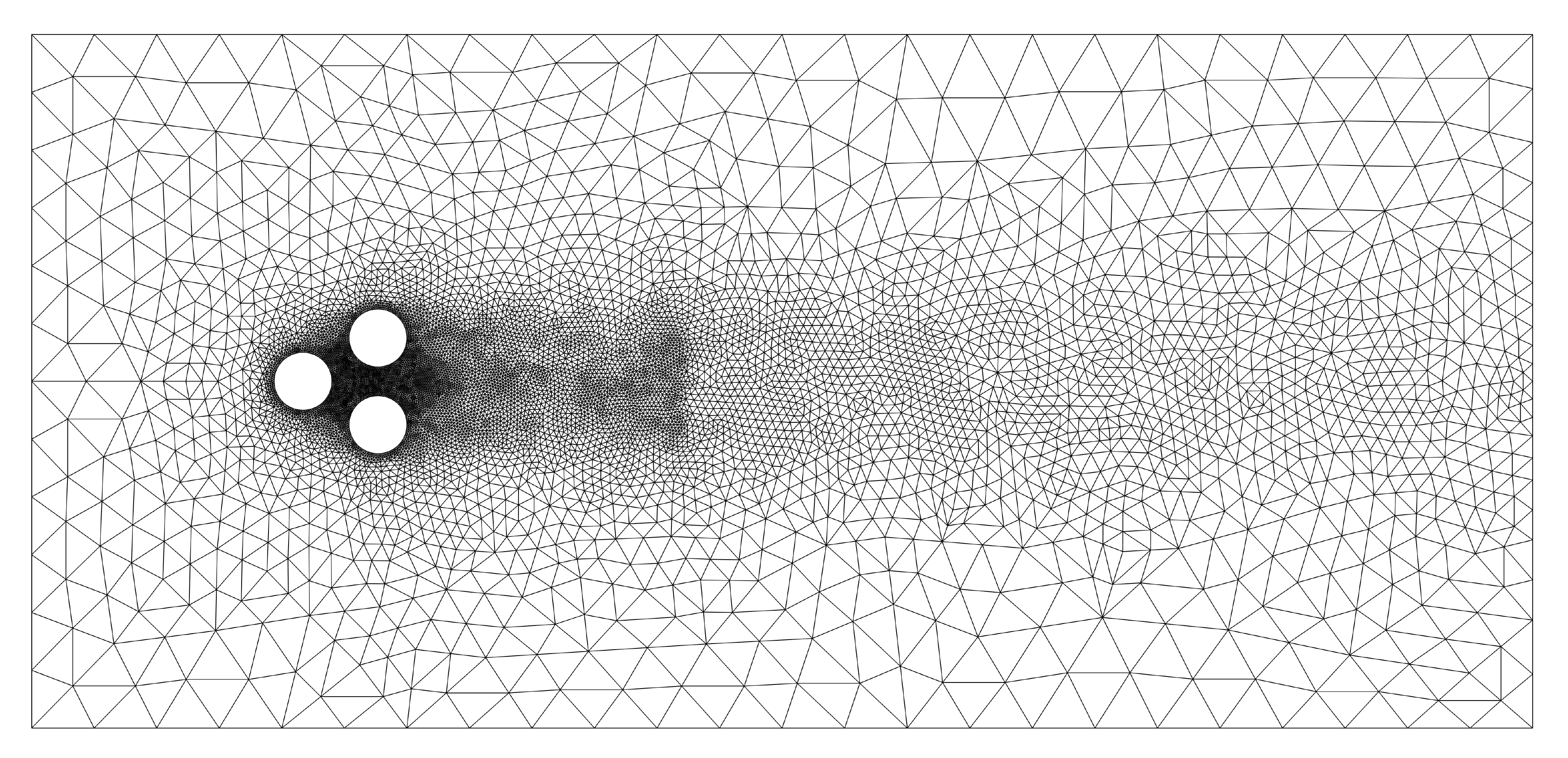}}	  
	\caption{Fluidic pinball (\textit{a}) configuration and (\textit{b}) grid.}
    \label{Fig:fluidic pinball:configuration}
\end{figure}

In figure \ref{Fig:fluidic pinball:configuration:a},
the flow is described in Cartesian coordinate system
where the $x$-axis points in the direction of the flow,
the $z$-axis is aligned with the cylinder axes
and the $y$-axis is orthogonal to both.
The origin $\bm{0}$ is placed in the center of the rightmost top and bottom cylinders.
Thus, the centers of the cylinders are described by
\begin{equation}
    \begin{array}{ll}
    x_1=x_F = -3 \> R \> \cos 30^\circ, & y_1 =y_F = 0,\\
    x_2=x_B = 0, & y_2 = y_B = -3R/2,\\
    x_3= x_T = 0, & y_3 = y_T = +3R/2.
    \end{array}
\end{equation}
Here, the subscripts ‘F’, ‘B’ and ‘T’refer to the front, bottom and top
cylinder. Alternatively, the subscripts '1', '2' and '3' are used for these cylinders
starting with the front cylinder and continuing in mathematically positive orientation.

The location is denoted by  $\bm{x} = (x, y) = x \> \bm{e}_x + y \> \bm{e}_y$,
where $\bm{e}_x$ and $\bm{e}_y$ are the unit vectors in $x$- and $y$-direction.
The flow velocity is represented  
by  $\bm{u} = (u, v) = u \> \bm{e}_x + v \> \bm{e}_y $.
The pressure and time symbols are $p$ and $t$, respectively.
In the following, 
all quantities are non-dimensionalized 
with cylinder diameter $D = 2R$,
the velocity $U_\infty$ and the fluid density $\rho$.

The corresponding Reynolds number reads $Re_D = U_\infty D/\nu = 100$.
This Reynolds number corresponds to asymmetric periodic vortex shedding.
\citet{Deng2020jfm} have investigated the transition scenario for increasing Reynolds number.
At $Re_1 \approx 18$, the steady flow becomes unstable in a Hopf bifurcation leading to periodic vortex shedding.
At $Re_2 \approx 68$, both the steady Navier-Stokes solutions 
and the limit-cycles bifurcate into two mirror-symmetric states.
\citet{ChenAlam2020jfm} performed a careful parametric analysis of the gap width
between the cylinders and associated this behaviour with the `deflected regime',
where base bleed through the rightmost cylinder are deflected upward or downward.
At $Re_3 \approx 104$ another Hopf bifurcation leads to quasi-periodic flow.
After $Re_4 \approx 115$, a chaotic state emerges.

The flow properties can be changed by the rotation of cylinders.
The corresponding actuation commands are denoted by 
\begin{equation}
\label{Eqn:ActuationCommands}
b_1 = U_F, \quad b_2 = U_B, \quad b_3 = U_T .
\end{equation}
Here, positive values denote the anti-clockwise direction.

Following \citet{Cornejo2019pamm},
we aim to  minimize of the averaged parasitic drag power $\bar{J_a}$ 
 penalizing the averaged actuation power $\bar{J_b}$.
 The resulting cost function reads
\begin{equation}
\label{Eqn:Pinball:Cost}
    \bar{J} = \bar{J}_a+\bar{J}_b.
\end{equation}
The first contribution $\bar{J}_a = c_D$ corresponds to drag coefficient
\begin{equation}
\label{Eqn:Pinball:DragCoefficient}
c_D = \frac{\bar{F}_D }{ (1/2) \rho D U_{\infty}^2}
\end{equation} 
for the chosen non-dimensionalization.
Here, $\bar{F}_D$ denotes total averaged drag force on all cylinders per unit spanwise length.
The second contribution arises from the necessary actuation torque 
to overcome the skin-friction resistance.

Following \citet{Deng2020jfm}, 
the flow is computed with direct numerical solution
in the computational domain
\begin{equation}
    \Omega = \{(x,y):-6 \leq x \leq 20 \wedge |y| \leq 6 \wedge (x-x_i)^2+(y-y_i)^2\geq1/4,i=1,2,3)\}.
\end{equation}
We use an in-house implicit finite-element method solver `UNS3'
which is of third-order accuracy in space and time.
The unstructured grid in figure \ref{Fig:fluidic pinball:configuration:b} contains 4225 triangles and 8633 vertices.
An earlier grid convergence study identified 
this resolution sufficient for up to 2 percent error in drag, lift and Strouhal number.


\subsection{Optimized actuation}
\label{ToC:fluidic pinball:EGM optimization}

\begin{table}
    \centering
    \def~{\hphantom{0}}
    \setlength{\tabcolsep}{5mm}		
    {
        \begin{tabular}{l|ccccccc}
            $m$    & $b_1$ & $b_2$  & $b_3$      & J \\
            \hline \\[-7pt]
            1      & 0     & -3     & 3          & 4.9579\\
            2      & 0.1   & -3     & 3          & 4.9695\\
            3      & 0     & -2.9   & 3          & 4.8979\\
            4      & 0     & -3     & 3.1        & 5.0702\\
        \end{tabular}
    }
    \caption{Fluidic pinball: Initial simplex ($m=1,2,3,4$) for the three-dimensional downhill simplex optimization.
        $b_i$ denotes the circumferential velocity of cylinders and $J$ corresponds to the net drag power (\ref{Eqn:Pinball:Cost}).}
    \label{Tab:fluidic pinball:Initial individuals}
\end{table}
\begin{figure}[htb]
    \subfloat[]{
      \centering
      \label{Fig:fluidic pinball:EGM:a}
      \includegraphics[width=0.9\textwidth]{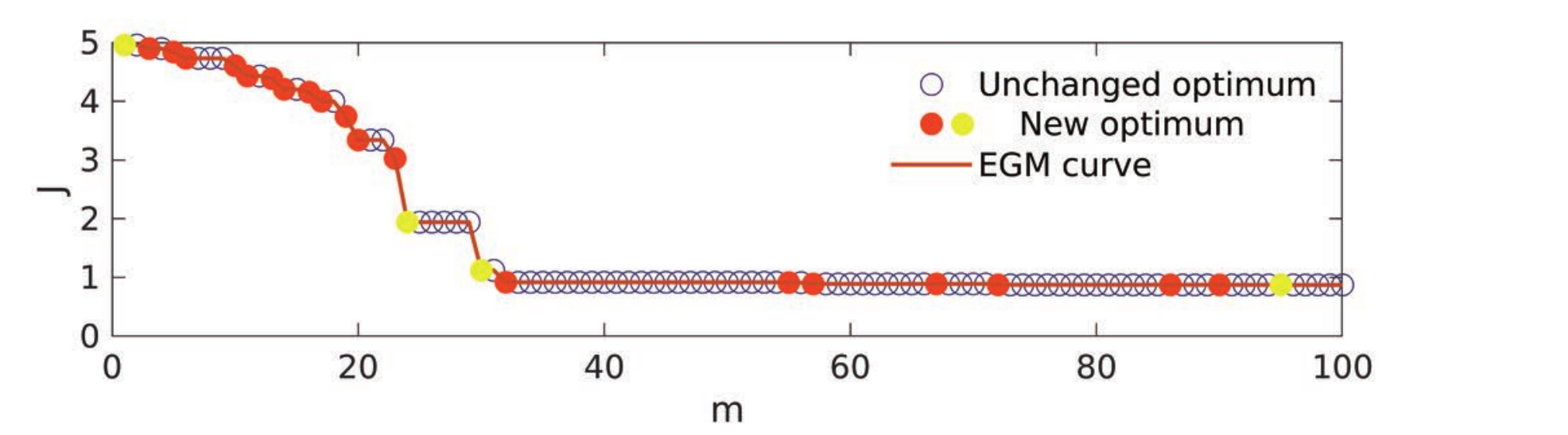}}

      \subfloat[]{
      \centering
      \label{Fig:fluidic pinball:EGM:b}
      \includegraphics[width=0.9\textwidth]{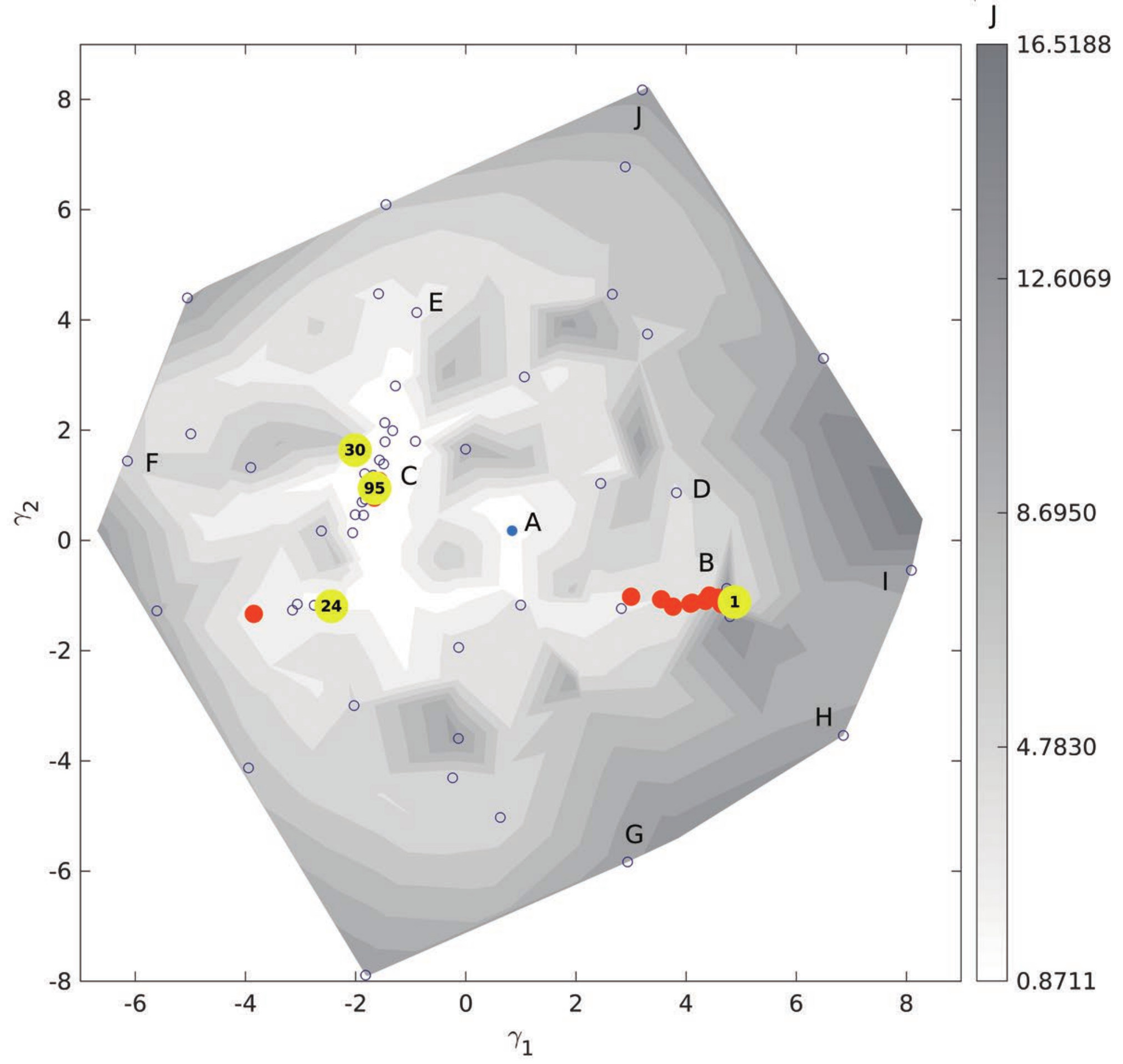}}  
    \caption{Optimization of the fluidic pinball  actuation with EGM.
    The actuation parameters and costs are visualized 
    like figure \ref{Fig:ToySystem:tested individuals} and \ref{Fig:ToySystem:learning curve}.
    For enhanced interpretability, selected new minima 
    are displayed as solid yellow circles in the learning curve (a) 
    and in the control landscape (b) corresponding to $m$. 
    $m$ counts the direct numerical simulations for net drag power computation. 
    The marked flows $A$-$J$ are explained in the text.}
    \label{Fig:fluidic pinball:EGM}
\end{figure}
\begin{figure}
	\centerline{\includegraphics[width=\textwidth]{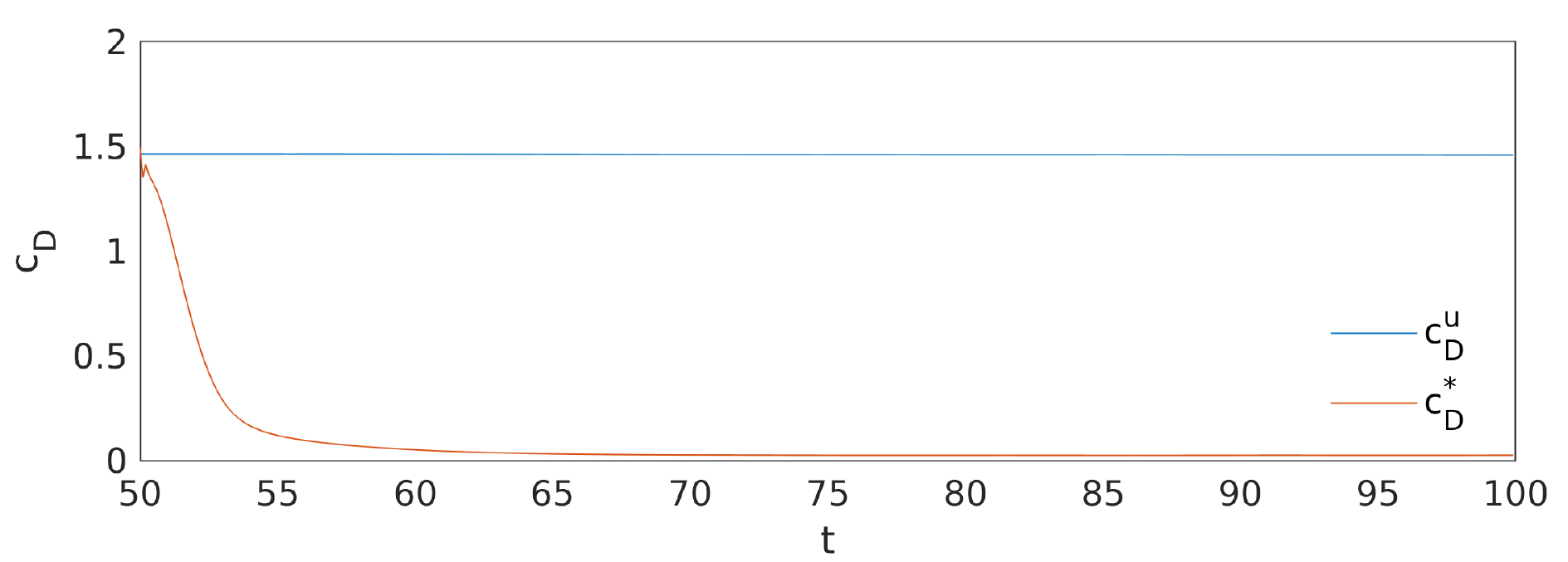}}
    \caption{Drag coefficient ($c_D$) of the fluidic pinball over time ($t$).
    $c_D^u$ denotes drag of unforced flow 
     while $c_D^\star$ represents the value for the best control.
     }
	\label{Fig:fluidic pinball:drag}
\end{figure}
\begin{figure}[htb]
	\centering
  
    \subfloat[]{
		\label{Fig:fluidic pinball:flow:a}
		\includegraphics[width=0.36\textwidth]{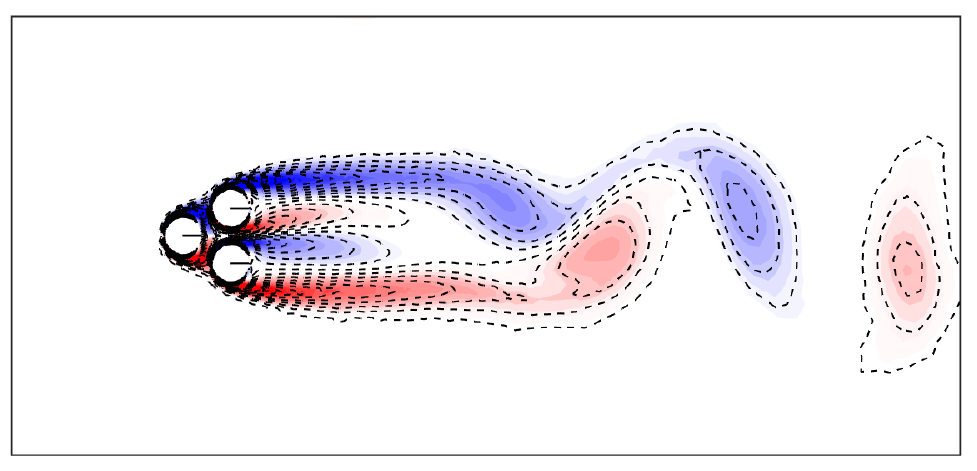}}
	\hfil
	\subfloat[]{
		\label{Fig:fluidic pinball:flow:b} 
		\includegraphics[width=0.36\textwidth]{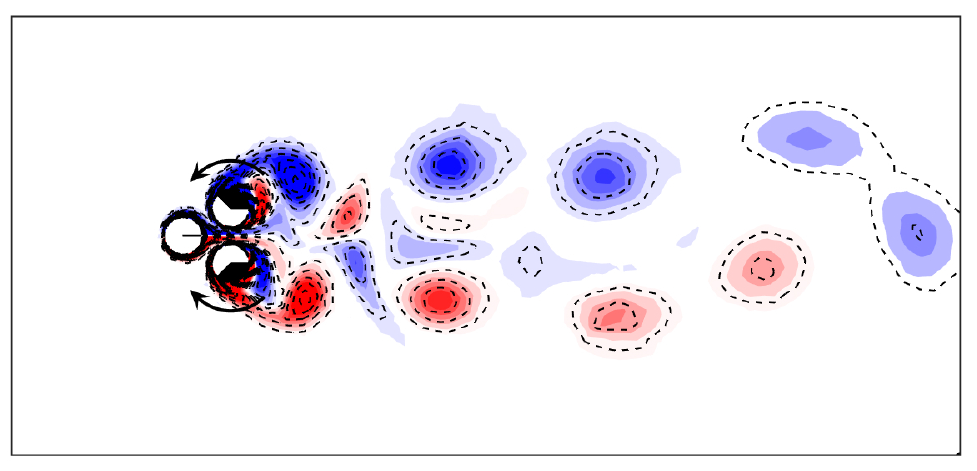}}
  
	\subfloat[]{
		\label{Fig:fluidic pinball:flow:c} 
		\includegraphics[width=0.36\textwidth]{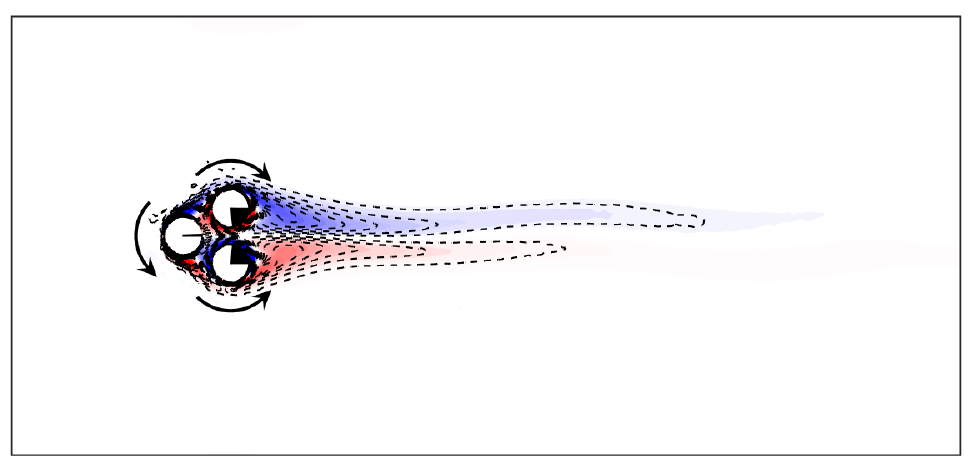}}
	\hfil
	\subfloat[]{
		\label{Fig:fluidic pinball:flow:d}  
		\includegraphics[width=0.36\textwidth]{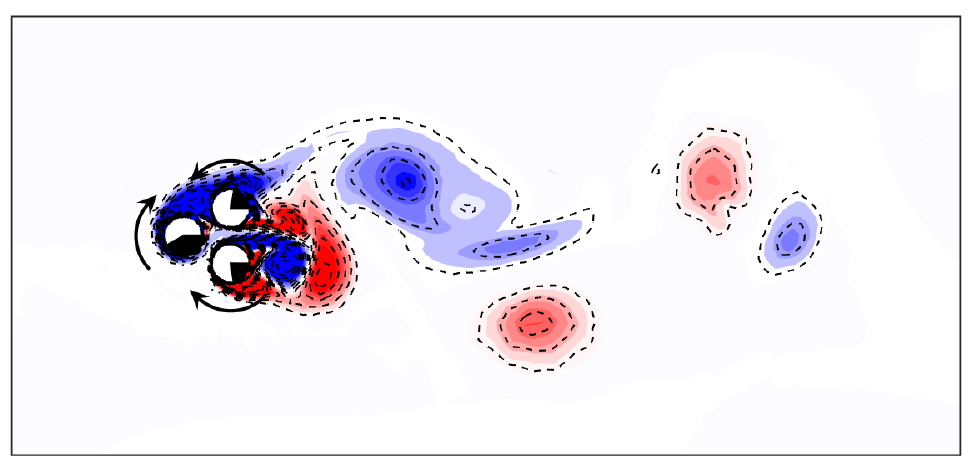}}

	\subfloat[]{
		\label{Fig:fluidic pinball:flow:e} 
		\includegraphics[width=0.36\textwidth]{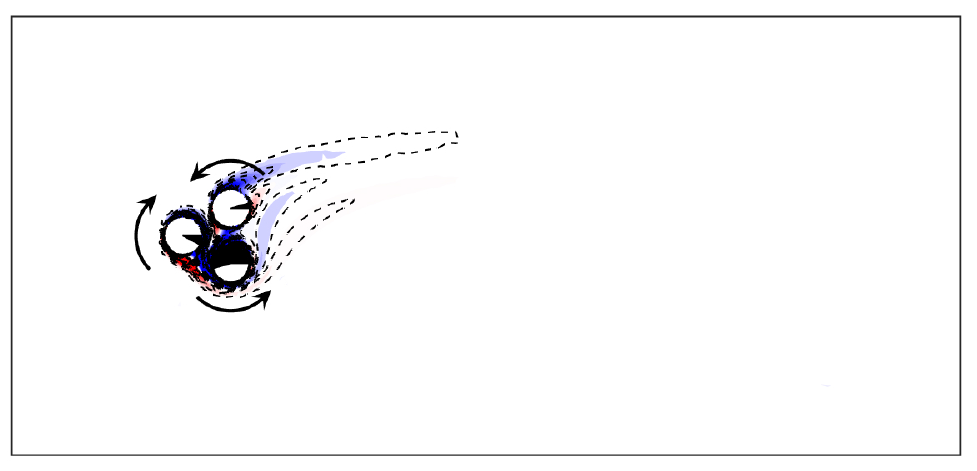}}
	\hfil
	\subfloat[]{
		\label{Fig:fluidic pinball:flow:f} 
        \includegraphics[width=0.36\textwidth]{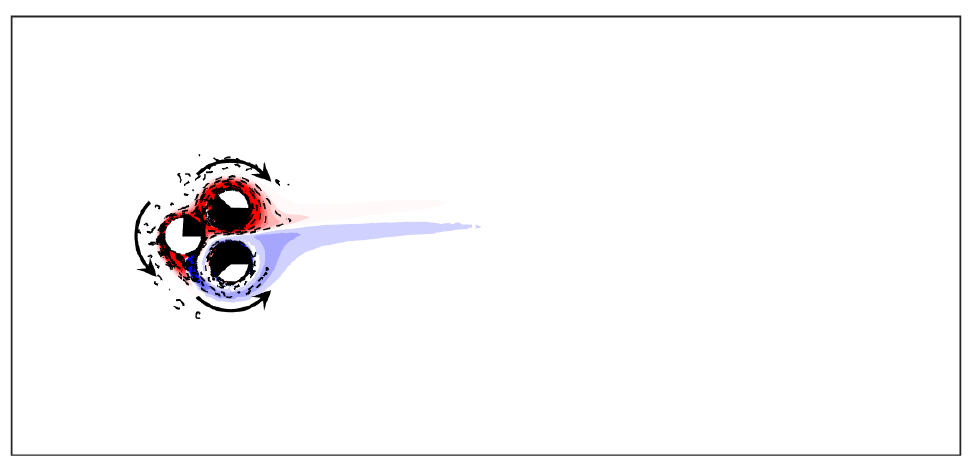}}	  
        
    \subfloat[]{
		\label{Fig:fluidic pinball:flow:g} 
		\includegraphics[width=0.36\textwidth]{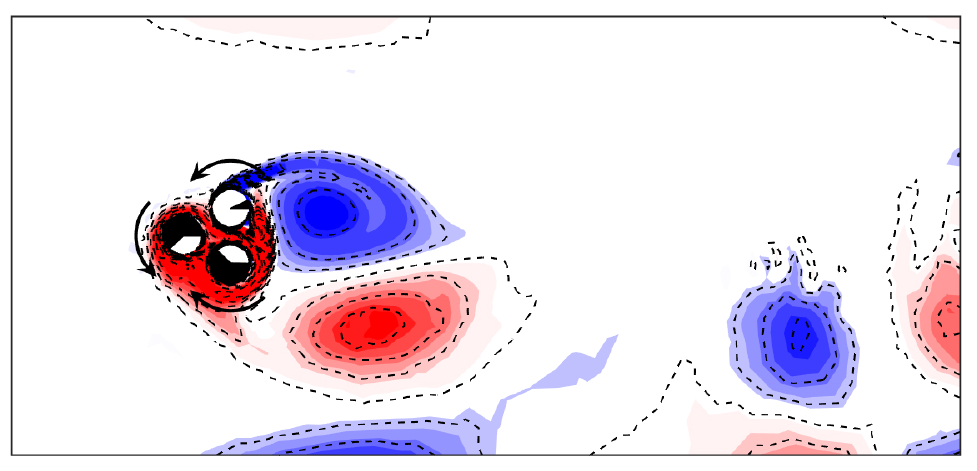}}
	\hfil
	\subfloat[]{
		\label{Fig:fluidic pinball:flow:h} 
		\includegraphics[width=0.36\textwidth]{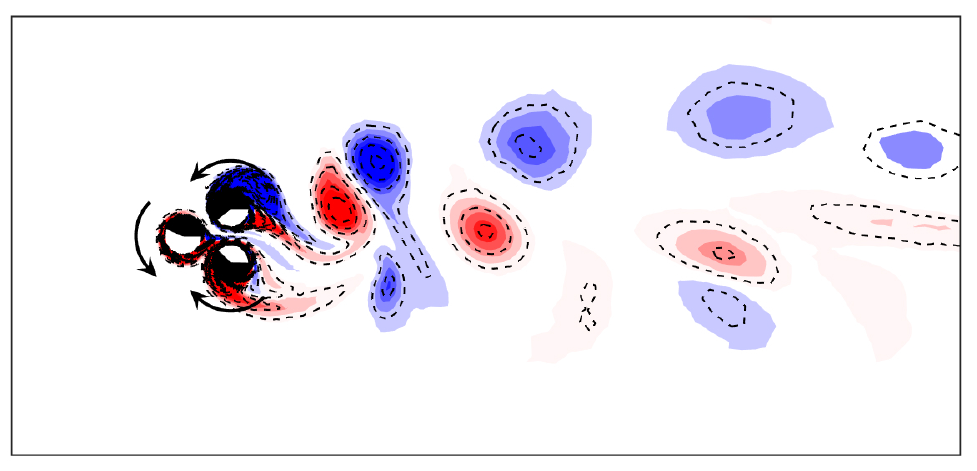}}	  

    \subfloat[]{
        \label{Fig:fluidic pinball:flow:i} 
        \includegraphics[width=0.36\textwidth]{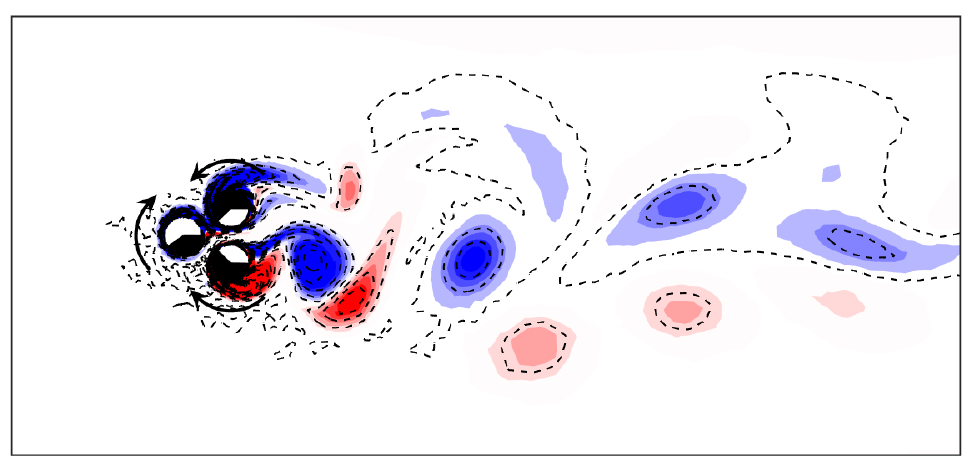}}
    \hfil
    \subfloat[]{
        \label{Fig:fluidic pinball:flow:j} 
        \includegraphics[width=0.36\textwidth]{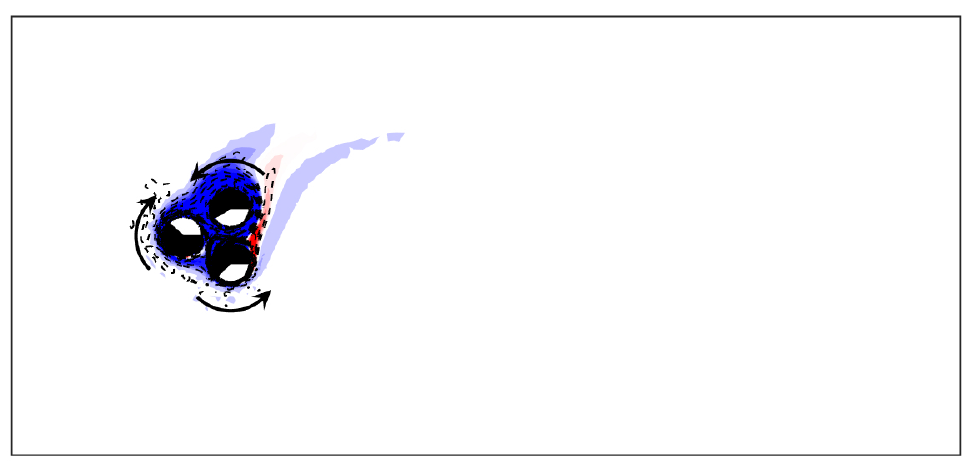}}	                             
	\caption{Fluidic pinball flows of different actuations 
	         of the control landscape (Figure \ref{Fig:fluidic pinball:EGM:b}). 
	         Subfigure $a-j$ corresponds to actuations with letters $A$--$J$, respectively,
	         and display the vorticity of the post-transient snapshot.
	         Positive (negative) vorticity is color-coded in red (blue). 
	         The dashed lines correspond to iso-contourlines of vorticity.
	         The orientation of the cylinder rotations is indicated by the arrows.
	         The cylinder rotation is proportional to the angle of the black sector inside.}
	\label{Fig:fluidic pinball:flow}
\end{figure}

In the subsequent study, the actuation commands 
$b_1 = U_F$, $b_2=U_B$ and $b_3=U_T$ are bounded by 5, 
i.e, the search space reads
\begin{equation}
\label{Eqn:Pinball:Domain}
\Omega := \left \{  [b_1,b_2,b_3]^{\rm T} \in R^3 \colon \vert b_i \vert \le 5 \quad \hbox{for} \quad i=1,2,3 \right \}.
\end{equation}
Previous symmetric parametric studies 
have identified symmetric Coanda forcing $b_1=0$, $b_2=-b_3$ around $2$ as optimal for net drag reduction,
both in low-Reynolds number direct numerical simulations \citep{Cornejo2019pamm} 
and in high-Reynolds number unsteady Reynolds Averaged Navier-Stokes (URANS) simulations \citep{Raibaudo2020pf}.
The chosen bound of $5$ adds a large security factor to these values,
i.e., the optimum can be expected to be in the chosen range.
Steady bleed into the wake region is reported as another means for wake stabilization
by suppressing the communication between the upper and lower shear layer.
This study starts from the base-bleeding control in search of a different actuation from boat tailing. 

The Latin hypercube sampling (LHS), downhill simplex method (DSM) and explorative gradient method  (EGM)
are applied to minimize the net drag power \eqref{Eqn:Pinball:Cost}
with steady actuation
in the three-dimensional domain \eqref{Eqn:Pinball:Domain}.
Following \S~\ref{ToC:Method:Simplex} and \ref{ToC:Method:EGM},
the initial simplex comprises four vertices:
the individual controlled by base-bleeding actuation ($b_1 = 0,b_2 = -3,b_3 = 3$),
the other three individuals are positively shifted by 0.1 for each actuation.
The individuals and their corresponding costs 
are listed in table \ref{Tab:fluidic pinball:Initial individuals}.
All the individuals have a larger cost than the unforced benchmark $J = 1.4611$.
The increase of the actuation amplitude($m = 2, m = 4$) indicates higher cost.
And the indivudual with a smaller bottom actuation($m = 3$) is associated with a smaller cost.
Thus, the initial condition seems to pose a challenge for optimization.

In this section, the optimization process of EGM is investigated.
Figure \ref{Fig:fluidic pinball:EGM:a}
shows the best cost found with $m$ simulation.
EGM is quickly and practically converged after $m=32$ evaluations
and yields the near-optimal actuation at the $95^{th}$ test

\begin{equation}
\label{Eqn:Pinball:Winner}
b^\star_1=0.1004, \quad  b^\star_2= 1.8645, \quad b^\star_3 =-1.8314, \quad  J^\star =0.8711.
\end{equation}
The cost function $J^\star=0.8711$ reveals a net drag power 
saving of $40\%$ with respect to the unforced value $J_u = 1.4611$.
The large amount of suboptimal testing is indicative for a complex control landscape.


As displayed in figure \ref{Fig:fluidic pinball:drag},
the drag coefficient falls from  $1.4597$ for unforced flow 
to $0.0253$ for the actuation \eqref{Eqn:Pinball:Winner}
within few convective time units.
This near-optimal actuation corresponds to 98\% drag reduction. 
This 98\% reduction of drag power requires 58\% investment in actuation energy.

The best actuation corresponds to nearly symmetric Coanda forcing 
with a circumferential velocity of 1.8.
This actuation deflects the flow towards 
the positive $x$-axis and effectively 
removes the dead-water region with reversal flow.
The slight asymmetry of the actuation is not a bug 
but a feature of the optimal actuation 
after the pitchfork bifurcation at $Re_2 \approx 68$.
This achieved performance and actuation 
is similar to the optimization feedback control
achieved by machine learning control \citep{Cornejo2019pamm},
comprising a slightly asymmetric Coanda actuation with small phasor control from the front cylinder.
Also, the optimized experimental stabilization of the high-Reynolds number regime
lead to asymmetric steady actuation \citep{Raibaudo2020pf}.
The asymmetric forcing may be linked to the fact 
that the unstable asymmetric steady Navier-Stokes solutions have a lower drag
than the unstable symmetric solution.

Figure  \ref{Fig:fluidic pinball:EGM:b} shows the control landscape,
i.e., two-dimensional proximity map of the three-dimensional actuation parameters.
Neighbouring points in the proximity map correspond similar actuation vectors.
The proximity map is computed with classical multi-dimensional scaling \citep{Cox2000book}.
This map shows all performed simulations for figure  \ref{Fig:fluidic pinball:EGM:a}
as solid red circles, when the evaluation improves the cost with respect to the iteration history
and as open blue circles otherwise.
Select new minima are highlighted with yellow circles:
The first run $m=1$ on the right side, 
the converged run $m=95$ on the left,
and the intermediate runs $m=24$ and $30$
when the explorative step jumps in new better territories. 
The colorbar represents interpolated values of the cost \eqref{Eqn:Pinball:Cost}.

The meaning of the feature coordinates $\gamma_1$ and $\gamma_2$ 
will be revealed by following analysis.
Ten of the individuals of the control landscape 
are selected and marked with letters between $A$ and $J$:
\begin{description}
\item[A) unforced flow ] in the center;
\item[B) base-bleeding flow ] $m=1$ as the initial individual;
\item[C) optimal actuation ] after $m=95$ evaluations;
\item[D) an asymmetric base-bleeding actuation ] $m=84$ showing a strong front actuation;
\item[E) an almost single actuation ] $m=37$ at the bottom cylinder;
\item[F--J) extreme actuations ] at the boundary of the control landscape
corresponding to $m=60$, $m=39$, $m=63$, $=15$, respectively.
\end{description}

The flows corresponding to actuations $A$--$J$ in figure \ref{Fig:fluidic pinball:EGM:b}
are depicted in figure \ref{Fig:fluidic pinball:flow}.
The optimized actuation ($C$) yields a partially stabilized flow,
like the machine learning feedback control by \citet{Cornejo2019pamm}.
Actuation $C$ corresponds to complete stabilizations with strong Coanda forcing,
located near $\gamma_2 \approx 0$ for small $\gamma_1$. 
In contrast, flow $B$ on the opposite side of the control landscapes represents strong base bleeding.
Actuations $J$, $G$ and $H$, located at the top and bottom of the control landscapes
correspond to Magnus effects. Large positive (negative) feature coordinates $\gamma_2$  
are associated with large positive (negative) total circulations and associated lift forces.
Summarizing, the analysis of these points 
reveals that the feature coordinate $\gamma_1$ corresponds to the strength of the Coanda forcing
and is hence related to the drag. 
In contrast,  $\gamma_2$ is correlated with the total circulation of the cylinder rotations
and thus with the lift.
\citet{Ishar2019jfm} arrives at a similar interpretation 
of the proximity map for differently actuated fluidic pinball simulations.

\subsection{Downhill simplex method}
\label{ToC:fluidic pinball:gradient optimization}

\begin{figure}[htb]
    \subfloat[]{
      \centering
      \label{Fig:fluidic pinball:simplex:a}
      \includegraphics[width=0.9\textwidth]{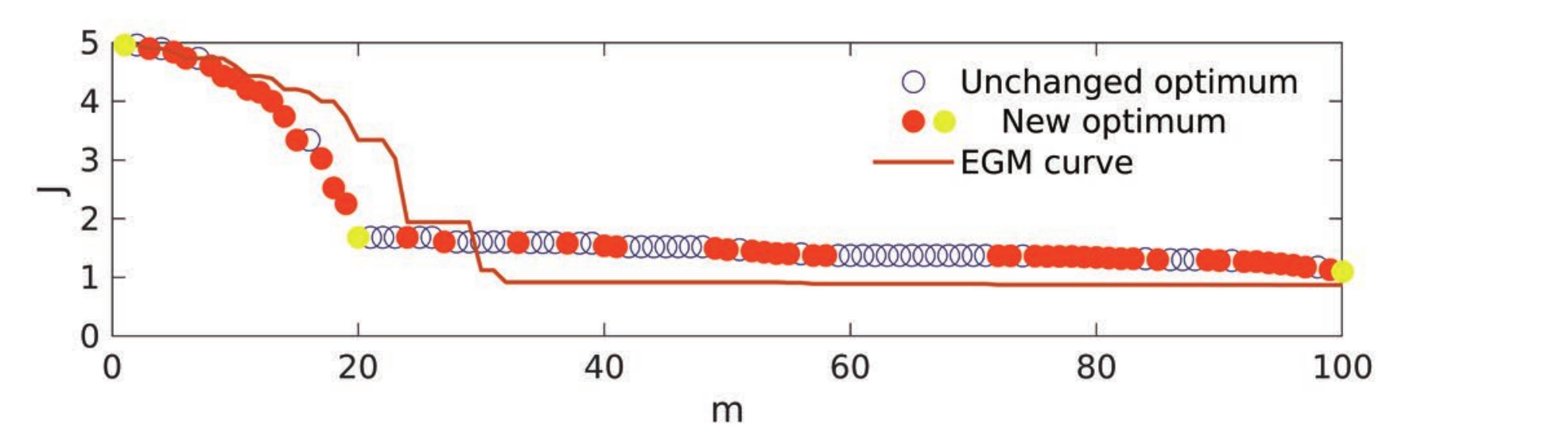}}

      \subfloat[]{
      \centering
      \label{Fig:fluidic pinball:simplex:b}
      \includegraphics[width=0.9\textwidth]{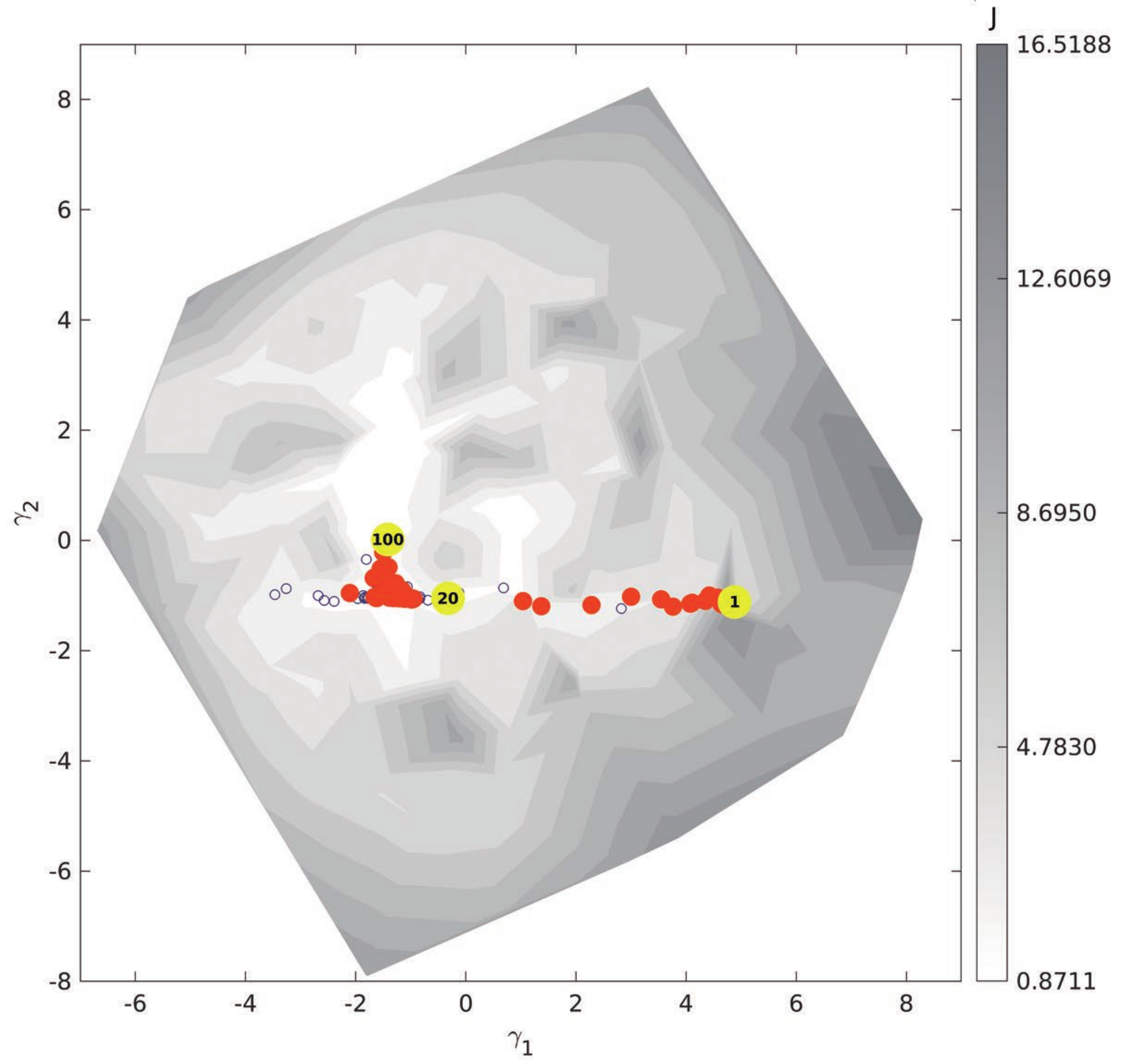}}  
    \caption{Same as figure 
    \ref{Fig:fluidic pinball:EGM}, but with DSM.}
    \label{Fig:fluidic pinball:simplex}
\end{figure}
Figure \ref{Fig:fluidic pinball:simplex:a} 
shows the optimization process of the downhill simplex method (DSM).
After step-by-step decends in the former 20 simulations,
the net drag cost decreases slowly during the following $80\%$ computation.
The optimal actuation after $100$ simulation is $b_1 =0.1632$, $b_2 = 1.5346$, $b_3 = -1.5378$) with cost $J = 0.8937$.

Figure \ref{Fig:fluidic pinball:simplex:b} reveals the optimization 
from a broad slope to a tortuous valley after the $20^{th}$ test.
In face of the complex landscape on the way to global optimum (from $\gamma = [4,-1]^{\rm T}$ 
to $\gamma = [-1,0]^{\rm T}$), 
DSM consumes relatively high computation resources.
EGM is seen to outperform DSM at $m \geq 30$ 
because of an explorative step.
For random initial conditions, DSM often performs better than EGM,
because the exploration as insurrance policy brings less return 
for this comparatively simple control landscape.

\subsection{Latin hypercube sampling}
\label{ToC:fluidic pinball:explorative optimization}

\begin{figure}[htb]
    \subfloat[]{
      \centering
      \label{Fig:fluidic pinball:LHS:a}
      \includegraphics[width=0.9\textwidth]{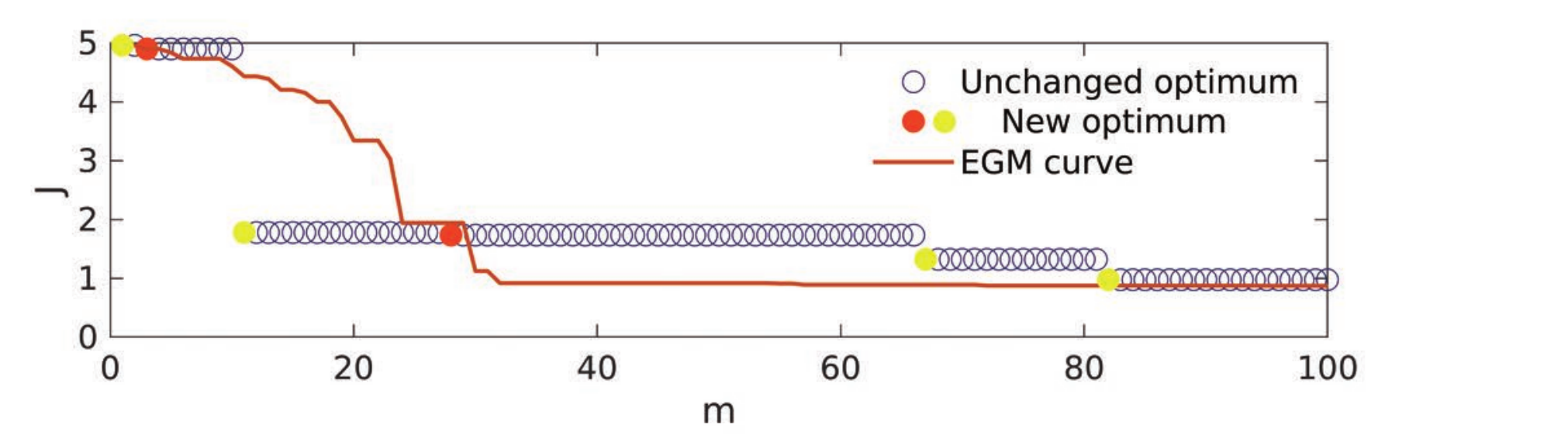}}

      \subfloat[]{
      \centering
      \label{Fig:fluidic pinball:LHS:b}
      \includegraphics[width=0.9\textwidth]{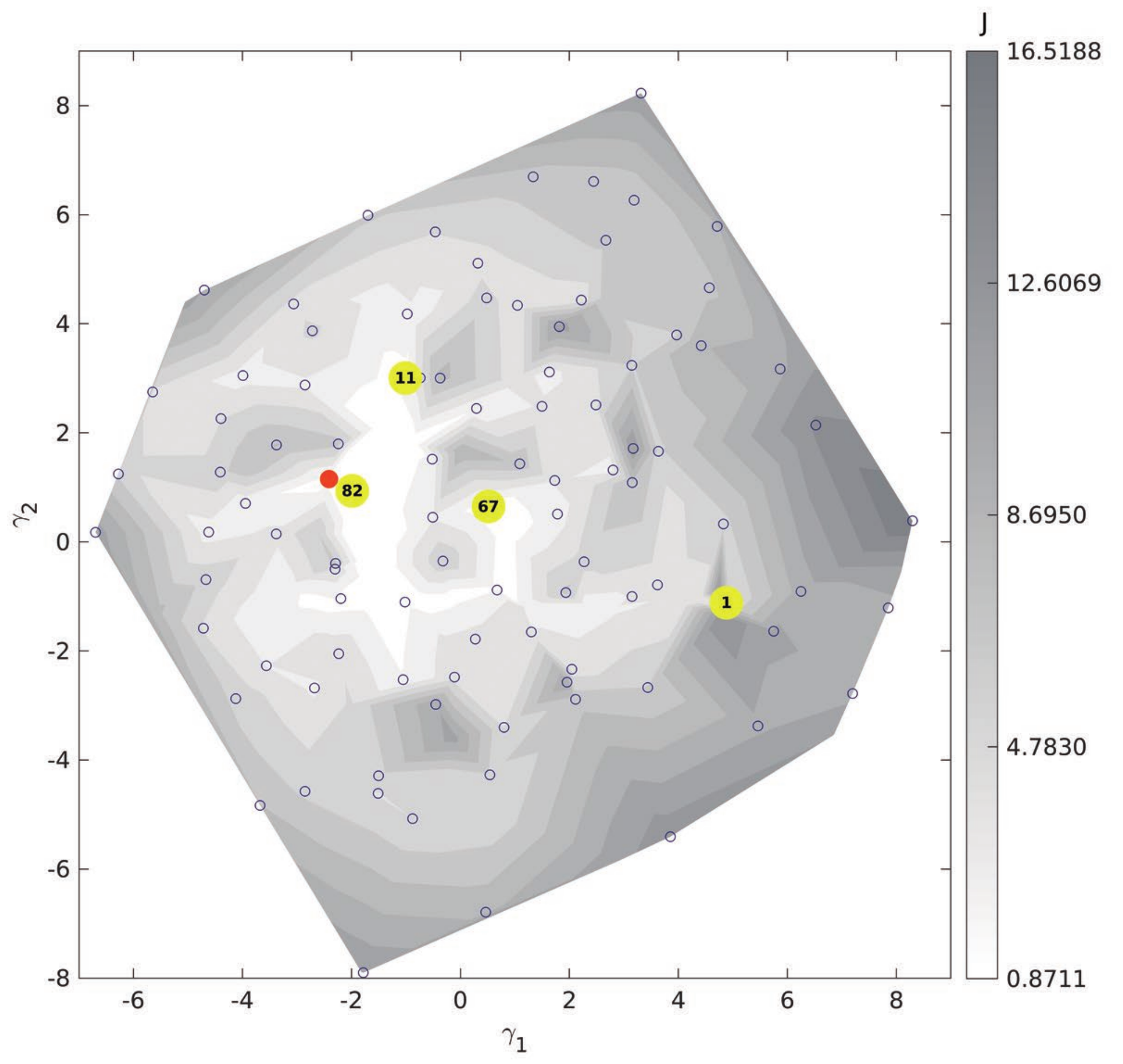}}  
    \caption{Same as figure 
    \ref{Fig:fluidic pinball:EGM}, but with  LHS.}
    \label{Fig:fluidic pinball:LHS}
\end{figure}
Figure \ref{Fig:fluidic pinball:LHS:a} shows the performance of Latin hypercube sampling (LHS).
The algorithm is low-efficient during the most computation 
with only 4 new optima found.
The jump at the $11^{th}$ DNS reduces the cost by more than $50\%$.
The new optimum at $28^{th}$ simulation does not bring significant improvement,
followed by a stagnation for about $40\%$ of the optimization period.
The $67^{th}$ and $82^{rd}$ simulation further reduces the cost to 0.9741 with actuation
$b_1 =0.3787$, $b_2 = 2.0935$, $b_3 = -2.0287$.

The global search is further illustrated by the uniform tested points in figure \ref{Fig:fluidic pinball:LHS:b}.
The algorithm starts near $\gamma = [4,-1]^{\rm T}$ 
and explore the feature space from the boundary to the central area.
Three better minimum are explored one by one before the $82^{rd}$ individual near to the global mimimum is found.
LHS outperforms EGM at $m=11$ before EGM leads at $m \geq 30$.
Exploration is seen to have advantages at the beginning 
but exploitation wins already in the mid-term.

%% file: S5.tex
\section{Drag optimization of an Ahmed body with 10 actuation parameters}
\label{ToC:Ahmed}
Starting point of the computational fluid dynamics plant 
is an experimental study of a low-drag $35^\circ$ Ahmed body \citep{Li2018icfm8}.
The investigated Ahmed body configuration (\S~\ref{ToC:Ahmed:Configuration}) 
has the same physical dimensions.
The effect of actuation is assessed with 
a Reynolds-Averaged Navier-Stokes (RANS) simulation (\S~\ref{ToC:Ahmed:RANS}).
In \S~\ref{ToC:Ahmed:01D}, 
a parametric drag study 
with a single tangential actuator on the top edge is performed.
In \S~\ref{ToC:Ahmed:05D}, 
the tangential blowing of all five actuator groups is optimized for drag reduction.
The explorative gradient method (EGM) 
is contrasted to the downhill simplex method (DSM) 
and Latin hypercube sampling (LHS).
In \S~\ref{ToC:Ahmed:10D}, 
the velocity and oriention of the five slot actuators 
are optimized with EGM, 
thus giving rise to a ten-dimensional search space.
As expected drag reduction increases
with the dimension of the search space, i.e., expanding actuation opportunities.
The corresponding physical drag reduction mechanisms are investigated.

\subsection{Configuration}
\label{ToC:Ahmed:Configuration}

\begin{figure}[htb]
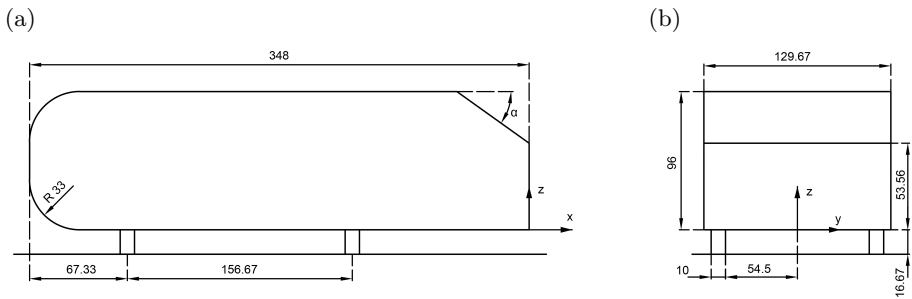

	\centering
	\subfloat[]{
		\label{Fig:Configuration:Model:a}
		\includegraphics[height=3.5cm]{./S4A_Model_a.pdf}}		
	\hfil
	\subfloat[]{
		\label{Fig:Configuration:Model:b}
		\includegraphics[height=3.5cm]{./S4A_Model_b.pdf}}	  
	\caption{Dimensions of the investigated 1:3-scaled Ahmed body.
			\textit{a}) Side view. \textit{b}) Back view. 
			The length unit is $\rm mm$ and the angle is specified in degrees.}
	\label{Fig:Configuration:Model}
\end{figure}

Point of departure is an experimentally investigated
1:3-scaled Ahmed body characterized 
by a slanted edge angle of $\alpha=35^\circ$
with  length $L$, width $W$ and height $H$ 
of $348\> \rm mm$, $130\> mm$ and $96 \> \rm mm$, respectively. 
The front edges are rounded with a radius of $0.344 \> H$. 
The model is placed on four cylindrical supports with a diameter equal to $10\> \rm mm$ 
and the ground clearance is $0.177\> H$. 
The origin  of the Cartesian coordinate system $(x, y, z)$,
is located in the symmetry plane on the lower edge of the model's vertical base (see figure \ref{Fig:Configuration:Model}). 
Here, $x$, $y$ and $z$ denote the streamwise, spanwise and wall-normal coordinate, respectively. 
The velocity components 
in the $x$, $y$ and $z$ directions
are denoted by $u$, $v$ and $w$, respectively.
The free-stream velocity is chosen to be $U_\infty =30 \> \rm m/s$.

\begin{figure}[htb]
	\centering
	\subfloat[]{
		\label{Fig:Configuration:Actuation:a}
		\includegraphics[height=4cm]{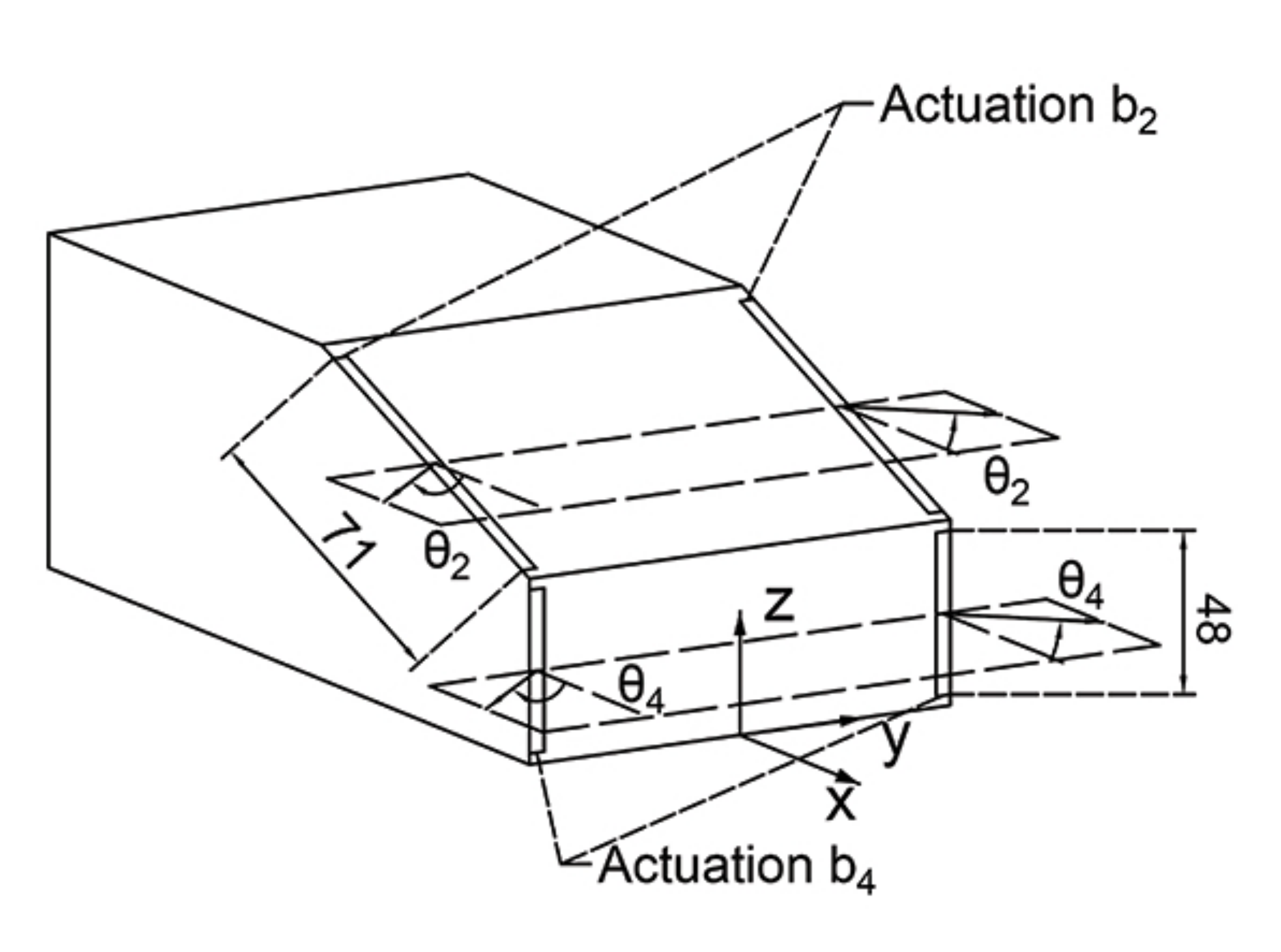}}
	\hfil
	\subfloat[]{
		\label{Fig:Configuration:Actuation:b}
		\includegraphics[height=4cm]{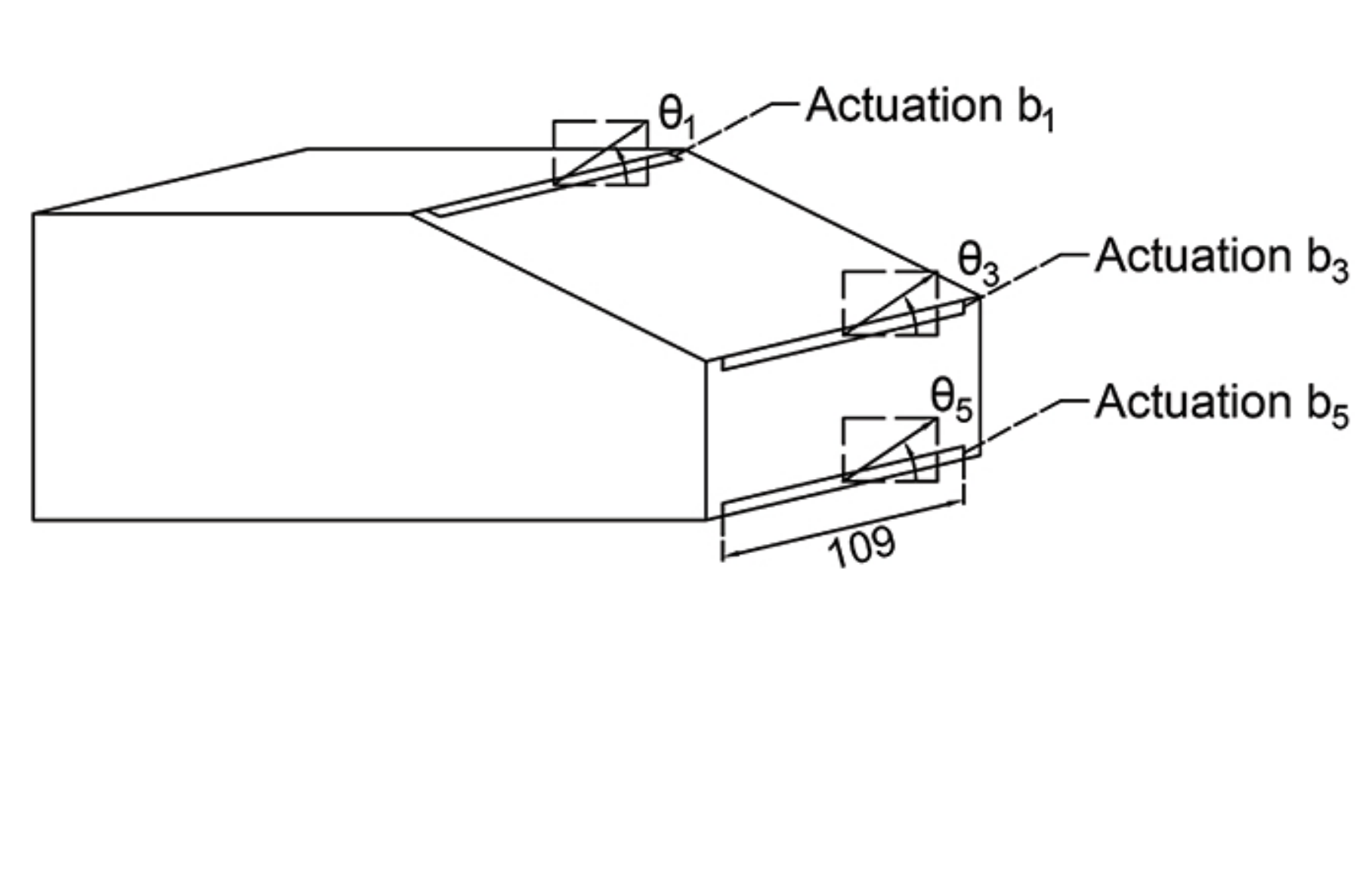}}	  
	\caption{Deployment and blowing direction of actuators on the rear window and  the vertical base.
			The angles $\theta_1$, $\theta_2$, $\theta_3$, $\theta_4$ and $\theta_5$ 
			are all defined to be positive when pointing outward (subfigure a) or upward (subfigure b).}
	\label{Fig:Configuration:Actuation}
\end{figure}

Five groups of steadily blowing slot actuators (figure \ref{Fig:Configuration:Actuation}) 
are deployed on all edges of the rear window and the vertical base. 
All slot widths are $2 \> \rm mm$.
The horizontal actuators at the top, middle and bottom side
have lengths of $109 \> \rm mm$.
The upper and lower sidewise actuators 
on the upper and vertical rear window
have a length of $71 \> mm$ and $48\> \rm mm$, respectively.
The actuation velocities $U_1, \ldots, U_5$ are independent parameters.
$U_1$ refers to the upper edge of the rear window, 
$U_3$ to the middle edge and
$U_5$ to the lower edge of the vertical base.
$U_2$ and $U_4$ correspond to the velocities
at the right and left sides of the upper and lower window, respectively.

Following the experiment by  \citet{Zhang2018jfm},
all blowing angles  can be varied 
as indicated in  figure  \ref{Fig:Configuration:Actuation}.
This study aims at minimizing drag 
as represented by the drag coefficient, $J=c_D$,
by varying the actuation control parameters. 
The actuation velocity amplitudes 
$U_i$, $i=1,\ldots,5$ are capped by twice 
of the single optimum value as discussed in \S~\ref{ToC:Ahmed:01D}.
The actuation angles $\theta_i$, $i=1,\ldots,5$  
are fixed to $0^\circ$, i.e., streamwise direction, in a 5-dimensional optimization. 
The actuation angles are later added into the input parameters 
in 10-dimensional optimization, 
with variable angles $\theta_1 \in[-35^\circ, 90^\circ]$, 
$\theta_2,\theta_3,\theta_4,\theta_5\in[-90^\circ, 90^\circ]$.

\subsection{RANS simulation}
\label{ToC:Ahmed:RANS}
\begin{figure}
	\centering
		\includegraphics[width=0.8\textwidth]{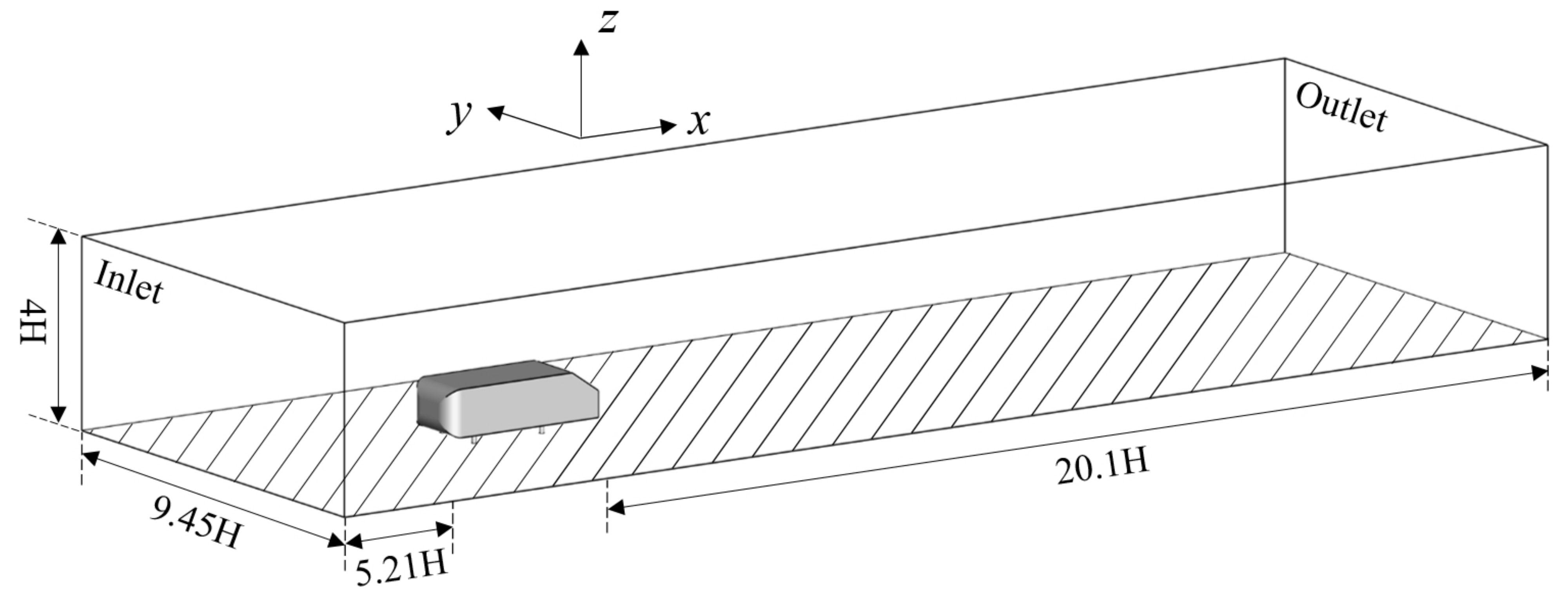}
	\caption{Computational domain of the RANS simulation.}
	\label{Fig:Simulation:domain}
\end{figure}
\begin{figure}
	\centering
		\includegraphics[width=0.7\textwidth]{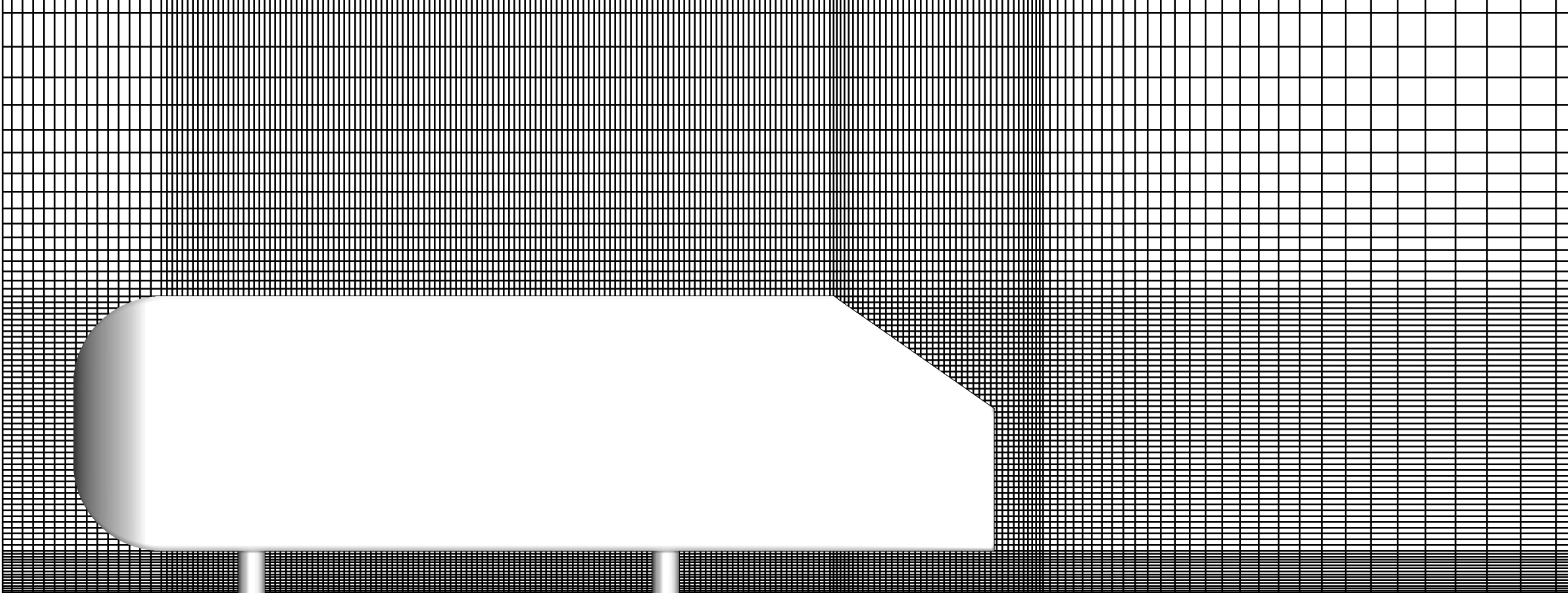}
	\caption{Side view of the part of the computational grid used for RANS.}
	\label{Fig:Simulation:Mesh}
\end{figure}

\begin{table}
	\begin{center}
		\def~{\hphantom{0}}
		\begin{tabular}{lccccccc}
		        Mesh grid points      & 2.5M   & 5M     & 10M\\[3pt]
			Drag Coefficient      & 0.294  & 0.313  & 0.318\\
		\end{tabular}
		\caption{Drag coefficient based on different mesh resolutions.}
		\label{Tab:S2:Mesh}
	\end{center}
\end{table}
A numerical wind tunnel (figure \ref{Fig:Simulation:domain}) 
is constructed using the commercial grid generation software Ansys ICEM CFD. 
The rectangular computational domain is bounded by  
$X_1 \le x \le X_2$, $0\le z \le H_T$, $ \vert y \vert \le W_T/2$.
Here, 
 $X_1 = -5.21 \> H, X_2 = 20.17 \> H$, $H_T = 4H$, and $W_T= 9.45H$. 
A coarse, medium and fine mesh 
using unstructured hexahedral computational grid are employed
in order to evaluate the performance of RANS method 
for the current problem with different mesh resolutions.
The statistics in Table \ref{Tab:S2:Mesh} show that using a finer mesh 
can be expected to have negligible improvement on the accuracy of the drag coefficient. 
Hence, the more economical medium mesh \ref{Fig:Simulation:Mesh} is used.
This mesh consists of 5 million elements and features dimensionless wall values
$\Delta x^+ = 20, \Delta y^+ = 3, \Delta z^+ = 30$.
In addition to resolving the boundary layer, 
the shear layers and the near-wake region, 
the mesh near the actuation slots is also refined.

Reynolds-Averaged Navier-Stokes (RANS) simulations 
using the realizable $k-\epsilon$ model 
with the constant parameters
 
\begin{equation*} 
\begin{aligned}
\sigma _k = 1.0, \sigma _\varepsilon = 1.2, C_2 = 1.9,
&\quad\quad
C_1 = \max  \left( 0.43, \frac{\eta}{\eta + 5 } \right),\\
\eta = \left ( 2\sum\limits_{i,j=1}^3 E_{ij} E_{ji} \right)^{1/2} \frac{k}{\varepsilon},
&\quad\quad
E_{ij}= \frac{1}{2} \left( \frac{\partial{u_i}}{\partial{x_j}} + \frac{\partial{u_j}}{\partial{x_i}} \right).
\end{aligned} 
\end{equation*}
$\sigma_k=1.0$, $\sigma_\epsilon=1.2$, $C_2=1.9$ are performed employing the commercial flow solver Ansys Fluent. 
The spatial discretization is based on a second-order upwind scheme in the form of SIMPLE scheme 
based on a pressure-velocity coupling method.
RANS simulation have been frequently and successfully 
been used to assess actuation effects from steady blowing \citep{ben2007generic,dejoan2005comparative,muralidharan2013numerical,viken2003flow}.
We deem RANS simulations to provide reasonable qualitative and approximately quantitative indications for actuator optimization and plan an experimental validation in the future.
Partially Averaged Navier-Stokes (PANS) simulations \citep{Han2013ijhff} 
and Large Eddy Simulation (LES) \citep{Krajnovic2009prsa,Brunn2006afc} 
are trusted higher-fidelity simulations 
for drag reduction with active flow control 
but are computationally orders of magnitudes more demanding.

\subsection{Formulation of an optimization problem based on streamwise blowing
at the top edge}
\label{ToC:Ahmed:01D}

\begin{figure}
	\centering
		\includegraphics[width=0.8\textwidth]{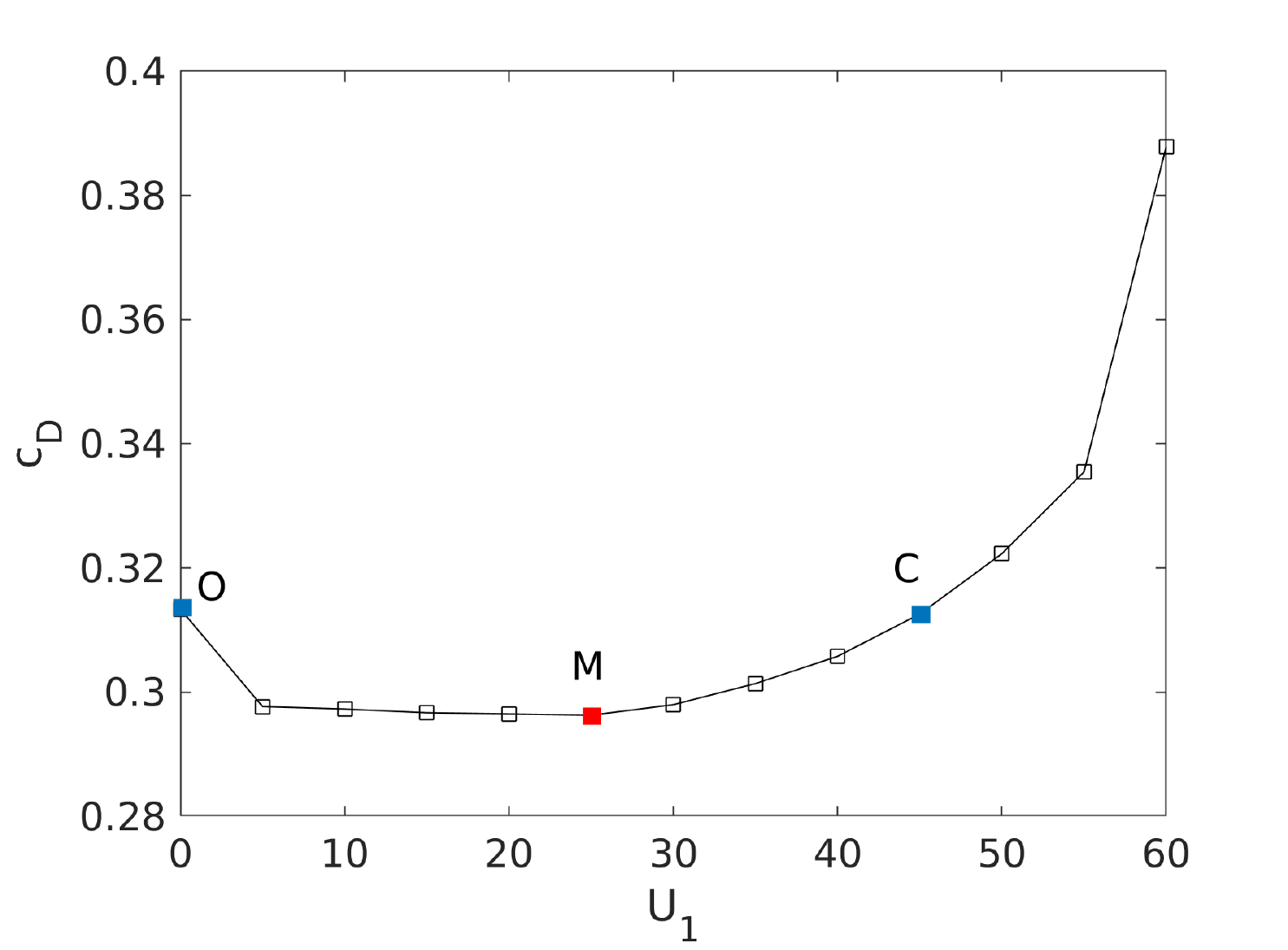}
	\caption{Drag coefficient as a function of the blowing velocity  $U_1$ of the streamwise-oriented top actuator. 
			Here, `$O$' marks the drag without forcing, `$M$' the best actuation, 
			and `$C$' the smallest actuation with worse drag than for unforced flow.}
	\label{Fig:01D:S4_Cd_U}
\end{figure}
The formulation and constraints 
of the optimization problem
is motivated by the drag reduction results 
from the top actuator blowing in streamwise direction.
Figure \ref{Fig:01D:S4_Cd_U} shows the drag coefficient
in dependency of streamwise blowing velocity,
all other actuators being off. 
The blowing velocity varies in increments of $5 \> \rm m/s$ 
from $0 \> \rm m/s$ to $60 \> \rm m/s$,
i.e., reaches twice the oncoming velocity.

The drag coefficient is quickly reduced by modest blowing,
has a shallow minimum near the actuation velocity $U_{b1}=25 \> \rm m/s$
before quickly increasing with more intense blowing.
This optimal value corresponds to $5/6$ of the oncoming velocity.
The best drag reduction is 5\% with respect to the unforced flow $C_d=0.3134$.
Near $U_1 = 45 \> \rm m/s$, the drag rapidly rises beyond the unforced value.

This behaviour motivates the choice of actuation parameters.
The first five actuation parameters 
are normalized jet velocities $b_i = U_i/U_{b1}$, $i=1,\ldots,5$ introduced in \S~\ref{ToC:Ahmed:Configuration}.
Thus, $b_1=1$ corresponds to minimal drag with a single streamwise-oriented top actuator.
All $b_i$ are capped by $2$:  $b_i \in [0,2]$, $i=1,\ldots,5$.
At $b_1=1.8$, point `$C$' in figure \ref{Fig:01D:S4_Cd_U}, 
actuation yields already drag increase.
The first vertex of the amoeba of the downhill simplex search is put at 
$b_1=b_2=b_3=b_4=b_5=1.8$.
From figure \ref{Fig:01D:S4_Cd_U}, 
we expect a drag minimum at lower values, 
hence the  next five vertices test the value $1.6$,
e.g. $(b_1,b_2,\ldots,b_5) = \left(1.8-0.2 \delta_{1,m-1}, 1.8-0.2 \delta_{2,m-1}, \ldots, 1.8-0.2\delta_{5,m-1} \right)$
for $m=2,\ldots,6$.
The downhill simplex algorithm 
may be expected to move to the outer border of the actuation domain,
if maximum drag reduction lies outside the domain,
thus indicating too restrictive constraints.
An example is drag reduction of wall turbulence \citep{Fernex2020prf}.

We refrain from starting already with a much larger actuation domain,
as the exploration with LHS and the proposed explorative gradient search
will consistently test too many large velocities.
An increase of the upper velocity bound by a factor 2, for instance, 
implies that only $2^{-5}$ or around  3\%
of uniformly distributed sampling points are in the original domain
and 97\% of the samples are outside.

The next five parameters characterize 
the deflection of the actuator velocity
with respect to the streamwise direction (see \S~\ref{ToC:Ahmed:Configuration}),
$b_{i+5} = \theta_i / (\pi/2) $, $i=1\ldots,5$,
and are normalized with $90^\circ$.
Now all $b_i$, $i=1,\ldots,10$ span an interval of width 2, 
except for the more limited deflection $b_6$ of the top actuator.
Summarizing, the  domain for the most general actuation reads
\begin{equation}
\Omega :=
\left\{  
\bm{b} \in {\cal R}^{10} \> \colon \>  
\begin{array}{ll}
b_i \in [0, 2]& \hbox{for } i=1,\ldots,5 \\
b_i \in \left[-35/90,1\right] &  \hbox{for } i=6\\
b_i \in [-1,1] & \hbox{for } i=7,\ldots,10  
\end{array}
\right\} .
\end{equation}
The choice of $\bm{b}$ as symbol shall remind about the control $B$-matrix in control theory
and is consistent with many earlier publications of the authors,
e.g.,  the review article by \citet{Brunton2015amr}.

\subsection{Optimization of the streamwise trailing edge actuation}
\label{ToC:Ahmed:05D}
The drag of the Ahmed body is
optimized with streamwise blowing from the five slot actuators.
We apply a simplex downhill search,
Latin hypercube sampling and the explorative gradient method
of \S~\ref{ToC:Ahmed:05D:Simplex}, \S~\ref{ToC:Ahmed:05D:LHS} and \S~\ref{ToC:Ahmed:05D:Combined},
respectively.

	\subsubsection{Downhill simplex algorithm}
	\label{ToC:Ahmed:05D:Simplex}

	\begin{table}
		\centering
		\def~{\hphantom{0}}
		\setlength{\tabcolsep}{5mm}		
		{
			\begin{tabular}{l|ccccccc}
			$m$    & $b_1$ & $b_2$ & $b_3$ & $b_4$ & $b_5$ & J \\ \hline \\[-8pt]
			1      & 1.8   & 1.8   & 1.8   & 1.8   & 1.8   & 0.4153\\
			2      & 1.6   & 1.8   & 1.8   & 1.8   & 1.8   & 0.4048\\
			3      & 1.8   & 1.6   & 1.8   & 1.8   & 1.8   & 0.4109\\
			4      & 1.8   & 1.8   & 1.6   & 1.8   & 1.8   & 0.3996\\
			5      & 1.8   & 1.8   & 1.8   & 1.6   & 1.8   & 0.4075\\
			6      & 1.8   & 1.8   & 1.8   & 1.8   & 1.6   & 0.4040\\
			\end{tabular}
		}
		\caption{Initial simplex ($m=1,\ldots,6$) for the five-dimensional downhill simplex optimization.
			$b_i$ are the normalized actuation velocities and $J$ corresponds to the  drag coefficient.}
		\label{Tab:05D:Simplex}
	\end{table}
	Following \S~\ref{ToC:Ahmed:01D}, 
	the downhill simplex algorithm is centered
	around $b_i=1.8$, $i=1,\ldots,5$ as first vertex
	and explores a lower actuation $b_{m-1}=1.6$ in all directions for vertices $m=2,\ldots,6$.
	Table \ref{Tab:05D:Simplex} shows the values of the individuals and corresponding cost.
	All vertices have a larger drag than for the unforced benchmark $C_d=0.3134$.
	And all vertices with $b_i=1.6$ are associated with a smaller drag 
	indicating a downhill slide to small actuation values
	consistent with the expectations from \S~\ref{ToC:Ahmed:01D}.

	\begin{figure}[htb]
		\subfloat[]{
		  \centering
		  \label{Fig:05D:S:a}
		  \includegraphics[width=0.9\textwidth]{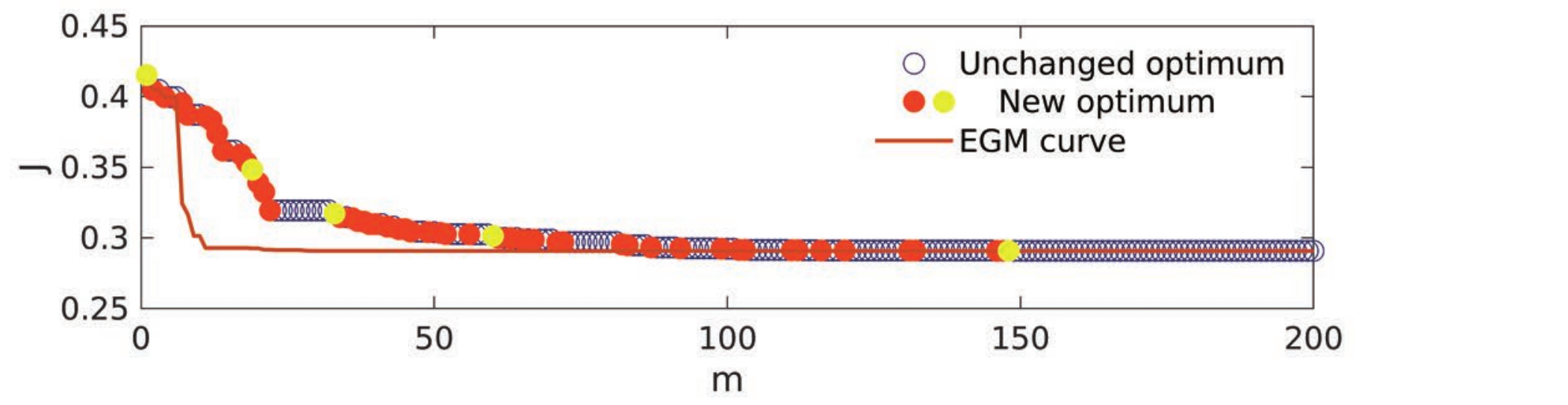}}
  
	  	\subfloat[]{
		  \centering
		  \label{Fig:05D:S:b}
		  \includegraphics[width=0.9\textwidth]{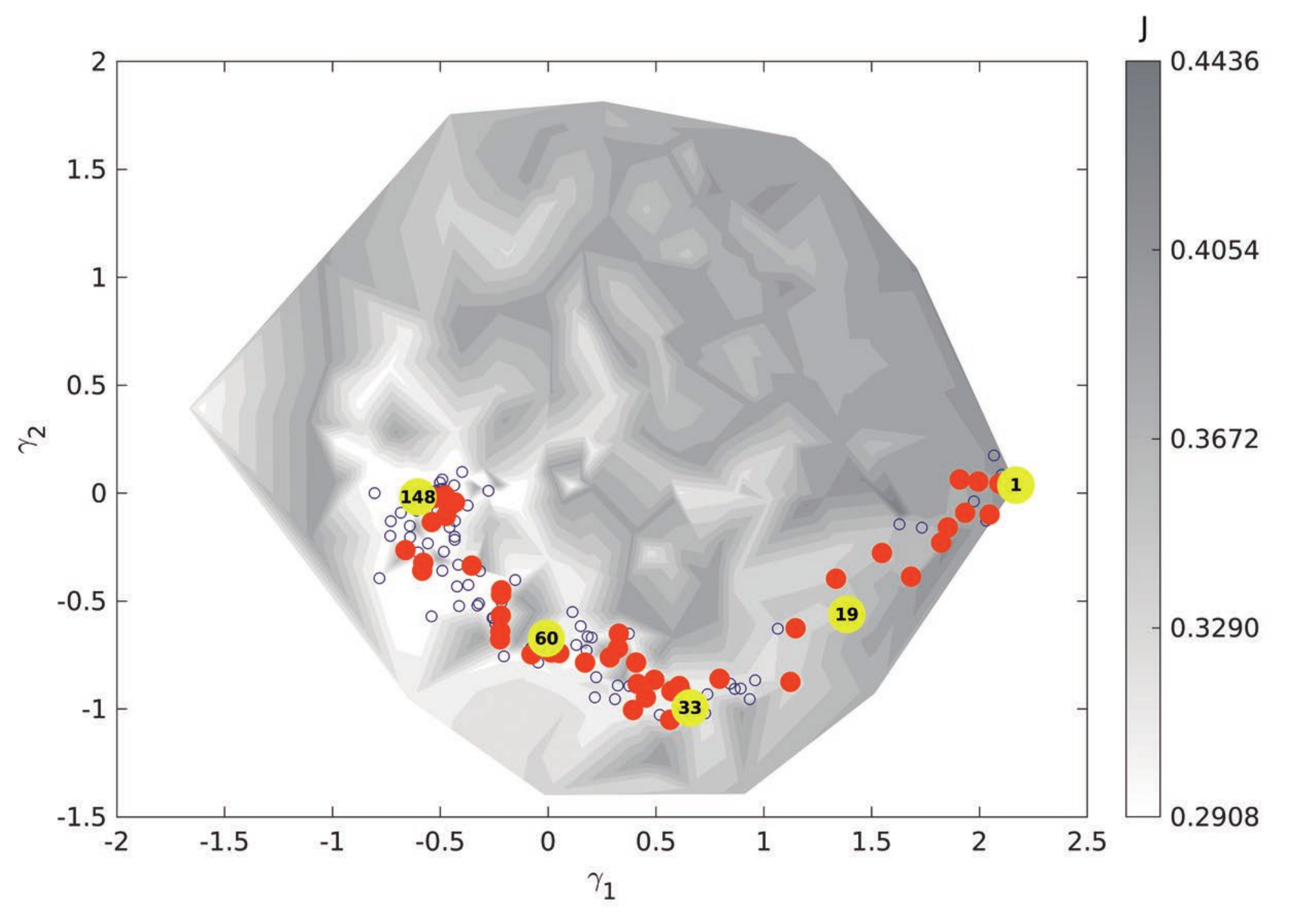}}  
		\caption{Optimization of five tangential jet actuator groups with DSM.
		  The top figure displays the best achieved drag reduction in terms of the number of evaluations (RANS simulations).
		  The bottom figure shows the 
		  proximity map of all evaluated actuations.
		  The contour plot corresponds to the interpolated cost function (drag coefficient) from all RANS simulations of this section.
		  As in \S~\ref{ToC:Method Comparison},
		  solid red circles mark newly find optima 
		  while open blue circles mark unsuccessful tests of cost functions.
		  For better interpretability,
		  select newly found optima are highlighted 
		  with by a yellow solid circle like in  \S~\ref{ToC:Fluidic Pinball}.
		  In the control landscape, 
		  these circles are marked with the index $m$.}
	  	\label{Fig:05D:S}
  	\end{figure}
	Figure \ref{Fig:05D:S} (top) shows the evolution of the downhill simplex algorithm 
	with 200 RANS simulations.
	Like in \S~\ref{ToC:Method Comparison}, solid red circles mark newly found optima 
	while open blue circles record the best actuation so far.
	The drag quickly descends after staying shortly on a plateau at $m \approx 20$.
	From there on, the descent becomes gradual.
	The optimal drag $J=0.2908$ 
	is reached with the 148th RANS simulation and corresponds to 7\% drag reduction.
	The optimal actuation reads 
	$b_1=0.7264$, 
	$b_2=0.5508$, 
	$b_3=0.1533$, 
	$b_4=0.6746$, 
	$b_5=0.7716$.
	While the middle horizontal jet has small amplitude,
	the other actuation velocities on the four edges of the Ahmed body 
	are $55$\% to $77$\% of the optimal value achieved 
	with single actuator.

	Figure \ref{Fig:05D:S} (bottom) illustrates the downhill search 
	in a control landscape $J(\gamma_1,\gamma_2)$ described in \S~\ref{ToC:Method}.
	Here $(\gamma_1,\gamma_2)$  feature vectors defining a proximity map
	of the five-dimensional actuation parameters $(b_1,\ldots, b_5)$.
	This landscape indicates a complex topology 
	of the five-dimensional actuation space
	by many local maxima and minima in the feature plane.
	This complexity may explain why most simplex steps
	did not yield a better cost.
	The feature coordinate $\gamma_1,\gamma_2$ 
	arise from a kinematic optimization process 
	and have no inherent meaning.
	The simplex algorithm is seen to crawl from right $\bm{\gamma} \approx (2,0)$
	to the assumingly global minimum at $\bm{\gamma} \approx [-0.6,0]^{\rm T}$ 
	through an elongated curved valley. 
        The simulation results for $m=1$, $19$, $33$, $60$ and $148$ are marked with yellow solid circles. 
	Note that the construction of this proximity map includes 
	also undisplayed actuation data from LHS and EGM
	so that the control landscapes 
	remain identical for all discussed five-dimensional actuations.

	\subsubsection{Latin hypercube sampling}
	\label{ToC:Ahmed:05D:LHS}
	
	\begin{figure}[htb]
		\subfloat[]{
		  \centering
		  \label{Fig:05D:LHS:a}
		  \includegraphics[width=0.9\textwidth]{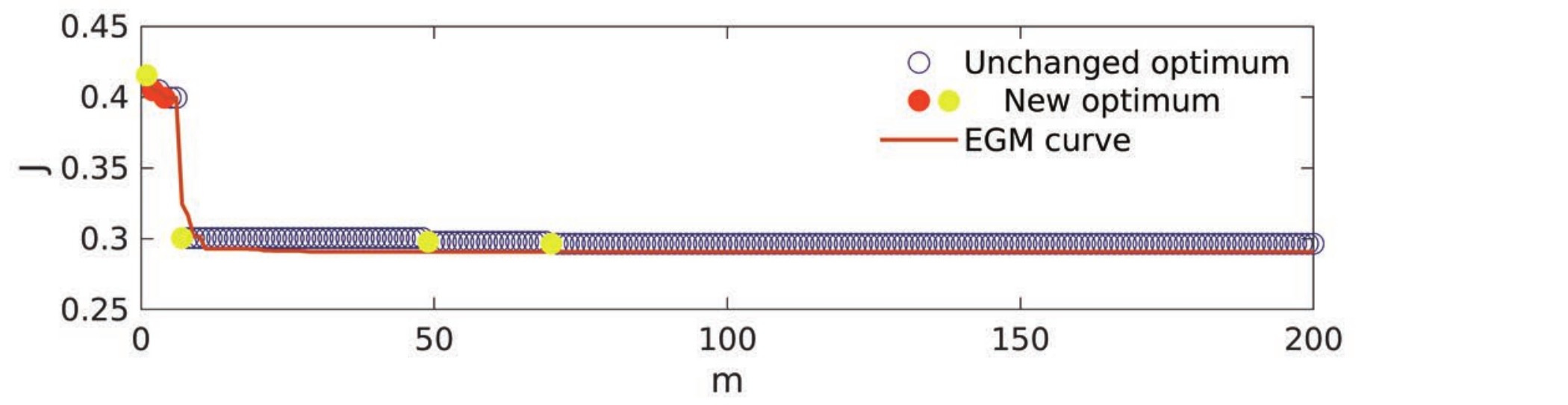}}
  
	  	\subfloat[]{
		  \centering
		  \label{Fig:05D:LHS:b}
		  \includegraphics[width=0.9\textwidth]{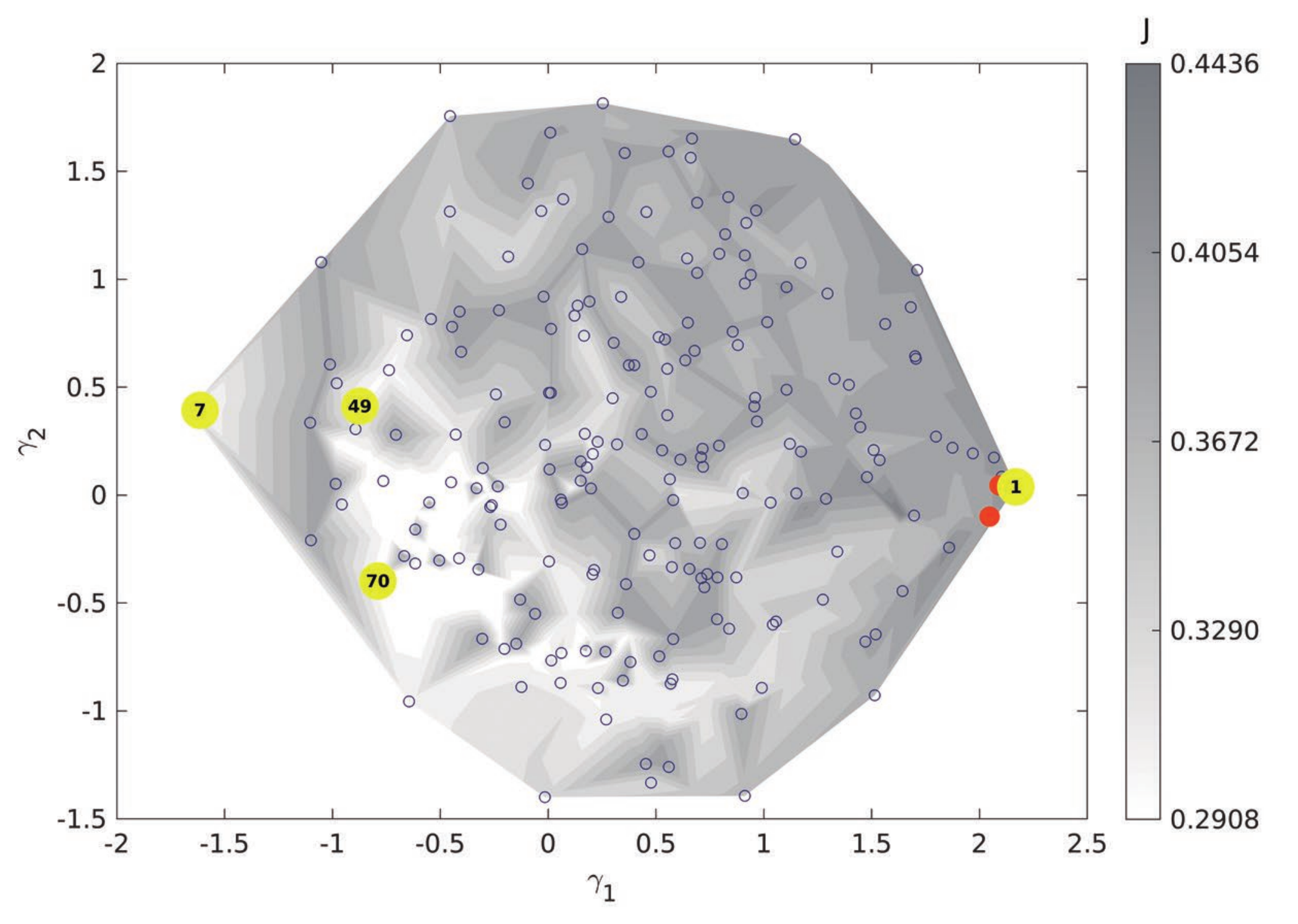}}  
		\caption{Same as figure \ref{Fig:05D:S}, but with LHS.}
		\label{Fig:05D:LHS}
  	\end{figure}
	Figure \ref{Fig:05D:LHS} (top) shows 
	the slow learning process associated with Latin hypercube sampling (LHS)
	starting with the simplex reference point $b_1=\ldots=b_5=1.8$.
	Apparently the optimization is ineffective.
	Only 3 new optima are successively obtained in 200 RANS simulations.
	The remaining simulations yield worse drags than the best discovered before.
	At the 70th RANS simulation, 
	the best drag coefficient of
	 $C_d=0.2928$, with 
	 $b_1=0.0994$, 
	 $b_2=0.9587$, 
	 $b_3=0.1276$, 
	 $b_4=0.0289$ and
	 $b_5=1.0393$. 
	corresponds to 5\% reduction like the one-dimensional top actuator 
	$b_1=1$, $b_2=b_3=b_4=b_5=0$.
	Intriguingly, only the upper side and bottom actuator have $b_i$ amplitudes near unity
	while remaining parameters are less than 13\% of the one-dimensional optimum.
	These results show that near optimal drag reductions 
	can be achieved with quite different actuations.
	Moreover, individual actuation effects are far from additive.
	Otherwise, the almost complimentary LHS optimum for actuators 2--5 
	and the one-dimensional optimum of \S~\ref{ToC:Ahmed:01D}
	should yield 10\% reduction with
	$b_1 \approx 1$, $b_2\approx 1$, $b_3\approx 0.13$, $b_4 \approx 1$ and $b_5\approx 1$.

	Figure \ref{Fig:05D:LHS} (bottom) shows the LHS in the control landscape.
	In the first iteration, LHS jumps to the opposite site of domain and finds better drag.
	The next successive two improvements are in a good terrain
	but the optimum  at $m=70$ is still far from the assumingly global minimum at $\bm{\gamma} \approx (-0.6,0)$ (see figure \ref{Fig:05D:S}).
	The exploratory steps uniformly cover the whole range of feature vectors.

	\subsubsection{Explorative gradient method}
	\label{ToC:Ahmed:05D:Combined}
	
	\begin{figure}[htb]
		\subfloat[]{
		  \centering
		  \label{Fig:05D:EGM:a}
		  \includegraphics[width=0.9\textwidth]{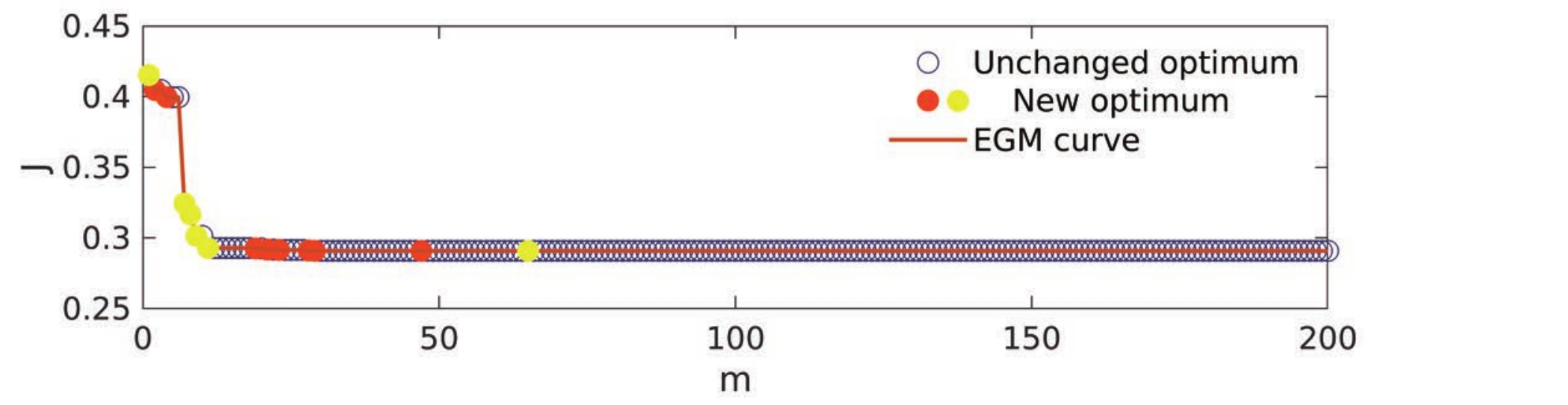}}
  
	  	\subfloat[]{
		  \centering
		  \label{Fig:05D:EGM:b}
		  \includegraphics[width=0.9\textwidth]{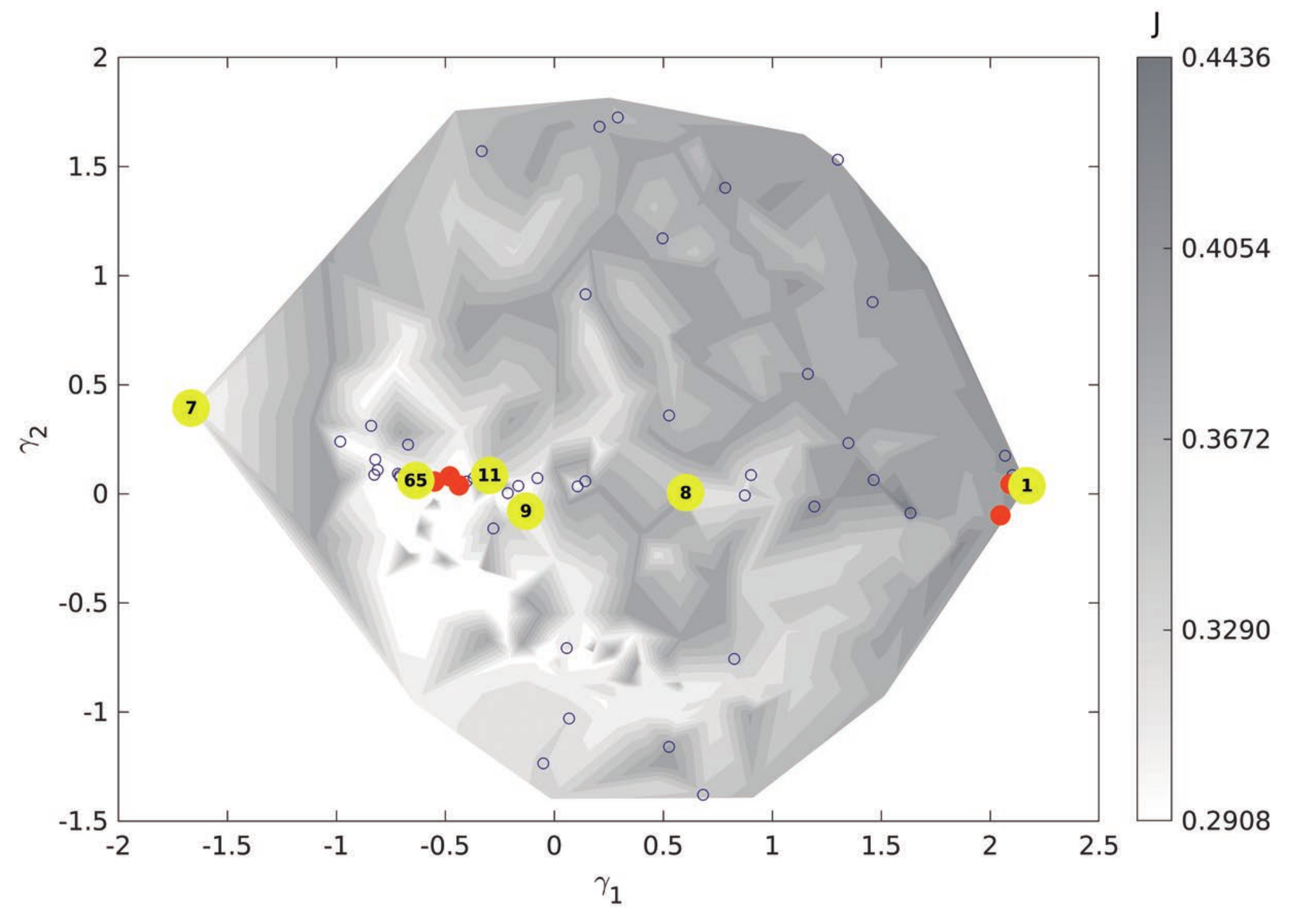}}  
		\caption{Same as figure \ref{Fig:05D:S}, but with EGM.}
		\label{Fig:05D:EGM}
  	\end{figure}
	From the  figure \ref{Fig:05D:EGM}, 
	the explorative gradient method is seen to converge much faster than the downhill simplex algorithm.
	The best actuation is found at the 65th RANS simulation
	yielding the same drag coefficient  $C_d=0.2908$ of the downhill simplex algorithm
	with only slightly different actuation parameters  
	$b_1=0.6647$, 
	$b_2=0.4929$, 
	$b_3=0.1794$, 
	$b_4=0.7467$ and
	$b_5=0.7101$.
	
	The fast convergence of the explorative gradient method is  
	initially surprising since up to 50\% of the steps are for explorative purposes,
	i.e., shall identify distant minima.
	However, the control landscape in figure \ref{Fig:05D:LHS} 
	reveals how the explorative LHS steps help the algorithm to prevent 
	the long and painful march through the long and curved valley.
	At $m=7$, an explorative step leads to the opposite side of control landscape with a better cost value.
	Then, the subsequent iterations quickly lead near global minimum at $m=11$.
	The proposed new algorithm operates like a visionary mountain climber,
	who performs not only local uphill steps 
	but sends drones  to the remotest location
	to find better mountains and terrains.

\subsection{Optimization of the directed trailing edge actuation}
\label{ToC:Ahmed:10D}

In this section, 
the actuation space is enlarged by the jet directions of all slot actuators.
The jets may now be directed inwards or outwards 
as discussed in \S~\ref{ToC:Ahmed:Configuration}.
The actuation optimization for drag reduction 
is performed with explorative gradient method (\S~\ref{ToC:Ahmed:10D:Combined}).
The unforced and three actuated Ahmed body wakes 
are investigated in  \S~\ref{ToC:Ahmed:10D:Discussion}. 

    \subsubsection{Explorative gradient method}
    \label{ToC:Ahmed:10D:Combined}

\begin{figure}[htb]
		\subfloat[]{
		  \centering
		  \label{Fig:10D:EGM:a}
		  \includegraphics[width=0.9\textwidth]{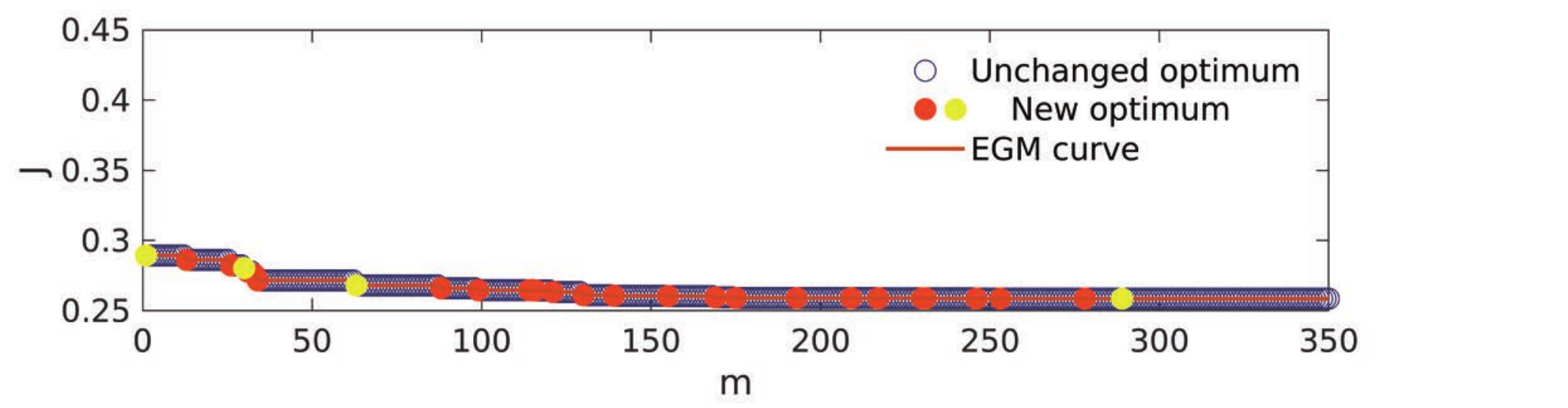}}
  
	  	\subfloat[]{
		  \centering
		  \label{Fig:10D:EGM:b}
		  \includegraphics[width=0.9\textwidth]{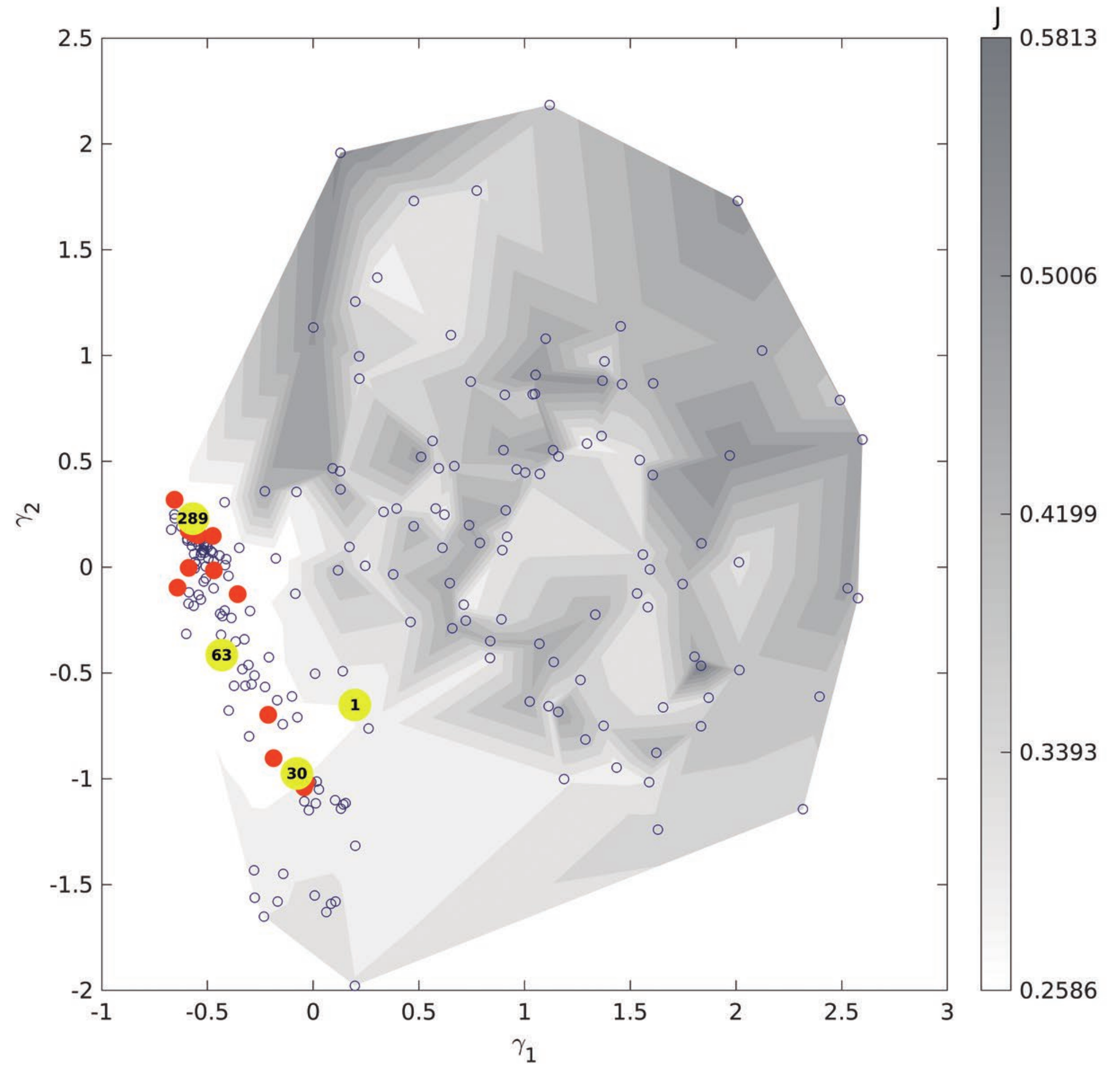}}  
		\caption{Same as figure \ref{Fig:05D:EGM}, but for the 10-dimensional optimization of the orientable trailing edge actuation with EGM.}
		\label{Fig:10D:EGM}
  	\end{figure}

    We employ the explorative gradient method 
    as best performing method of \S~\ref{ToC:Ahmed:05D}
    for the 10-dimensional actuation optimization problem.
    The search is accelerated by starting with a simplex centered 
    around the optimal actuation of the five-dimensional problem.
    The first vertex of table \ref{tab:initial_10D} contains this optimal solution.
    The cost is $4\permil$ lower than the previous section
    as the RANS integration for the first flow is not fully converged.
    The next five vertices represent isolated actuations at the optimal value 
    but directed $45^\circ$ outwards for the side edges 
    and upwards for the middle horizontal actuator.
    The corresponding drag values are larger.
    The next five vertices deflect the jets in opposite direction by $45^\circ$ or the maximum $35^\circ$ of the top actuator,  
    giving rise smaller drag than the previous deflection. 
    The drag of middle horizontal actuator remains close to the unforced benchmark
    because the jet velocity is small.
    \begin{table}
        \begin{center}
            \def~{\hphantom{0}}
            \begin{tabular}{l|cccccccccccc}
                $m$  & $b_1$    & $b_2$    & $b_3$    & $b_4$    & $b_5$    & $b_6$   & $b_7$ & $b_8$ & $b_9$ & $b_{10}$ & $J$ \\ \hline \\[-6pt]
                1      & 0.6647   & 0.4929   & 0.1794   & 0.7467   & 0.7101   & 0       & 0     & 0     & 0     & 0      & 0.2895\\
                2      & 0.6647   & 0        & 0        & 0        & 0        & 1/2     & 0     & 0     & 0     & 0      & 0.3268\\
                3      & 0        & 0.4929   & 0        & 0        & 0        & 0       & 1/2  & 0     & 0     & 0      & 0.3226\\
                4      & 0        & 0        & 0.1794   & 0        & 0        & 0       & 0     & 1/2   & 0     & 0      & 0.3168\\
                5      & 0        & 0        & 0        & 0.7467   & 0        & 0       & 0     & 0     & 1/2   & 0      & 0.3476\\
                6      & 0        & 0        & 0        & 0        & 0.7101   & 0       & 0     & 0     & 0     & 1/2    & 0.3060\\
                7      & 0.6647   & 0        & 0        & 0        & 0        & -35/90  & 0     & 0     & 0     & 0      & 0.3091\\
                8      & 0        & 0.4929   & 0        & 0        & 0        & 0       & -1/2  & 0     & 0     & 0      & 0.3085\\
                9      & 0        & 0        & 0.1794   & 0        & 0        & 0       & 0     & -1/2  & 0     & 0      & 0.3187\\
                10     & 0        & 0        & 0        & 0.7467   & 0        & 0       & 0     & 0     & -1/2  & 0      & 0.3001\\
                11     & 0        & 0        & 0        & 0        & 0.7101   & 0       & 0     & 0     & 0     & -1/2   & 0.3354\\
                
            \end{tabular}
            \caption{Initial individuals in the optimization of the directed trailing edge actuation. 
    $b_i$, $i=1,2,3,4,5$ represent the actuation amplitudes $U_i$ of the $i$th actuator.
    $b_i$, $i=6,7,8,9,10$ denotes the actuation angle $\theta_i$ 
    of the $(i-5)$th actuator. $J$ is the drag coefficient.}
            \label{tab:initial_10D}
        \end{center}
    \end{table}
    Figure \ref{Fig:10D:EGM} (top) illustrates the convergence of the explorative gradient method.
    After 289 RANS simulations,
    a drag coefficient of $0.2586$ is achieved corresponding to a 17\% drag reduction.
    The optimal actuation values read 
    $b_1=0.8611$, 
    $b_2=0.9856$, 
    $b_3=0.0726$, 
    $b_4=1.0089$, 
    $b_5=0.8981$, 
    $b_6=-0.3000$ corresponding to $\theta_1 = -27^\circ$, 
    $b_7=-0.4666$ ($\theta_2 = -42^\circ$), 
    $b_8=0.7444$ ($\theta_3=67^\circ$), 
    $b_9=-0.4888$ ($\theta_3=-44^\circ$), and
    $b_{10}=0.2444$  ($\theta_3=-22^\circ$).
    All outer actuators have velocity amplitudes near unity
    and are directed inwards, i.e., emulate Coanda blowing.
    The third middle actuator blows upward with low amplitude.
    The strong inward blowing seems to be related to the additional 10\% drag reduction
    as compared to the 7\% of streamwise actuation.

    Figure \ref{Fig:10D:EGM} (bottom) shows the search process in a proximity map.
    It should be noted that this control landscape is based on data in a ten-dimensional actuation space
    and hence different from the 5-dimensional space in \S~\ref{ToC:Ahmed:05D}.
    The algorithm quickly descends in the valley 
    while many exploration steps probe suboptimal terrain.
    One reason for this quick landing in good terrain
    is  the chosen initial simplex around the optimized actuation in the five-dimensional subspace.

    \begin{figure}
        \centering
            \includegraphics[width=0.9\textwidth]{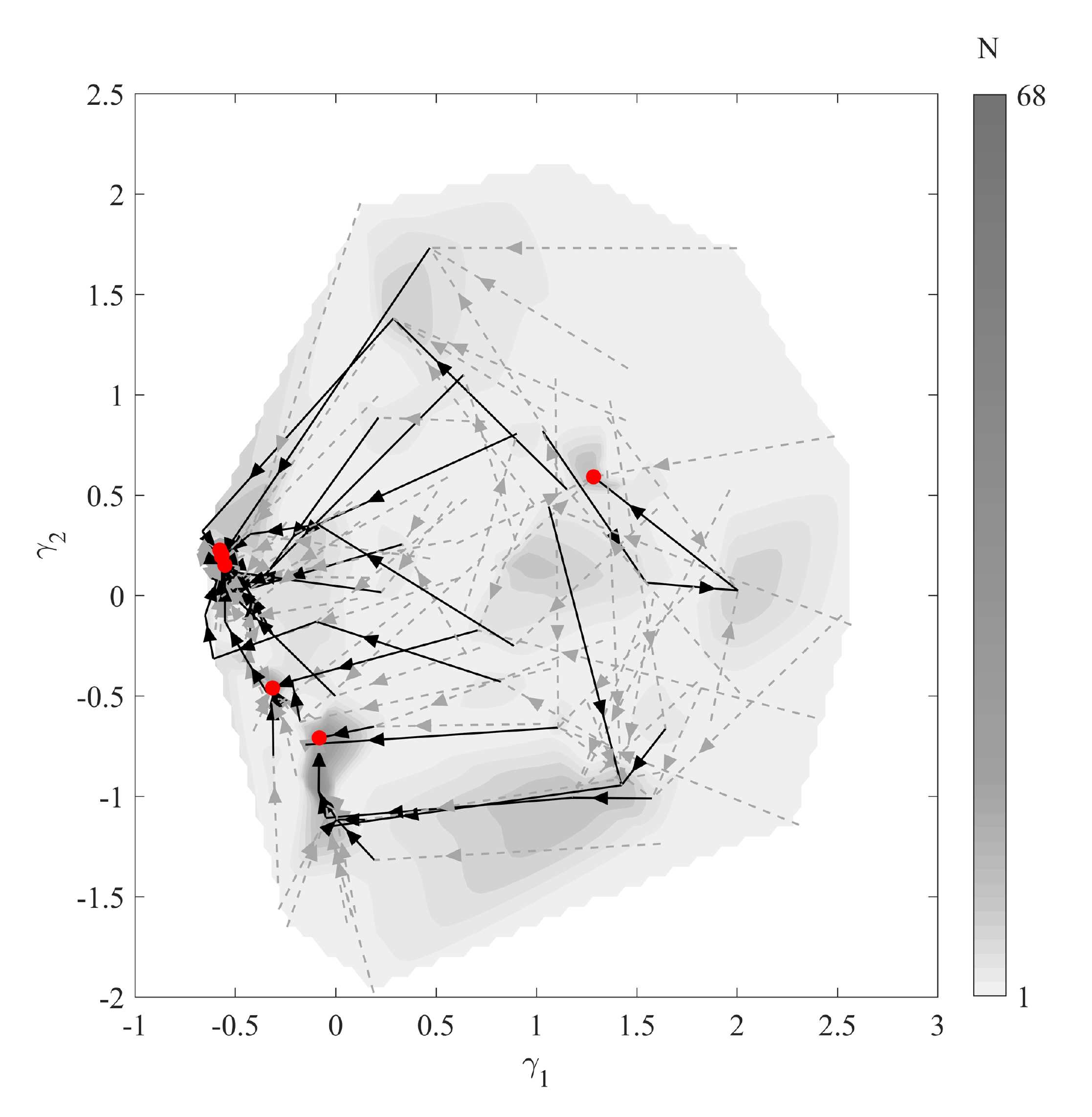}
        \caption{Steepest decent lines of the control landscape depicted in  figure \ref{Fig:10D:EGM}.
    For details see text.
        }
        \label{Fig:10D:Steepest descent lines}
    \end{figure}
    The topology of the control landscape of figure \ref{Fig:10D:EGM}
    is investiged with discrete steepst descent lines connecting neighboring data points in figure \ref{Fig:10D:Steepest descent lines}.
    For each investigated actuation vector, 
    the nearest five neighbours are considered.
    If all neighbours have higher drag, the vector is considered 
    as a local minimum and marked by a red point.
    Otherwise, 
    a gray dashed arrow is plotted to the best of these neighbour.
    This steepest descent is continued until a local minimum is reached.
    The corresponding path is called (discrete) steepest descent line.
    Line segments shared by at least 10 of these pathes
    may be considered as important valleys towards the minimum and highlighted as black solid arrow.
    The visiting times of each individual are marked by colorcoding.
    The global minimum of all data points is visited most, 54 steepest descend lines end here.
    The resulting pathways of `mountain trails' to `expressways' 
    may give an indication of the directions to be expected from local search algorithms.
    Moreover, crossing steepest descent lines indicate that the two-dimensional proximity maps 
    oversimplifies a higher-dimesional landscape structure.
    Intriguingly, the steepest descent line become more aligned to each other 
    in the valleys leading to global data minimum, i.e., the most relevant regions for optimization.

    \subsubsection{Discussion of streamwise and directed jet actuators}
    \label{ToC:Ahmed:10D:Discussion}
    In the following, 
    the physical structures associated 
    with the optimized one-, five- and ten-dimensional actuation are discussed.
    Evidently, more degrees of freedom are associated 
    with more opportunities for drag reduction. 
    Expectedly, the drag reduces by 5\% to 7\% to 17\%  
    as the dimension of the actuation parameters increase from 1 to 5 to 10, respectively.
    Intriguingly, 
    the increase of drag reduction from the optimized top actuator 
    to the best 5 streamwise actuators is only 2\%.
    For the square-back Ahmed body, \citet{Barros2015phd} experimentally observed
    that the individual drag reductions 
    from the streamwise blowing actuators on the four trailing edges
    roughly add up to the total drag reduction of 10\% with all actuators on. 
    This additivity of actuation effects is not corroborated for the slanted low-drag Ahmed body.
    Intriguingly, the inward deflection of the jet-slot actuators substantially decreases drag by 10\%.
    This additional drag reduction of 10\% has also been observed for the square-back Ahmed body
    when the horizontal jets were deflected inward with Coanda surfaces on all four edges \citep{Barros2016jfm}.
    Improved drag reduction with inward 
    as opposed to tangential blowing
    was also observed for the $35^\circ$ high-drag Ahmed body
    \citep{Zhang2018jfm} and the square back version \citet{Schmidt2015ef}.

    Table \ref{Tab:ActuationParameters} summarizes the discussed flows,
    associated drag reduction and actuation parameters.
    For brevity, we refer to flows 
    with no, one-dimensional, five-dimensional and ten-dimensional actuation spaces
    as case A, B, C and D, respectively.
    The actuation energy may be conservatively estimated by 
    the energy flux through all jet actuators:
    $ \sum_{i=1}^5 \int d\! A_i \rho U_i^3/2$.
    Here,  the actuation jet fluid is assumed 
    to be accelerated from $0$ to the actuation jet velocity $U_i$
    and then deflected after the outlet, e.g.,  via a Coanda surface.
    In this case, the actuation energy of cases $B$, $C$ and $D$
    would correspond to $3.2 \%$, $3.0 \% $ and $7.9 \% $ of the parasitic drag power, respectively. 
    This expenditure is significantly less than the saved drag power.
    The ratio from saved drag power to actuation energy is comparable
    for a truck model where steady Coanda blowing with 7\% energy expenditure 
    yields a 25\% drag reduction \citep{Pfeiffer2014aiaa}.
    This estimate should not be taken too literally 
    as actuation energy strongly depends on the realization of the actuator.
    It would be less, more precisely
    $ \sum_{i=1}^5 \int d \! A_i  \> \rho \cos (\theta_i) \> U_i^3/2$,
    when the actuation jet fluid leaves the Ahmed body 
    through a slot directed with the jet velocity, 
    and can be  expected much less
    when this fluid is taken from the oncoming flow,
    e.g.,  from the front of the Ahmed body.

\begin{figure}[htb]
        \centering
        \subfloat[]{
            \label{Fig:Discussion:Q:a}
            \includegraphics[width=0.45\textwidth]{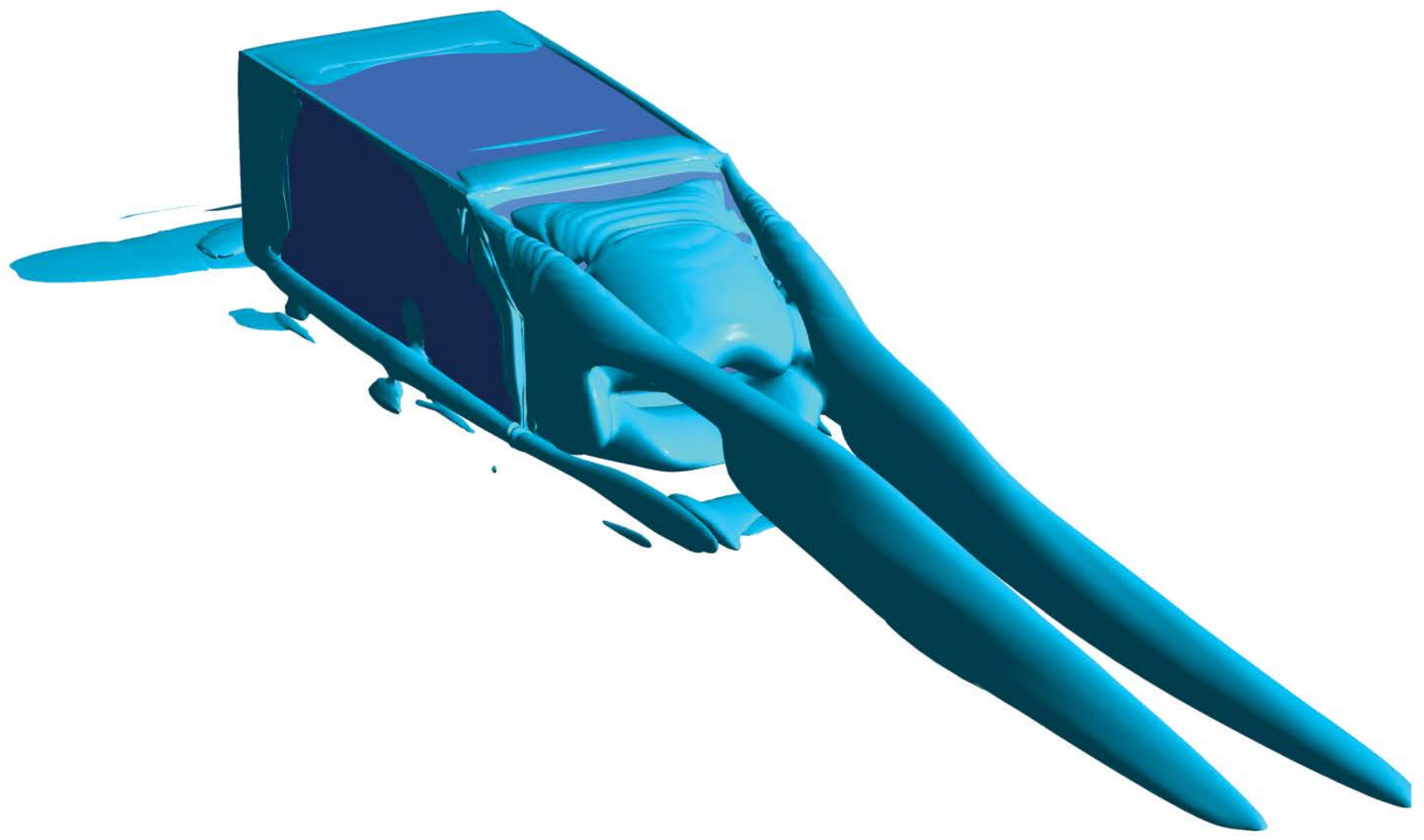}}
        \hfil
        \subfloat[]{
            \label{Fig:Discussion:Q:b}
            \includegraphics[width=0.45\textwidth]{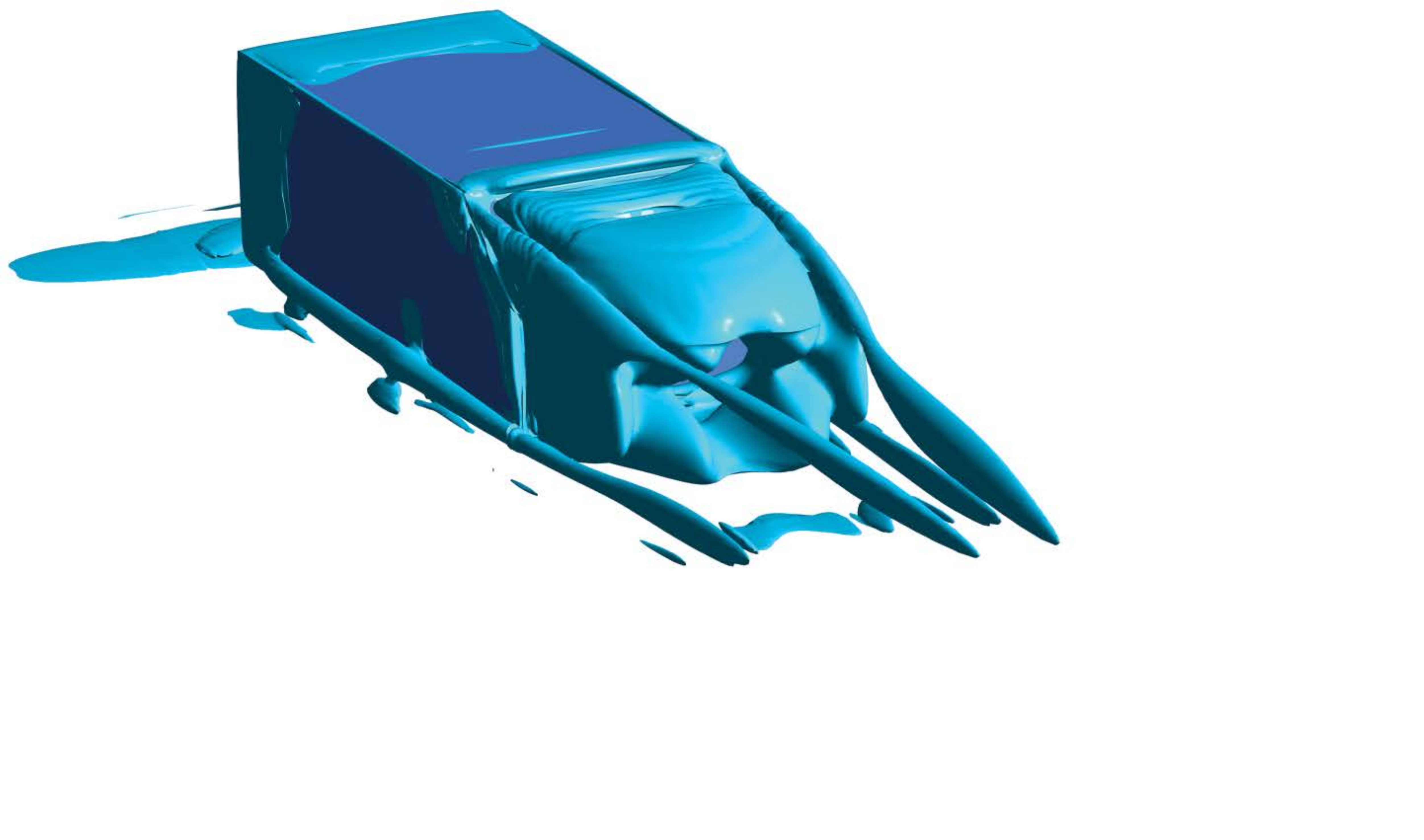}}

        \subfloat[]{
            \label{Fig:Discussion:Q:c}
            \includegraphics[width=0.45\textwidth]{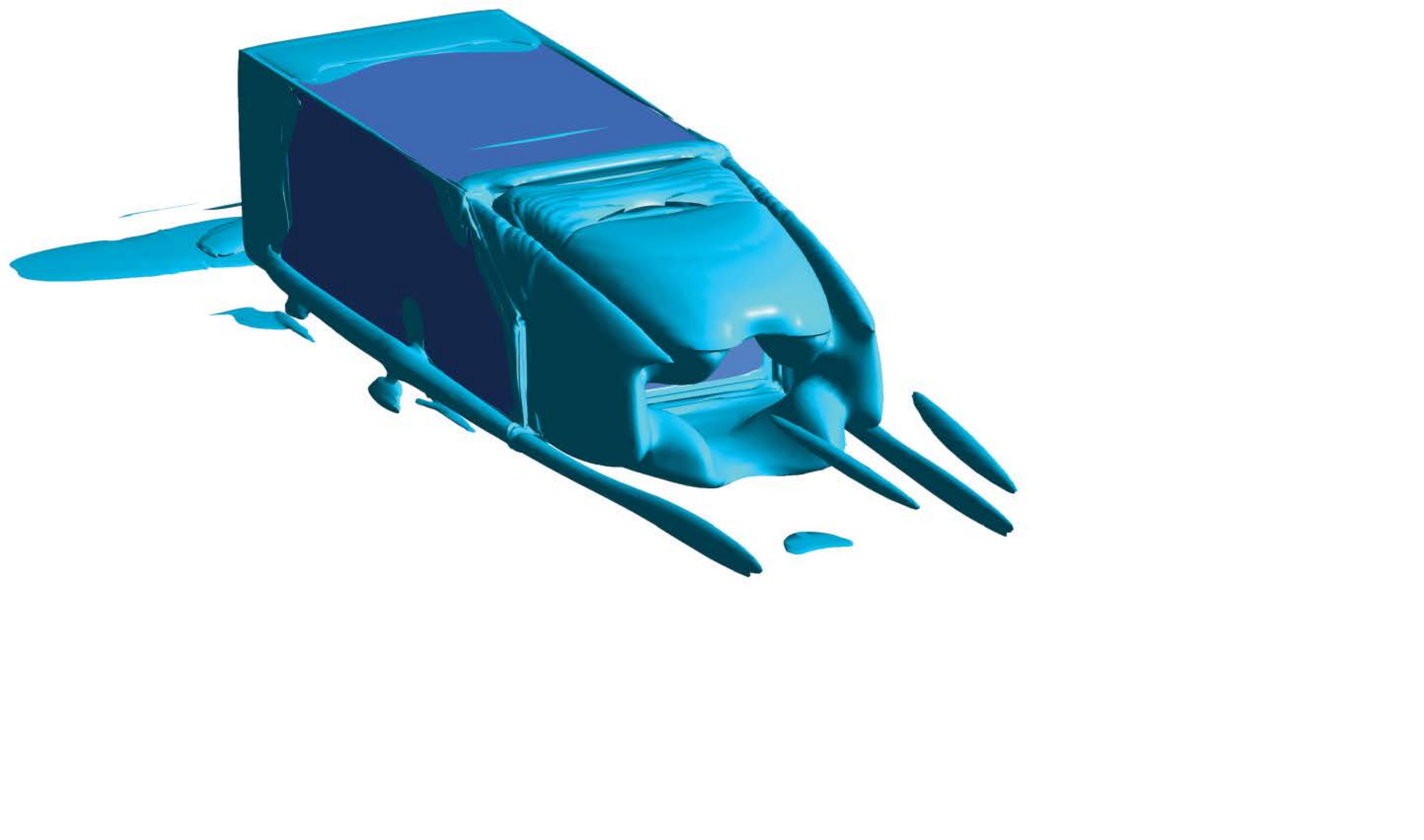}}
        \hfil
        \subfloat[]{
            \label{Fig:Discussion:Q:d}
            \includegraphics[width=0.45\textwidth]{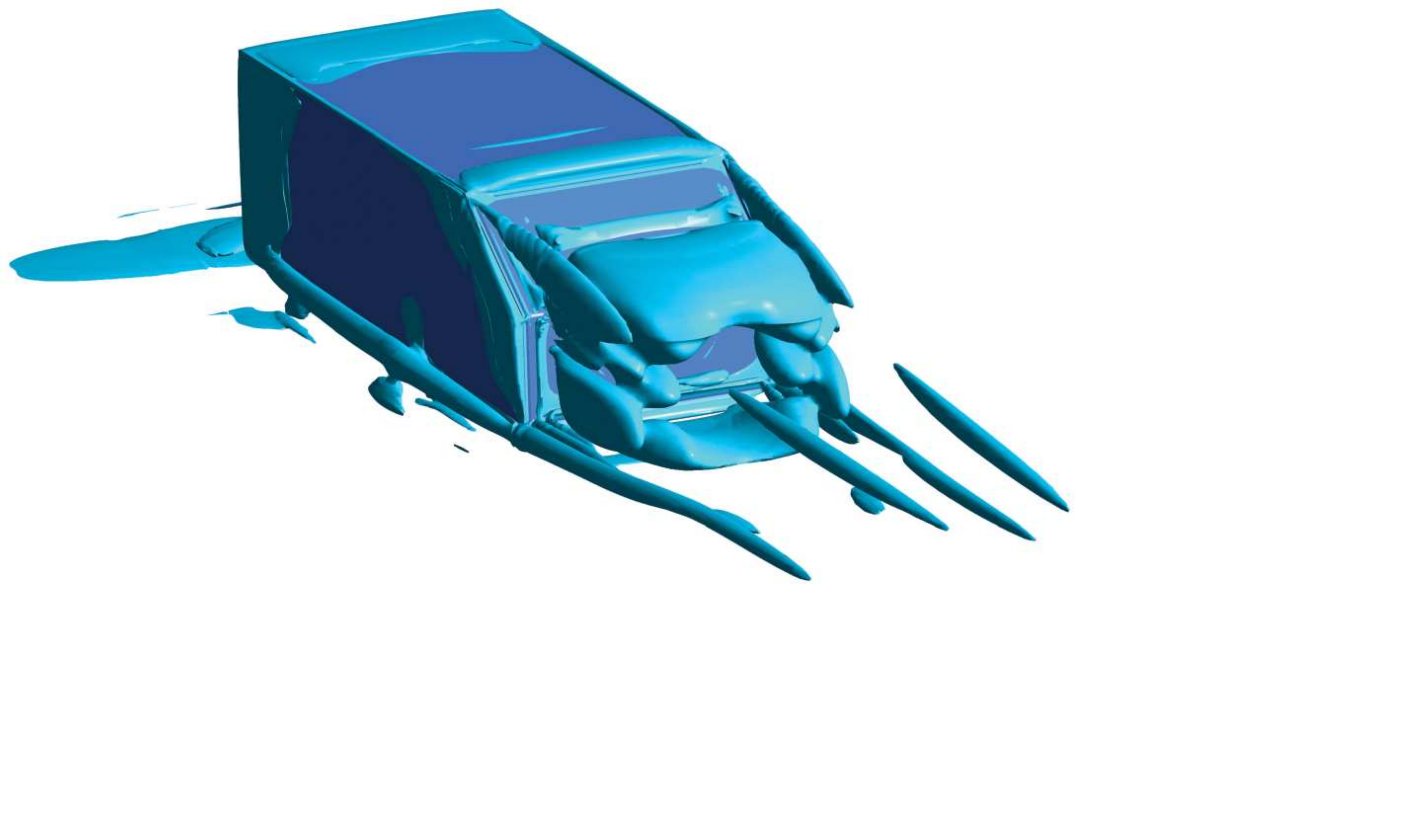}}
        \caption{Okubo-Weiss parameter $Q$  of flow. 
                \textit{a}) without control and under 
                \textit{b}) 1D, 
                \textit{c}) 5D and 
                \textit{d}) 10D control respectively, where $Q=15000 /s^2$.}
        \label{Fig:Discussion:Q}
    \end{figure}

    \begin{table}
    \begin{center}
    \begin{tabular}{ll|l|l|l|l|l}
    \multirow{2}*{Case} & \multirow{2}*{Drag reduction}   & \multicolumn{5}{l}{Actuation parameters}  \\
    \cline{3-7}
        ~ & ~ & Top & Upper side & Middle & Lower side & Bottom 
    \\ \hline 
    A) & \hbox to 3em{\rule{0pt}{0pt}\hfill $0\%$} 
    &  --- & --- & --- & --- & --- 
    \\
    B) & \hbox to 3em{\rule{0pt}{0pt}\hfill $5\%$}
    & $b_1=1$ 
    & $b_2=0$ 
    & $b_3=0$ 
    & $b_4=0$ 
    & $b_5=0$
    \\
    C) & \hbox to 3em{\rule{0pt}{0pt}\hfill $7\%$}
    & $b_1=0.6647$ 
    & $b_2=0.4929$ 
    & $b_3=0.1794$ 
    & $b_4=0.7467$ 
    & $b_5=0.7101$
    \\
    D) & \hbox to 3em{\rule{0pt}{0pt}\hfill $17\%$}
    & $b_1=0.8611$
    & $b_2=0.9856$ 
    & $b_3=0.0726$ 
    & $b_4=1.0089$ 
    & $b_5=0.8981$
    \\ & 
    &  $\theta_1=-27^\circ$ 
    & $\theta_2=-42^\circ$
    & $\theta_3=67^\circ$ 
    & $\theta_3=-44^\circ$
    & $\theta_3=-22^\circ$
    \end{tabular}
    \end{center}
    \caption{Investigated optimized actuations in comparison to the unforced benchmark.
    The table shows the achieved drag reduction and corresponding actuation parameters for
    A) the unforced benchmark, and for the optimized B) top streamwise actuator, C) all streamwise actuators, D) all deflected actuators.} 
    \label{Tab:ActuationParameters}
    \end{table}
    Figure \ref{Fig:Discussion:Q} displays iso-surfaces for the same Okubo-Weiss parameter value $Q$
    for all four cases.
    The unforced case A (figure \ref{Fig:Discussion:Q}\emph{a}) 
    shows a pronounced C-pillar vortices extending far into the wake.
    Under streamwise top actuation (case B, figure \ref{Fig:Discussion:Q}\emph{b}),
    the C-pillar vortices significantly shorten.
    The next change with all streamwise actuators optimized (case C) is modest
    consistent with the small additional drag decrease.
    The C-pillar vortices are slightly more shortened  (see figure \ref{Fig:Discussion:Q}\emph{c}).
    The inward deflection of the actuation (case D) is associated with aerodynamic boat tailing 
    as displayed in figure \ref{Fig:Discussion:Q}\emph{d}.
    The separation from the slanted window is significantly delayed
    and the sidewise separation is vectored inward.

\begin{figure}[htb]
        \centering
        \subfloat{		
            \includegraphics[width=0.45\textwidth]{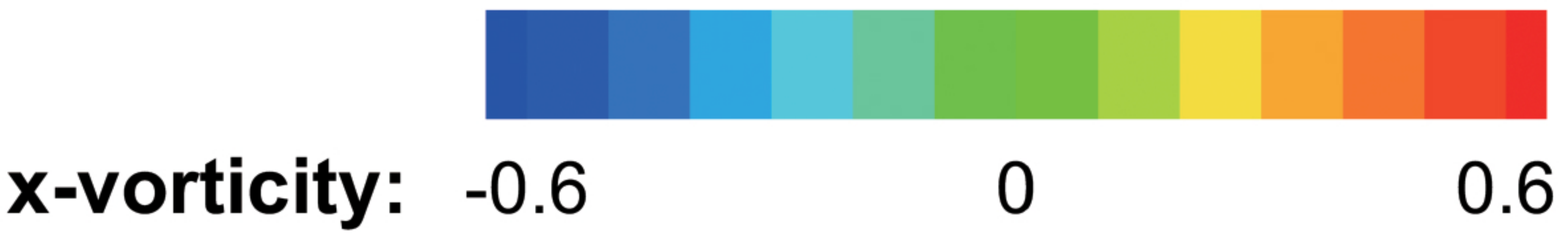}}

        \setcounter{subfigure}{0}
        \subfloat[]{
            \label{Fig:Discussion:vorticity:a}
            \includegraphics[width=0.4\textwidth]{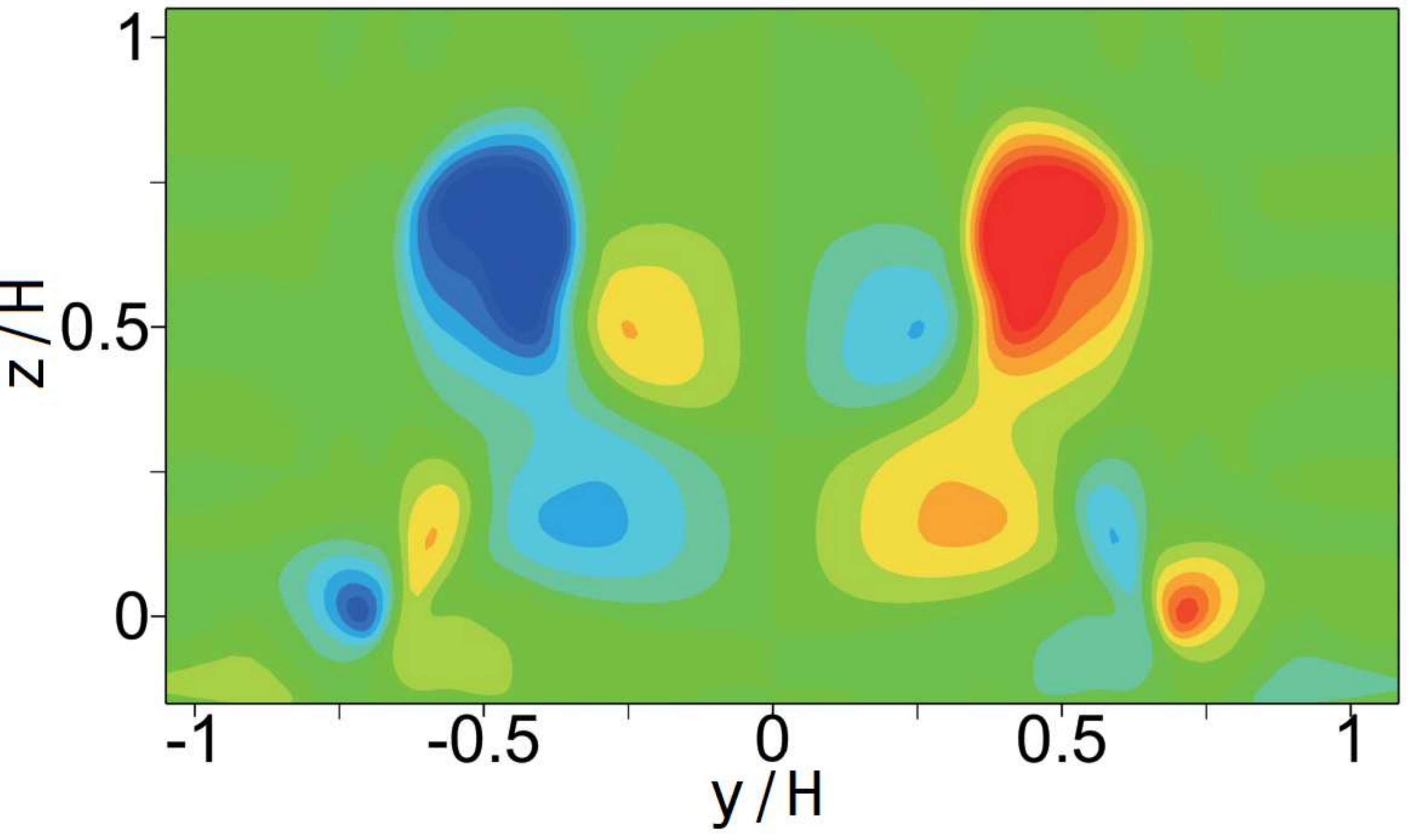}}
        \hfil
        \subfloat[]{
            \label{Fig:Discussion:vorticity:b}
            \includegraphics[width=0.4\textwidth]{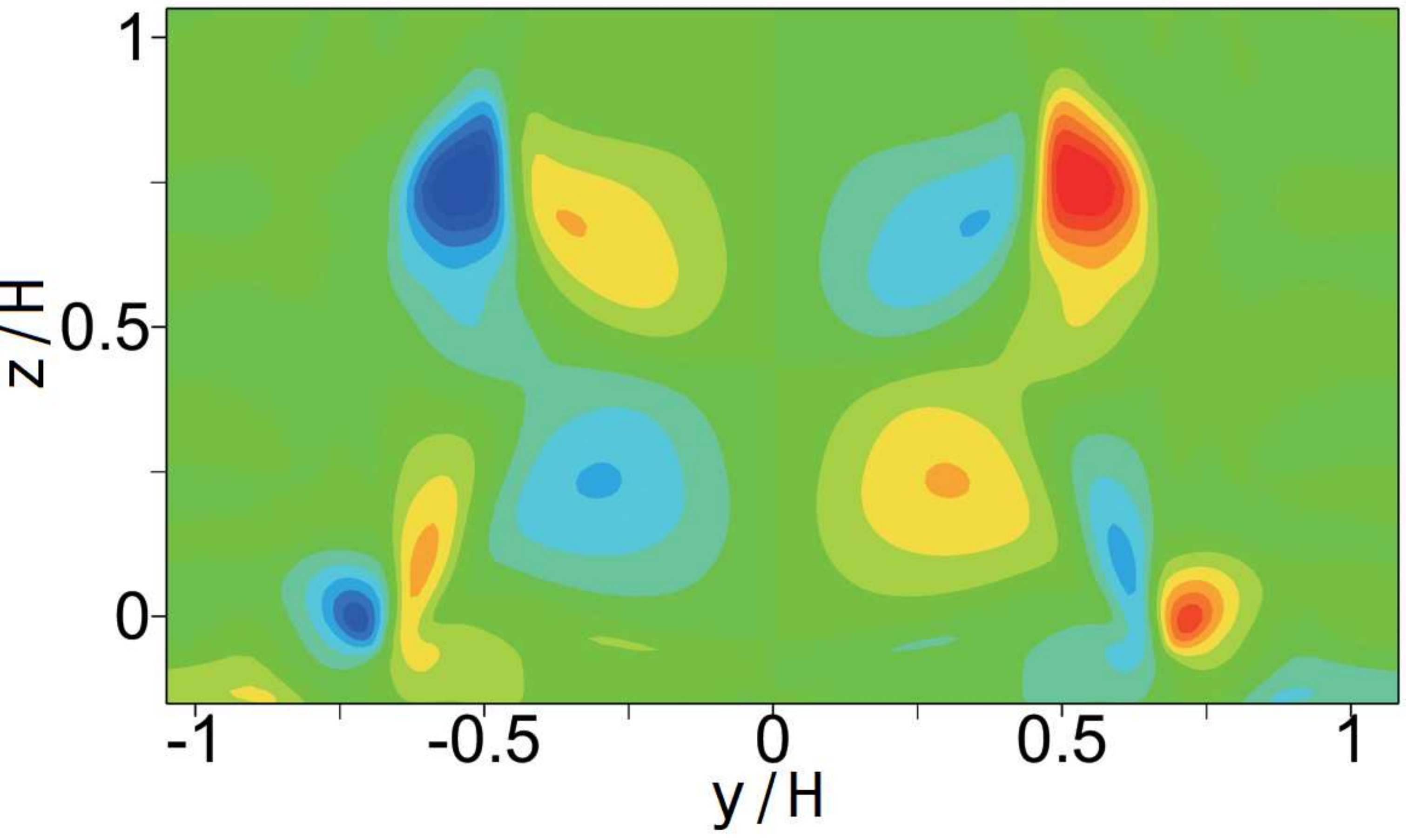}}

        \subfloat[]{
            \label{Fig:Discussion:vorticity:c}
            \includegraphics[width=0.4\textwidth]{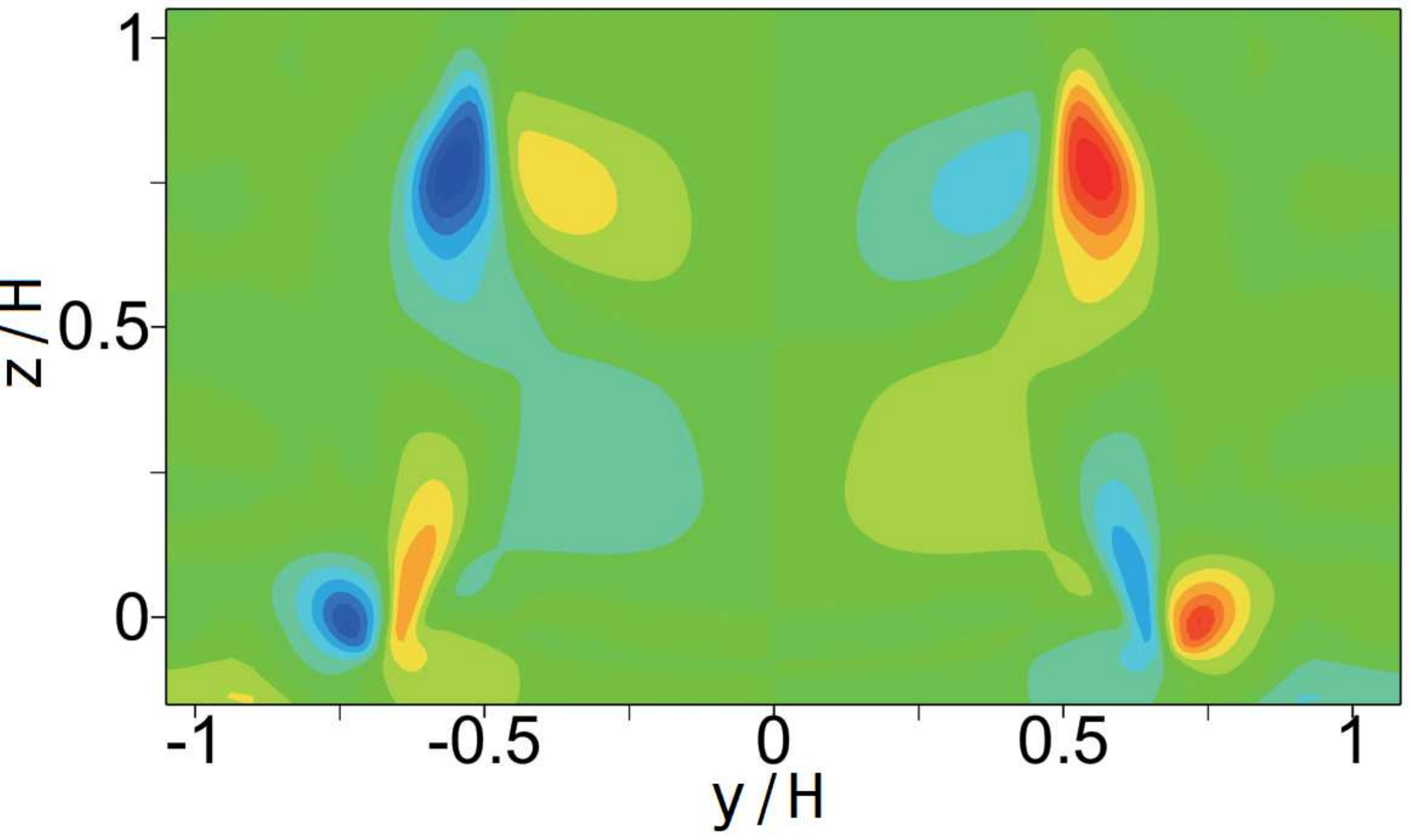}}
        \hfil
        \subfloat[]{
            \label{Fig:Discussion:vorticity:d}
            \includegraphics[width=0.4\textwidth]{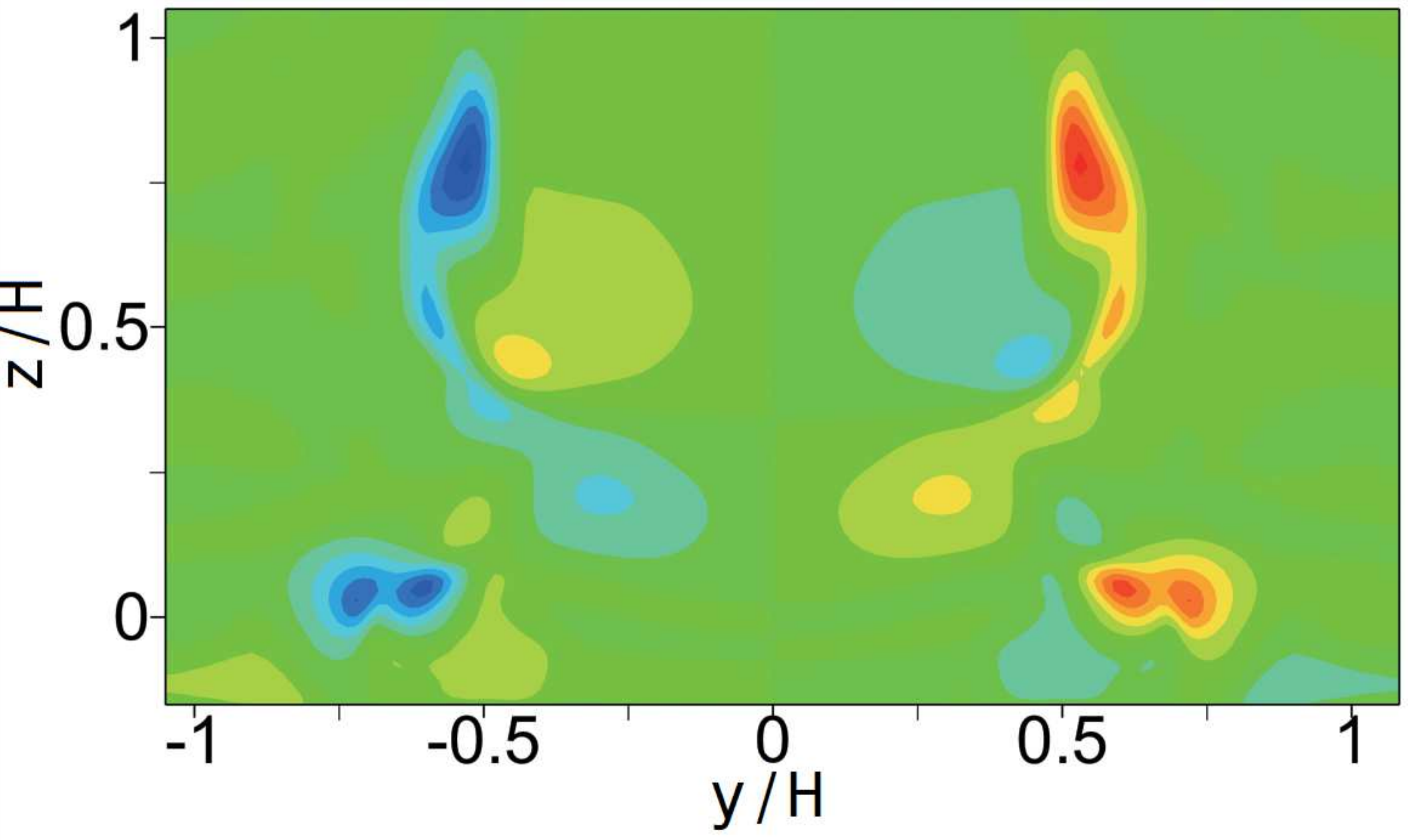}}
        \caption{Streamwise vorticity component in near-wake plane $x/H=1$.
                \textit{a}) without forcing and under 
                \textit{b}) 1D, 
                \textit{c}) 5D and 
                \textit{d}) 10D control respectively.}
        \label{Fig:Discussion:vorticity} 
    \end{figure}

    This actuation effect on the C-pillar vortices 
    is corroborated by the streamwise vorticity contours
    in a transverse plane on body height downstream ($x/H=1$).
    Figure \ref{Fig:Discussion:vorticity} shows this averaged vorticity component 
    for case A--D in subfigure \emph{a}--\emph{d}, respectively.
    The extension of the C-pillar vortices clearly shrink with increasing drag reduction.

\begin{figure}[htb]
        \centering
        \subfloat{		
            \includegraphics[width=0.45\textwidth]{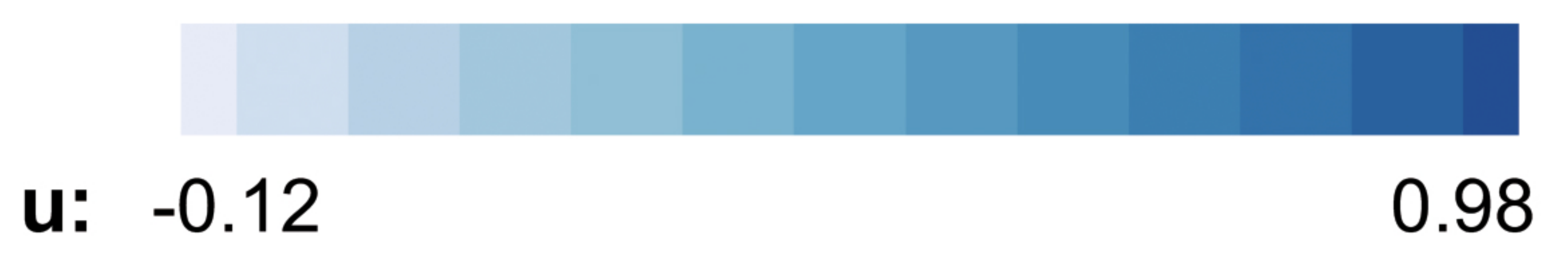}}

        \setcounter{subfigure}{0}
        \subfloat[]{
            \label{Fig:Discussion:Streamlines_X:a} 
            \includegraphics[width=0.4\textwidth]{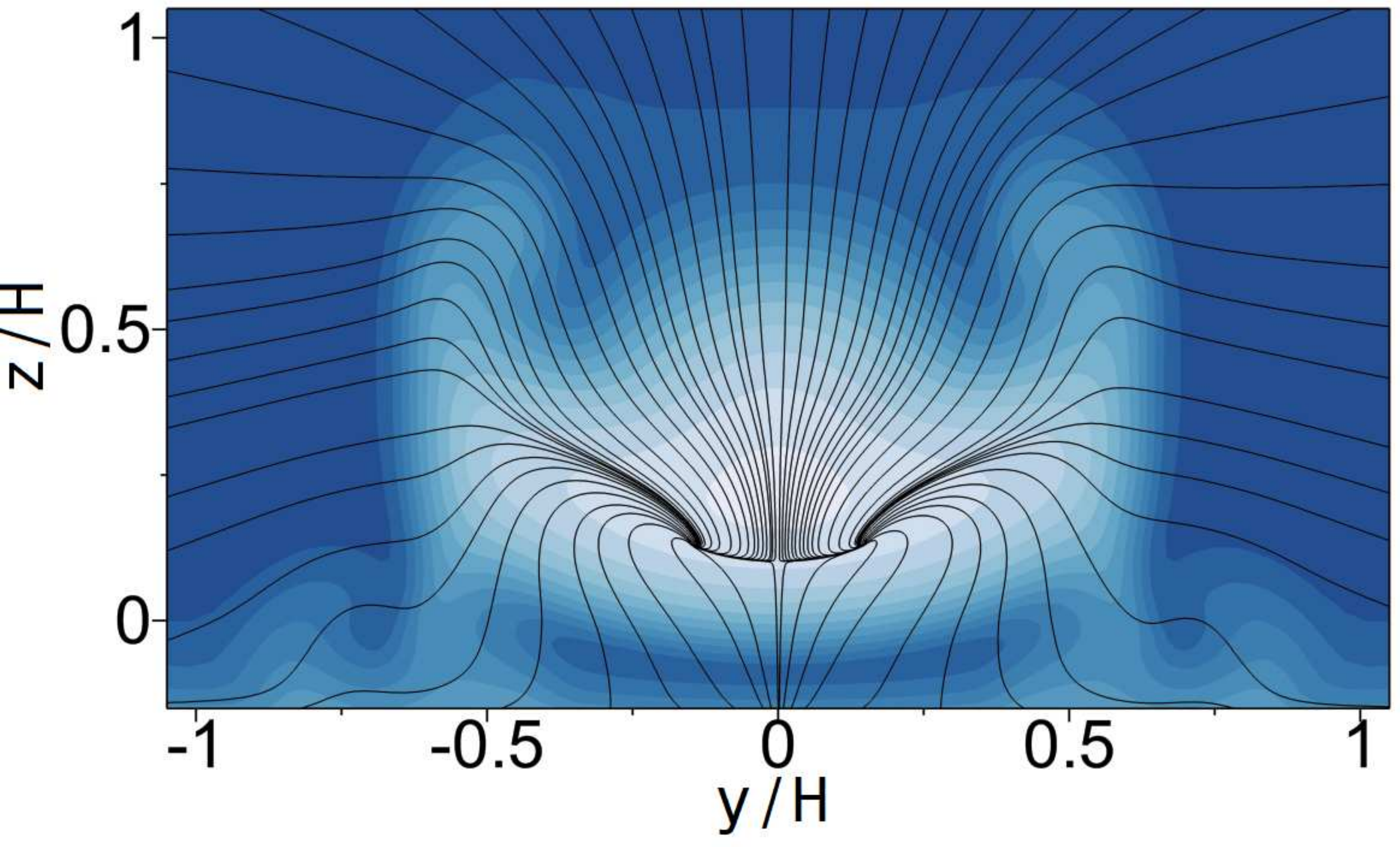}}
        \hfil
        \subfloat[]{
            \label{Fig:Discussion:Streamlines_X:b}
            \includegraphics[width=0.4\textwidth]{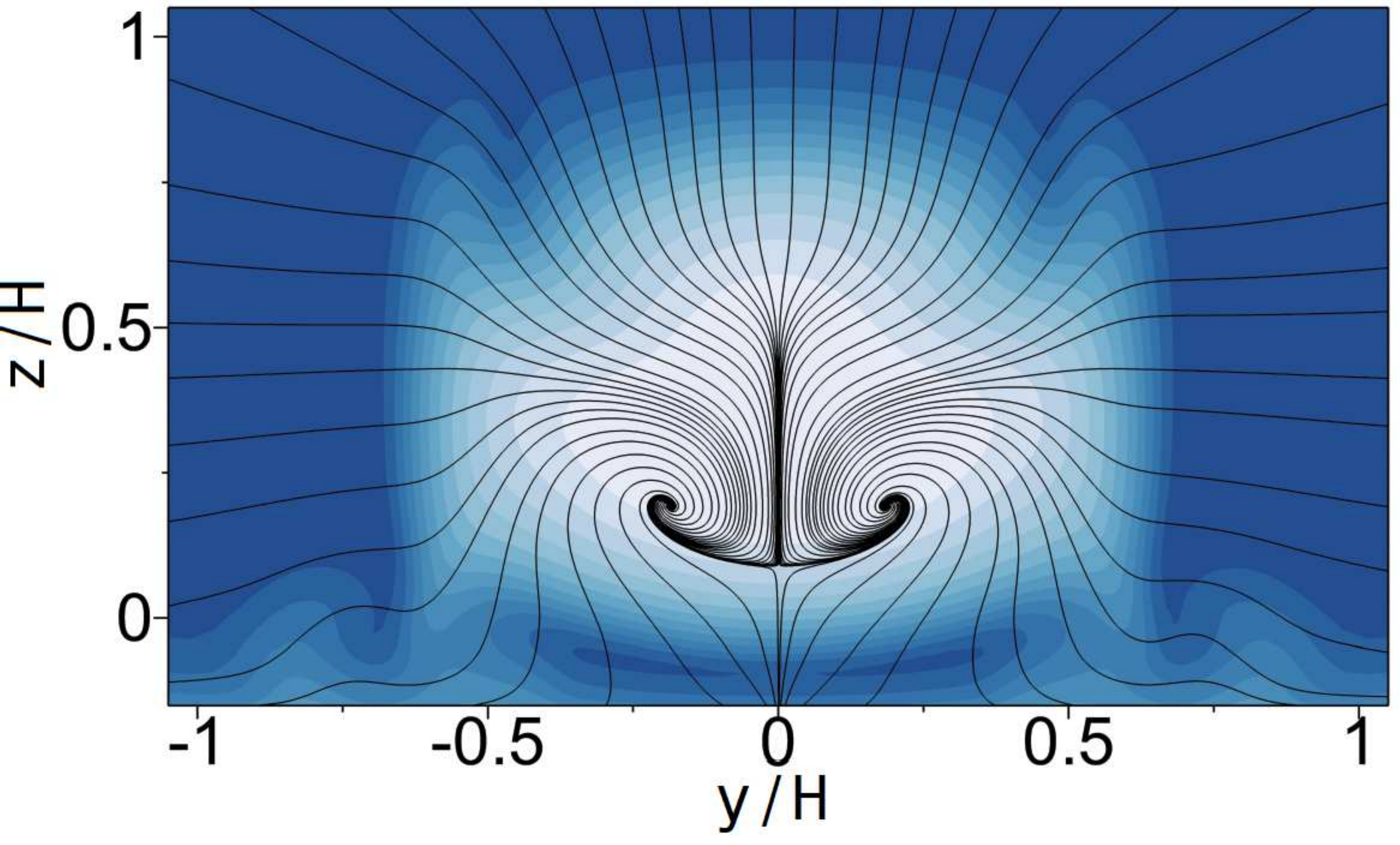}}

        \subfloat[]{
            \label{Fig:Discussion:Streamlines_X:c}
            \includegraphics[width=0.4\textwidth]{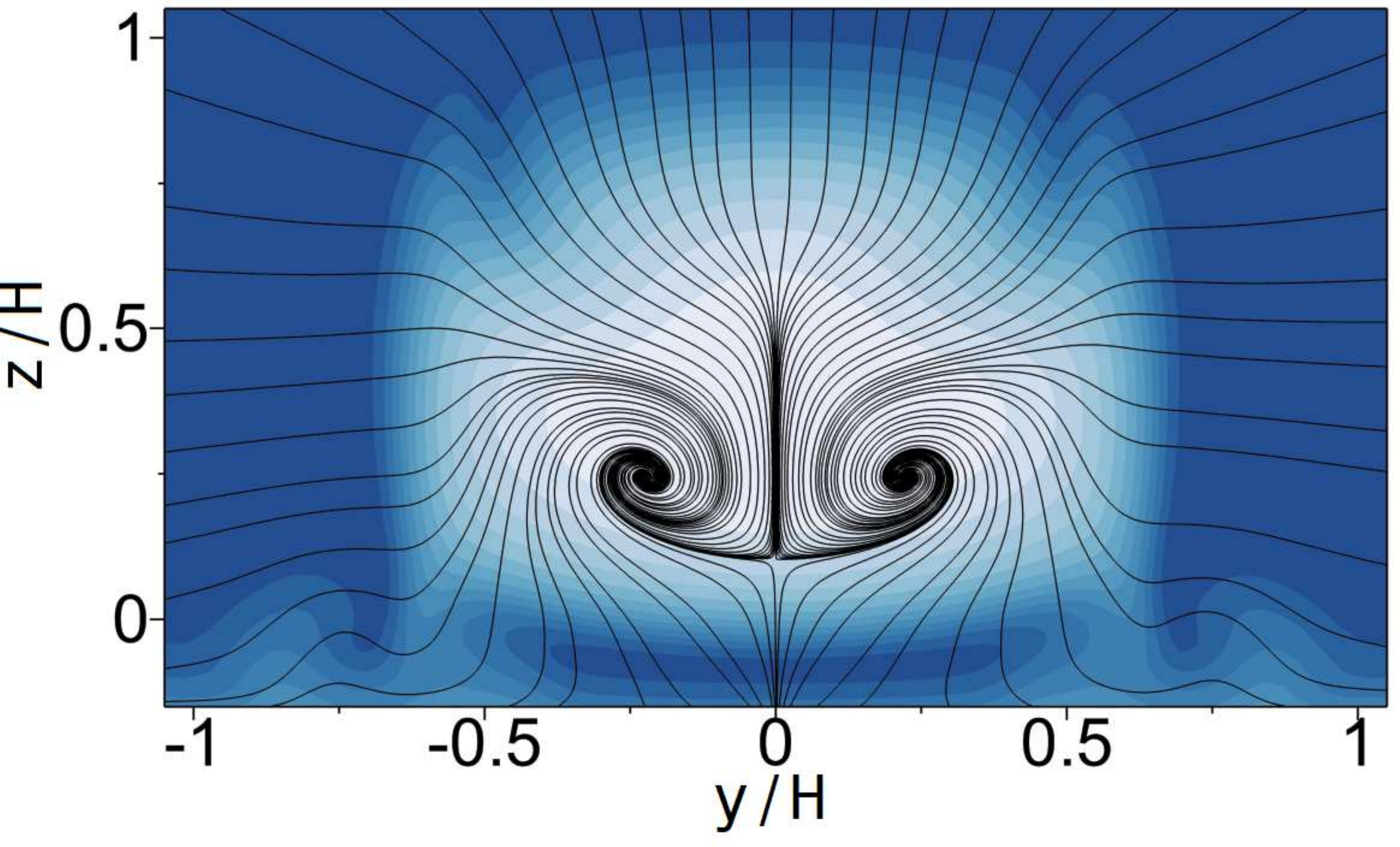}}
        \hfil
        \subfloat[]{
            \label{Fig:Discussion:Streamlines_X:d} 
            \includegraphics[width=0.4\textwidth]{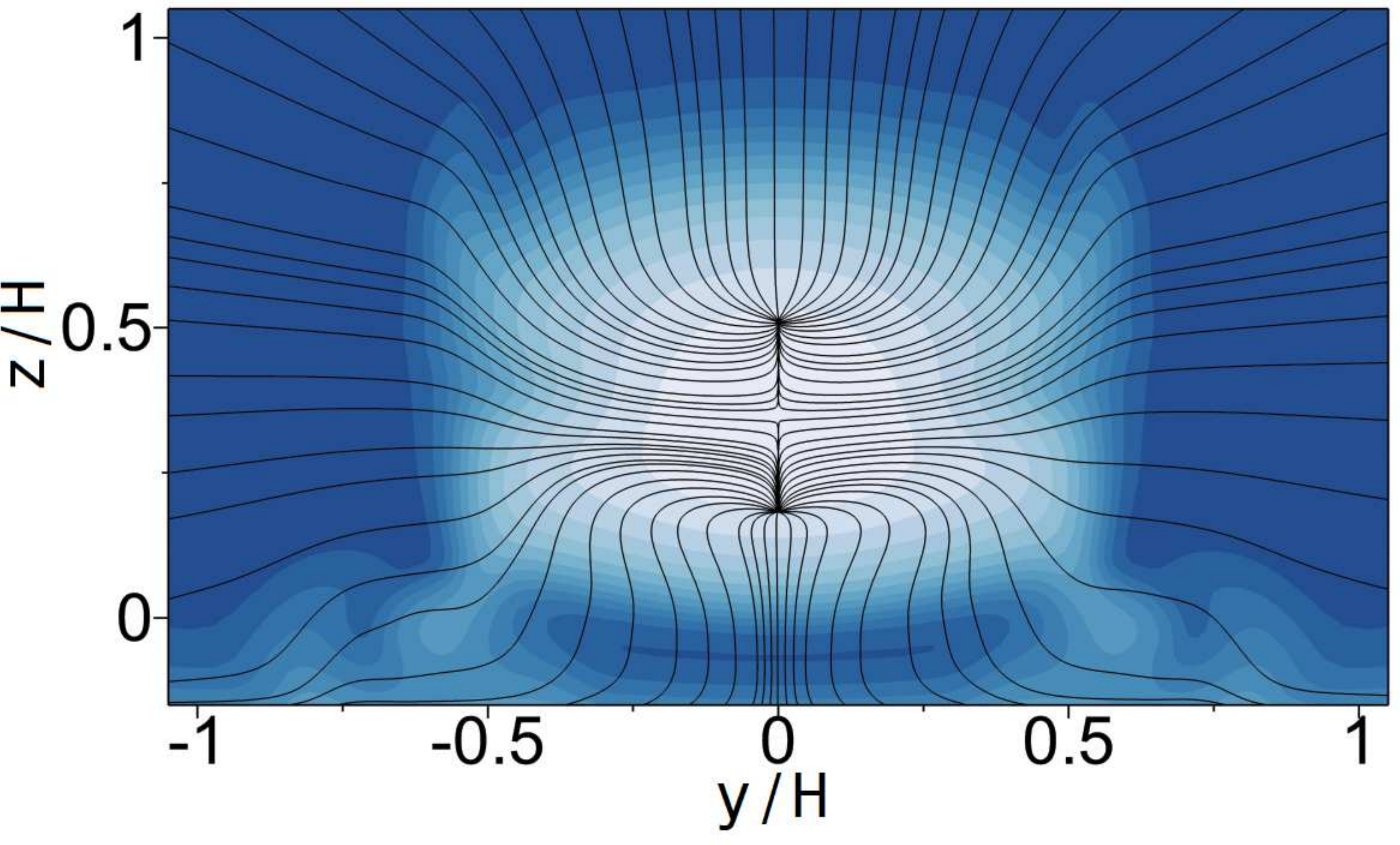}}
        \caption{Streamwise velocity component in the near-wake tranversal plane $x/H=1$ 
                and streamlines from the in-plane velocity components.
                \textit{a}) without forcing and under 
                \textit{b}) 1D, 
                \textit{c}) 5D and 
                \textit{d}) 10D control respectively.}
        \label{Fig:Discussion:Streamlines_X}
    \end{figure}

    Figure \ref{Fig:Discussion:Streamlines_X} shows the streamwise velocity component 
    and streamlines of the transversal velocity in the same plane for the same cases.
    Cases B and C feature a larger region of upstream flow 
    while case D has a narrowed region of backflow.
    From these visualizations, one may speculate 
    that the drag reduction from streamwise actuation (cases B and C)
    is due to a wake elongation towards the Kirchhoff solution
    while the inward directed actuation (case D) is associated with drag reduction from aerodynamic boat-tailing.

\begin{figure}[htb]
        \centering
        \subfloat{	
            \includegraphics[width=0.45\textwidth]{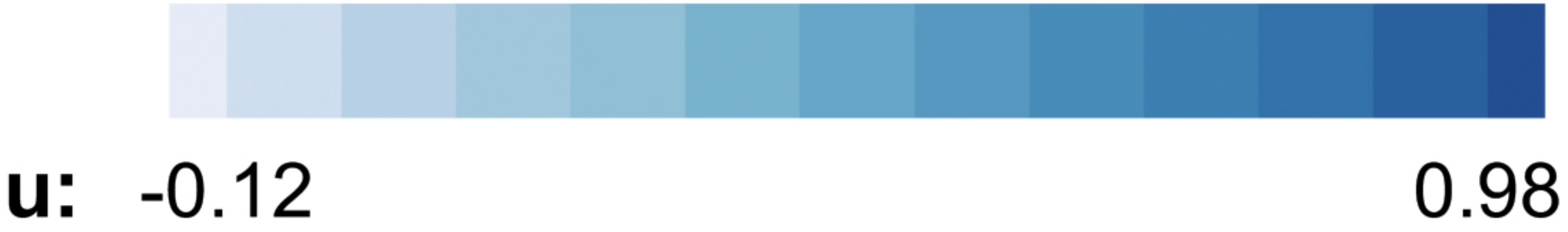}}

        \setcounter{subfigure}{0}
        \subfloat[]{
            \label{Fig:Discussion:Streamlines_Y:a}
            \includegraphics[width=0.4\textwidth]{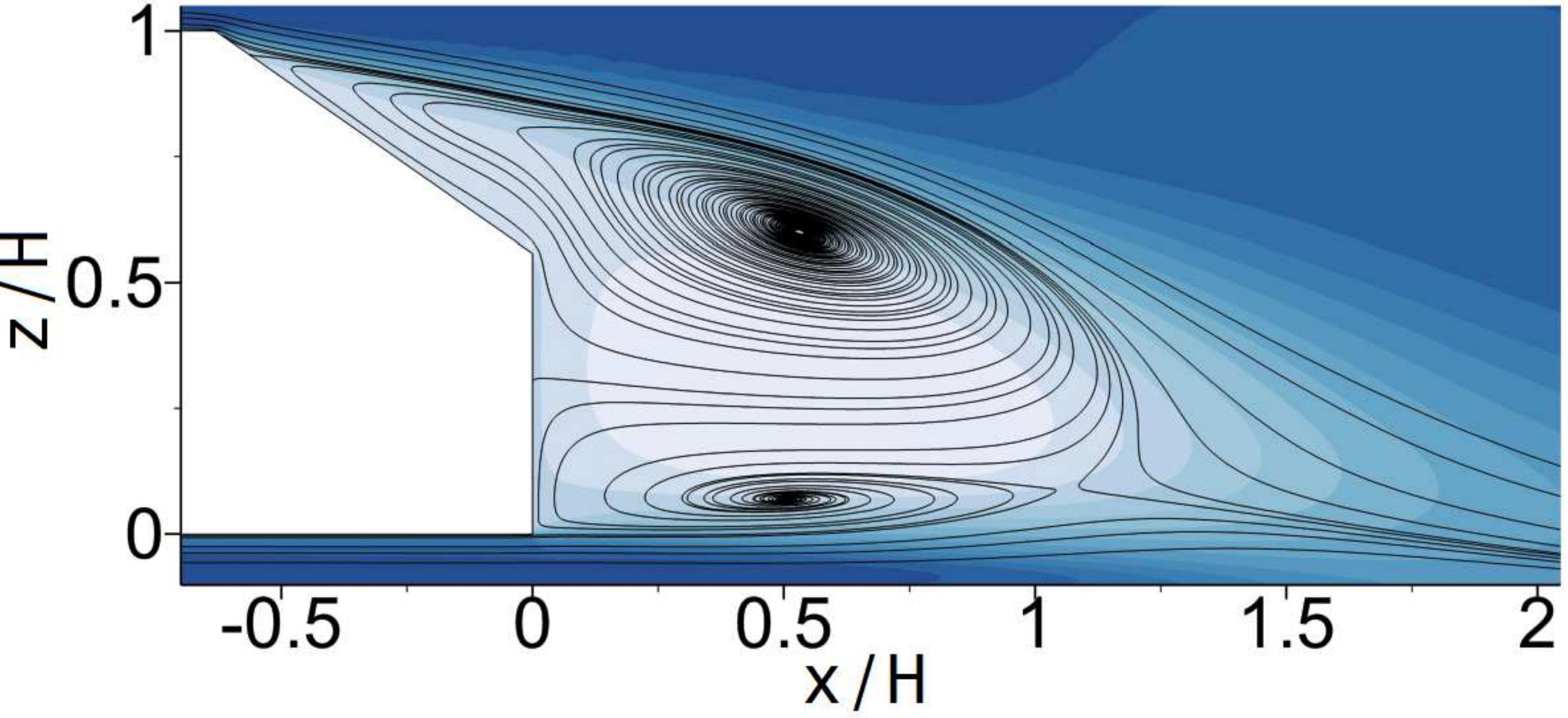}}
        \hfil
        \subfloat[]{
            \label{Fig:Discussion:Streamlines_Y:b}
            \includegraphics[width=0.4\textwidth]{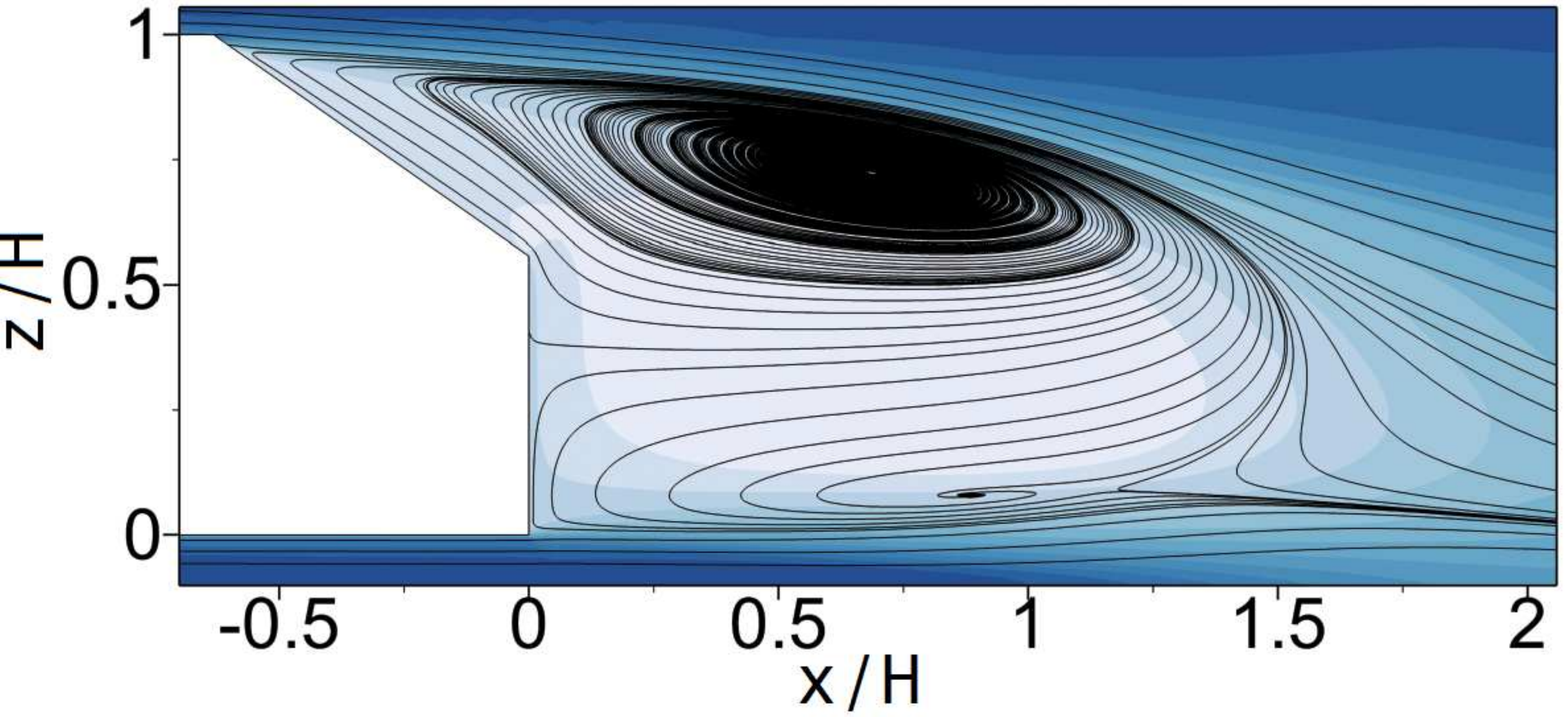}}

        \subfloat[]{
            \label{Fig:Discussion:Streamlines_Y:c}
            \includegraphics[width=0.4\textwidth]{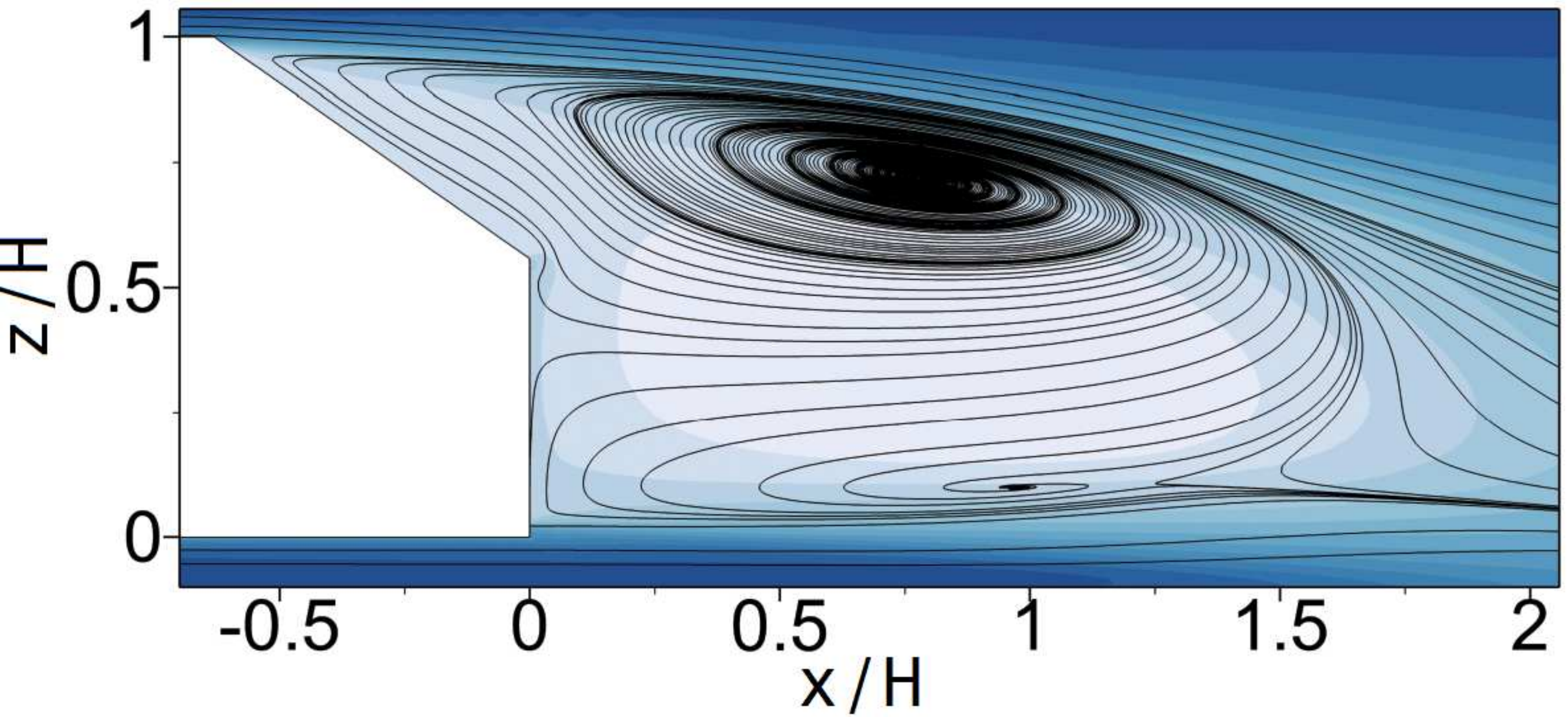}}
        \hfil
        \subfloat[]{
            \label{Fig:Discussion:Streamlines_Y:d}
            \includegraphics[width=0.4\textwidth]{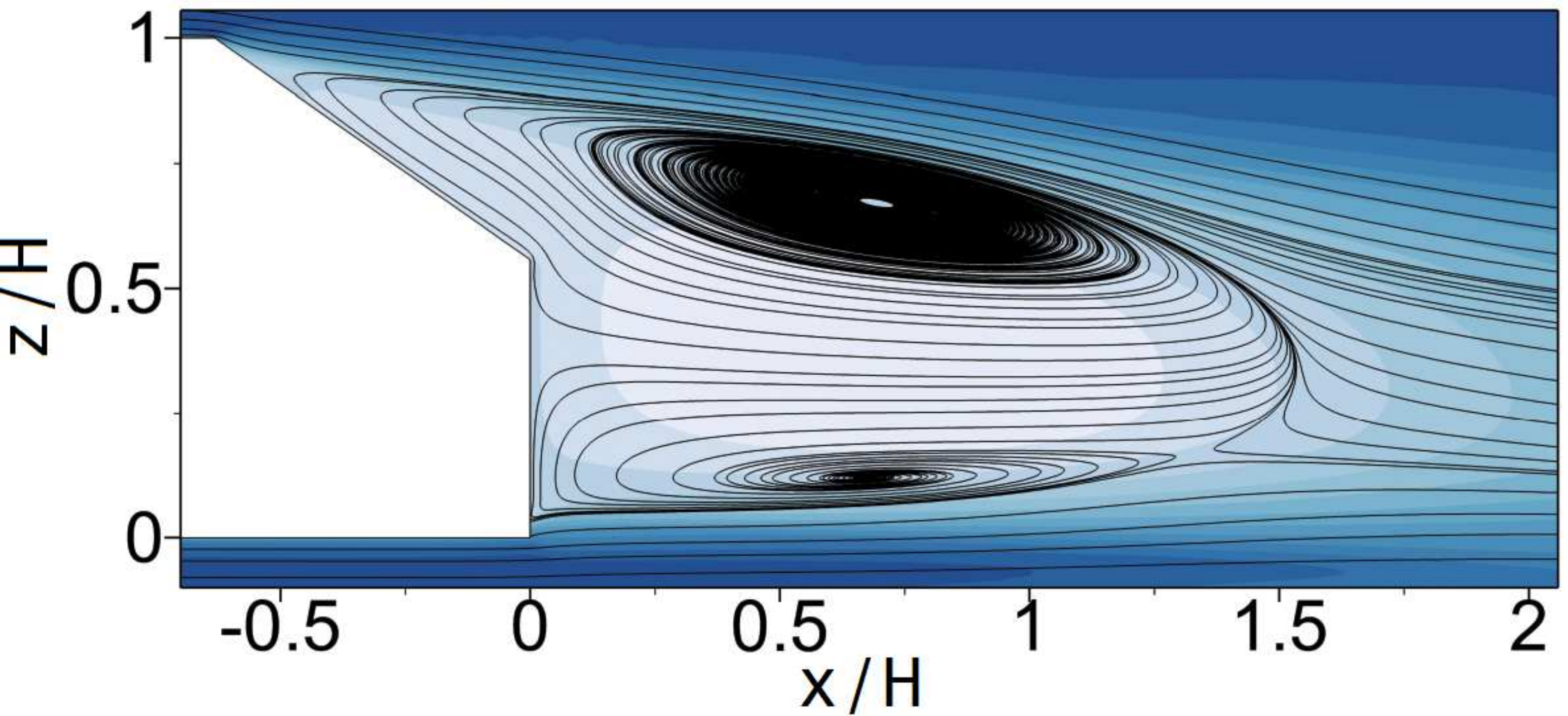}}
        \caption{Streamwise velocity component in the symmetry plane $y=0$ 
                and streamlines from the in-plane velocity components.
                \textit{a}) without forcing and under 
                \textit{b}) 1D, 
                \textit{c}) 5D and 
                \textit{d}) 10D control respectively.}
        \label{Fig:Discussion:Streamlines_Y}
    \end{figure}

    This hypothesis about different mechanisms of drag reduction
    is corroborated from the streamlines in the symmetry plane $y=0$
    in figure \ref{Fig:Discussion:Streamlines_Y}.
    The tangential blowing (see subfigures \emph{b}, \emph{c} 
    leads to an elongated fuller wake
    as compared to the unforced benchmark (subfigure \emph{a}). 
    The top shear-layer is oriented more horizontal under streamwise actuation---consistent with the Kirchhoff wake solution. 
    The inward-directed actuation (see subfigure  \emph{d}) also elongates the wake
    but gives rise to a more streamlined shape.
    The top and bottom shear-layers are vectored inward.

\begin{figure}[htb]
        \centering
        \subfloat{
            \includegraphics[width=0.45\textwidth]{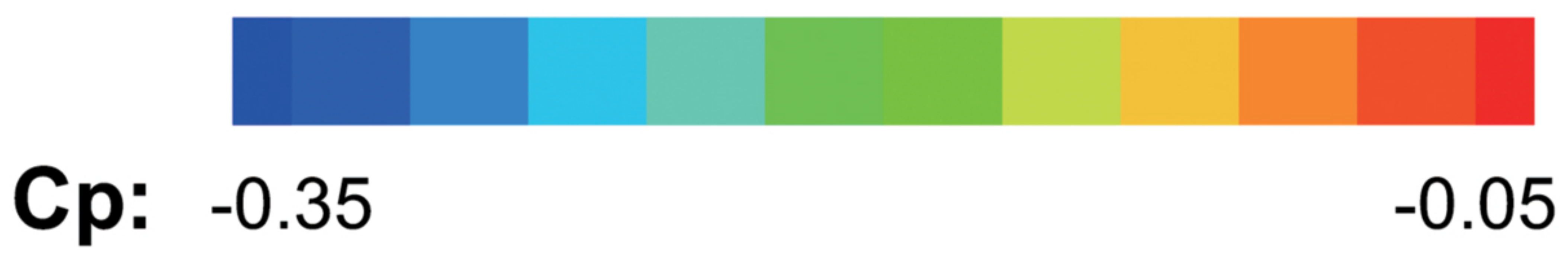}}

        \setcounter{subfigure}{0}
        \subfloat[]{
            \label{Fig:Discussion:Cp:a}
            \includegraphics[width=0.3\textwidth]{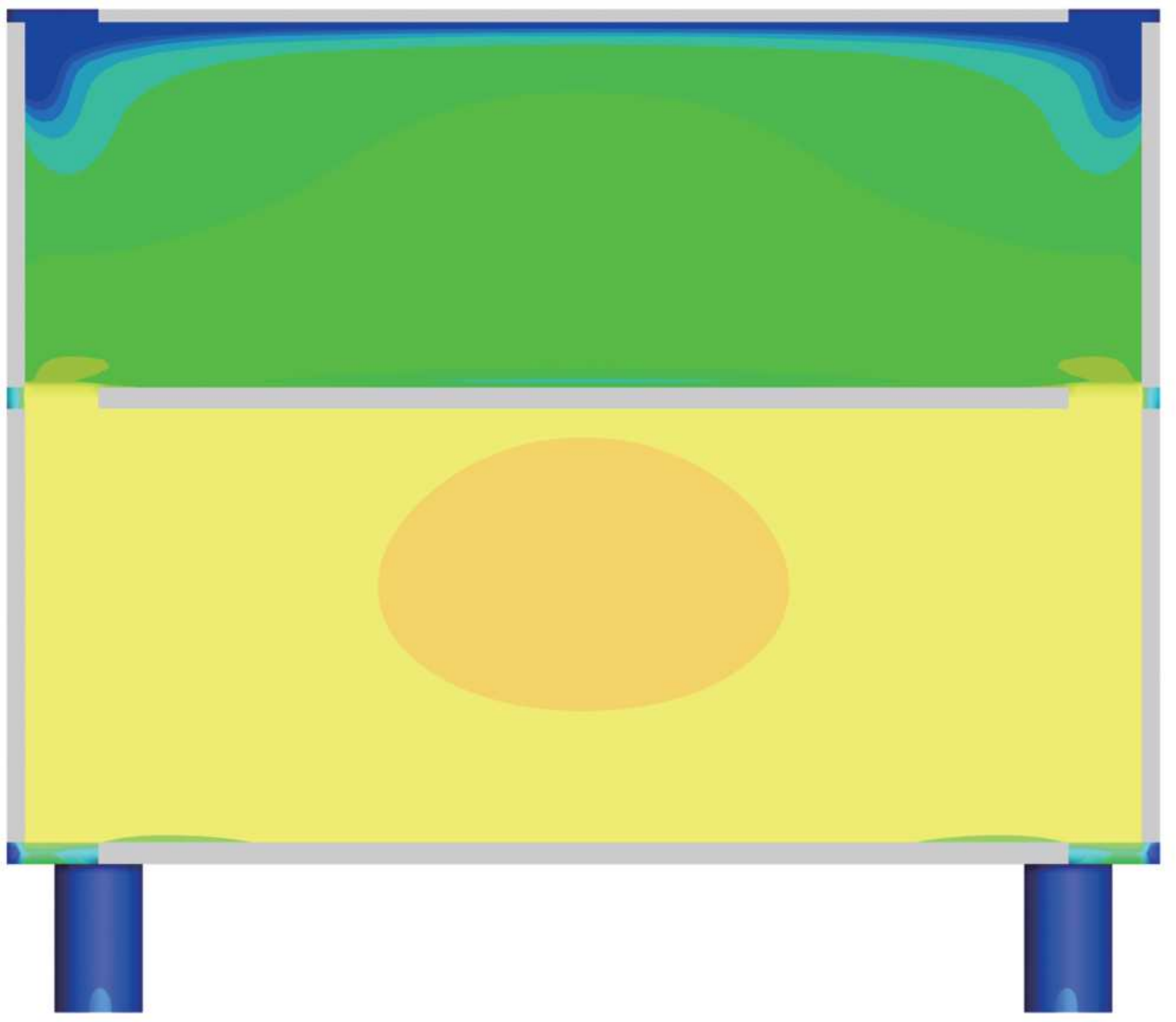}}
        \hfil
        \subfloat[]{
            \label{Fig:Discussion:Cp:b}
            \includegraphics[width=0.3\textwidth]{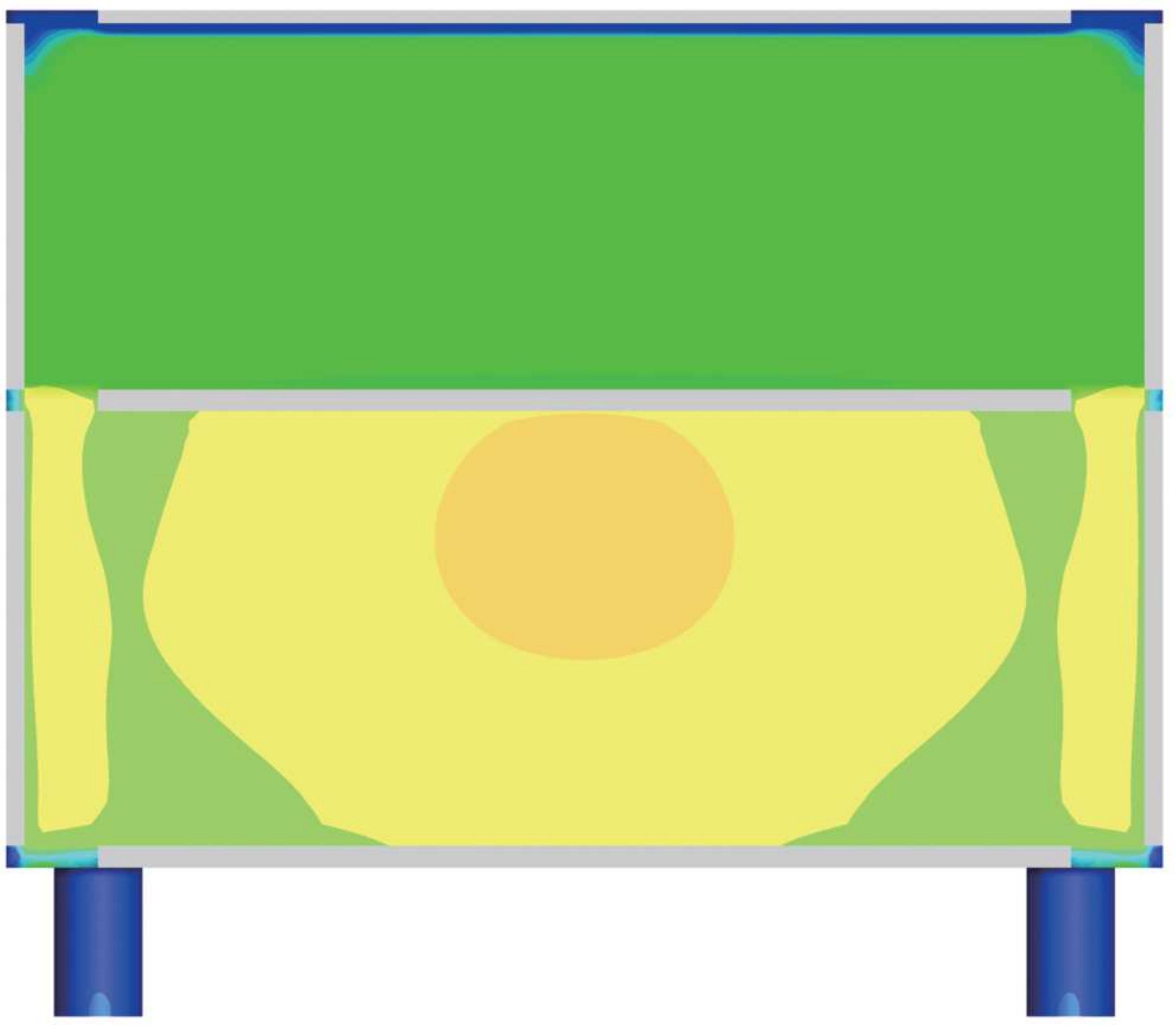}}

        \subfloat[]{
            \label{Fig:Discussion:Cp:c} 
            \includegraphics[width=0.3\textwidth]{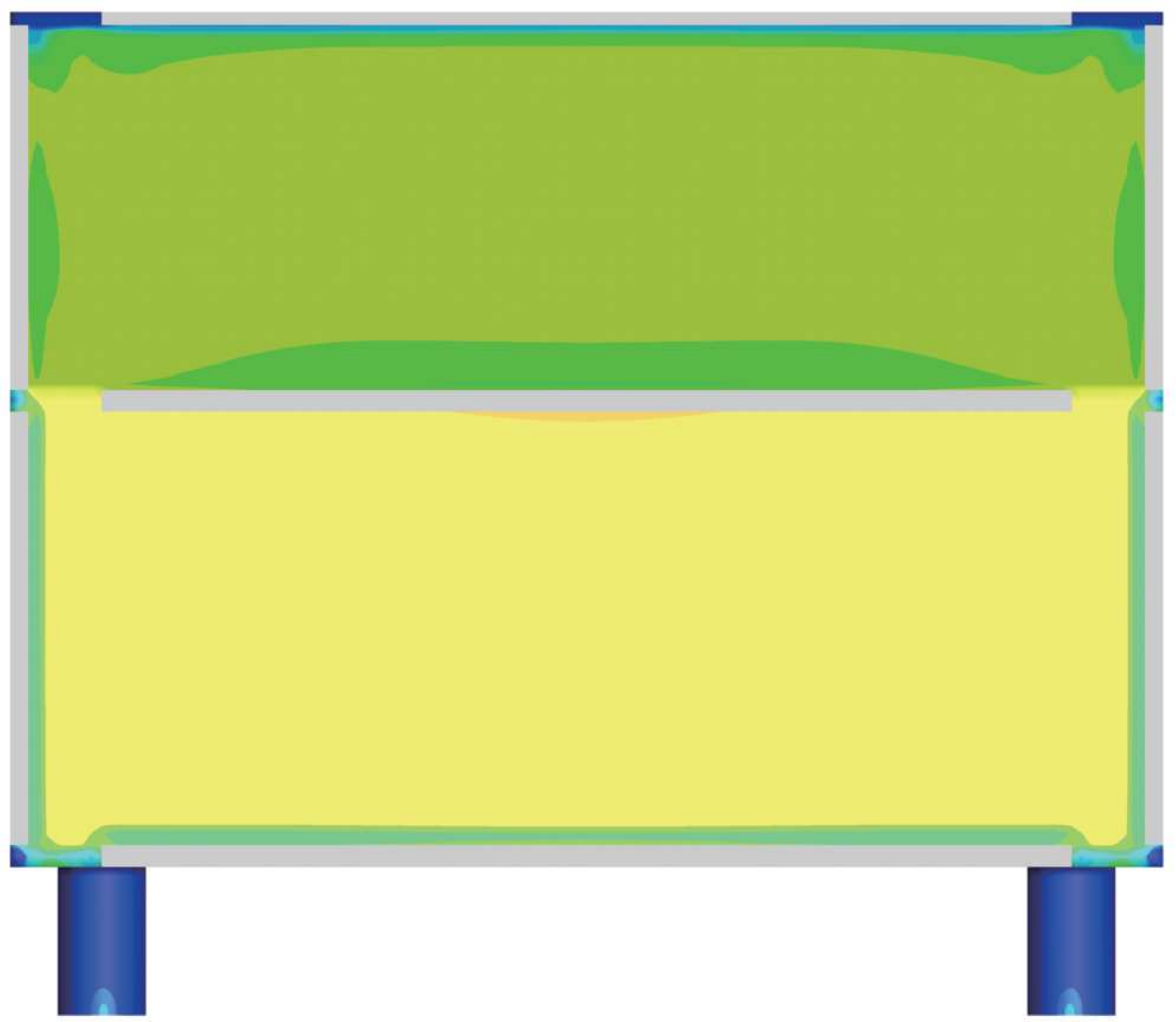}}
        \hfil
        \subfloat[]{
            \label{Fig:Discussion:Cp:d}
            \includegraphics[width=0.3\textwidth]{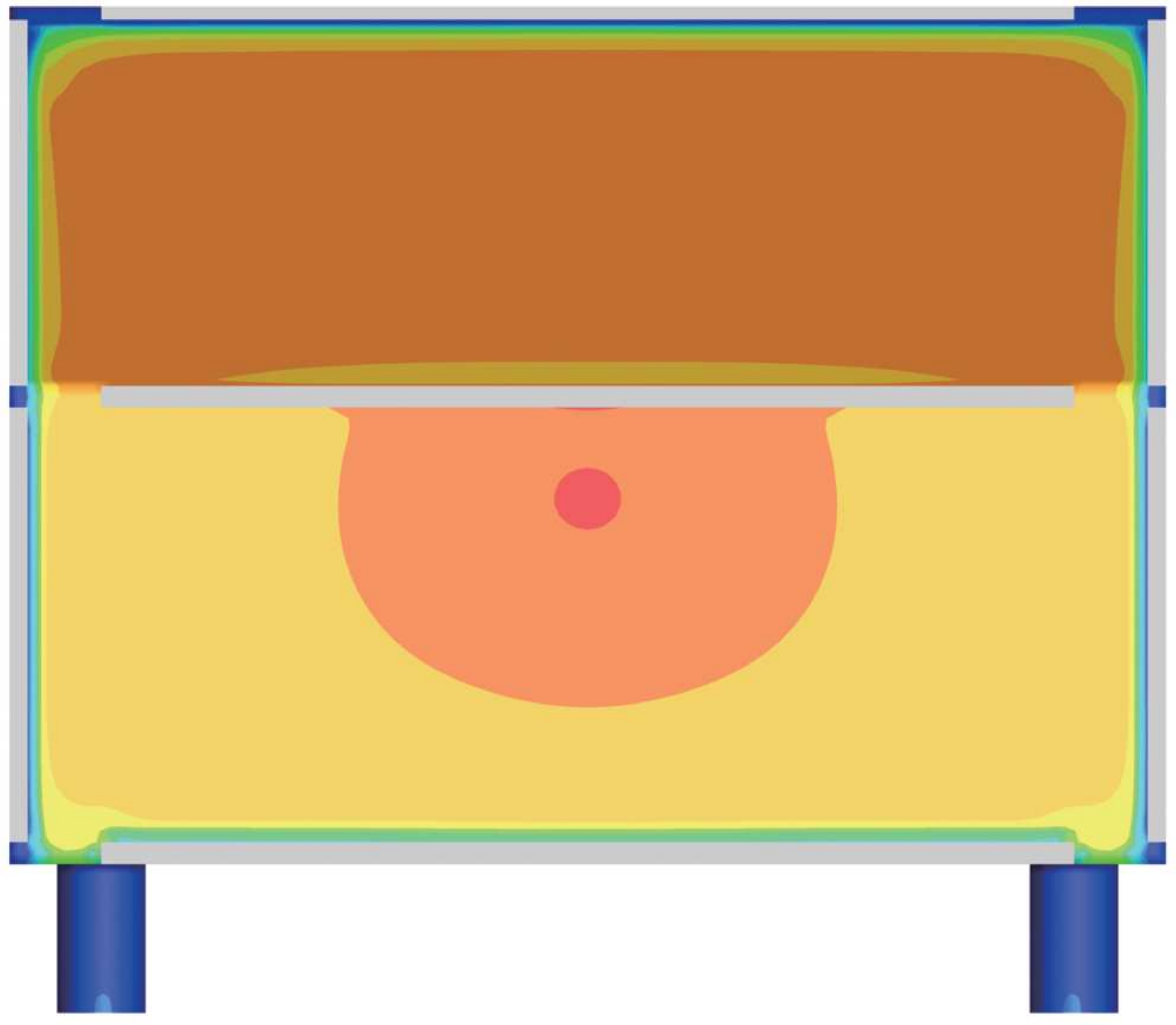}}
        \caption{Pressure coefficient on the slant and vertical base of flow.
                \textit{a}) without forcing and under 
                \textit{b}) 1D, 
                \textit{c}) 5D and 
                \textit{d}) 10D control respectively.}
        \label{Fig:Discussion:Cp}
    \end{figure}

    The drag reduction can more directly be inferred 
    from the $C_p$ distribution of the rearward windows
    in figure \ref{Fig:Discussion:Cp}.
    The 5\% drag reduction in subfigure \emph{b}) for case B
    is associated with a pressure increase of the vertical surface. The additional 2\% drag decrease for case C in subfigure \emph{c}
    is accompanied by an increase over vertical and slanted surface.
    The aerodynamic boat-tailing of case D with 17\% drag reduction alleviates significantly the pressures on both surfaces.

%% file: S6.tex
\section{Conclusions}
\label{ToC:Conclusions}

We propose a novel optimization approach 
for active bluff-body control  exploiting 
local gradients with a downhill simplex algorithm
and exploring new better minima with Latin hypercube sampling (LHS).
This approach is called \emph{explorative gradient method (EGM)}
as the iterations alternate between downhill simplex iteration 
as a robust gradient method 
and LHS as the most explorative step.
A distinguishing feature of EGM is that it performs an `aggressive' exploitation
in one step and the arguably most optimal exploration in another step.
Thus, both, exploitation and exploration 
come with optimizing principles 
and with an a priori evaluation investment 
which is determined upfront.
This policy has distinct advantages.
In some cases, the exploitation will be pointless because
the best minimum still needs to be found.
In other cases, the exploration will be ineffective,
because the dimension or complexity of the search space is too large.
EGM hedges against both scenarios of inefficiency
because high-dimensional search spaces
typically have unknown characteristics.

This policy may be contrasted with genetic algorithms 
which can be remarkably effective in high dimensions,
but the goal of explorative operations, like mutation, 
and exploitative operations, like crossover, 
come with no optimizing principle, 
like gradient-based convergence or geometric coverage of the search space.
A similar observation applies to other biologically inspired optimization methods \citep[see, e.g., ][]{Wahde2008book},
like ant colony or particle swarm optimization.
As another example, simulated annealing explores good minima before it increasingly exploits them.
Here, again, exploration and exploitation come with no optimizing principle
and the switch between exploration and exploitation is a design parameter.
We argue that the radical alternation between gradient-based exploitation and maximal exploration
is one of the most promising strategies in an unknown search space.

EGM is compared with other optimizers 
for an analytical test function
with one global and for local minima.
The study includes the failure rate in finding the global optimum
and the convergence rate.
EGM is found to be distinctly superior in both aspects
in comparison with
(1) Latin hypercube sampling (LHS), 
(2) Monte Carlo sampling, 
(3) a genetic algorithm,
(4) a downhill simplex method, and 
(5) a random restart or shotgun downhill optimization.
This behaviour is made physically plausible for smooth cost functions
with few mininima,
i.e., a typical case for active flow control.
 
 As first flow control example, 
EGM is applied to the minimization of the parasitic net drag power
of the multi-input fluidic pinball problem.
It yields a slightly asymmetric Coanda forcing with 40$\>$\% net drag power reduction
comprising 98$\>$\% drag reduction penalized by 58$\>$\% actuation energy.
As very similar actuation has been found with machine learning control 
for feedback law ansatz \citep{Cornejo2019pamm}.
This Coanda actuation foreshadows the optimization result 
of the subsequent drag minimization of the Ahmed body.
Intriguingly, EGM also probed base bleed 
and circulation control as options for drag power minimization.

EGM reduces the drag of  a $35^\circ$ slanted Ahmed body 
by 17\% with independent steady blowing at all trailing edges
at Reynolds number $Re_H=1.9 \times 10^5$.
The 10-dimensional actuation space 
includes 5 symmetric jet slot actuators or corresponding actuator groups
with variable velocity and variable blowing angle.
The resulting drag is computed 
with a Reynolds-Averaged Navier-Stokes (RANS) simulation.

The approach is augmented by auxiliary methods 
for initial conditions, 
for accelerated learning and 
for a control landscape visualization.
The initial condition for a RANS simulation with a new actuation 
is computed by the 1-nearest neighbour method.
In other words, the RANS simulation starts with 
the converged RANS flow 
of the closest hitherto examined actuation.
This cuts the computational cost by 60\% as it accelerates RANS convergence.
The actuation velocities are quantized to prevent testing of too similar control laws.
This optional element reduces the CPU time by roughly 30\%. 
The learning process is illustrated in a control landscape.
This landscape depicts the drag  in a proximity map---a two-dimensional feature space from the high-dimensional actuation response.
Thus, the complexity of the optimization problem can be assessed.

The slanted Ahmed body with 1, 5 and 10 actuation parameters
constitutes a more realistic plant for an optimization algorithm.
First, only the upper streamwise jet actuator is optimized.
This yields drag reduction of 5\%  
with pronounced global minimum for the jet velocity.
Second, the drag can be further reduced to 7\%
with 5 independent streamwise symmetric actuation jets.
Intriguingly, the actuation effects of the actuator are far from additive---contrary
to the experimental observation for the square-back Ahmed body \citep{Barros2015phd}.
The optimal parameters of a single actuator are not closely indicative 
for the optimal values of the combined actuator groups.
The control landscape depicts a long curved valley 
with small gradient leading to a single global minimum.
Interestingly, the explorative step is not only a security policy for the right minimum.
It also helps to accelerate the optimization algorithm 
by jumping out of the valley to a point closer to the minimum.

A significant further drag reduction of 17\% is achieved when, 
in addition to the jet velocities, 
also the jet angles are included in the optimization. 
Intriguingly, all trailing edge jets are deflected inward 
mimicking the effect of Coanda blowing
and leading to fluidic boat tailing.
The C-pillar vortices are increasingly weakened 
with one-, five- and ten-dimensional actuation.
Compared with the pressure increase at C-pillar in one- and five-dimensional control,
the ten-dimensional control brings a substantial pressure recovery over the entire base. 
The achieved 17\% drag decrease with constant blowing 
is comparable with the experimental 20\% reduction with high-frequency forcing by
\citet{Bideaux2011ija,Gillieron2013ef}.

For the $25^\circ$ high-drag Ahmed body,
\citep{Zhang2018jfm} have achieved 29\% drag reduction with steady blowing at all sides,
thus significantly outperforming all hitherto existing active flow control studies cited therein.
The actuation has only been investigated for few selected actuation values.
Hence, even better drag reductions are perceivable.
Yet, the unforced high-drag Ahmed body 
has a significantly higher drag coefficient of $0.361$
than the low-drag version and is hence not fully  comparable.
Their reduced drag coefficient of $0.256$ 
is almost identical with the one of this study.

We expect that our RANS-based active control optimization 
is widely applicable for virtually all multi-input steady actuations
or combinations of passive and active control \citep{Bruneau2010cf}.
The explorative gradient method mitigates 
the chances of sliding down a suboptimal minimum
at an acceptable cost.
The 1-nearest neighbour method for initial condition
and the actuation quantization accelerate the simulations and learning processes.
And the control landscape provides the topology of the actuation performance,
e.g., the number of local minima, nature and shape of valleys, etc.

The current performance benefits of EGM over other commonly used optimizers 
has also been corroborated in preliminary bluff-body drag reduction experiments
in Europe and China.
Evidently, the optimizer can also be employed for cost function minimization
for design parameters of passive devices 
or parameters of closed-loop control schemes.
An exciting new avenue is the generalization of EGM
from a parameter optimizer to a regression problem solver:
EGM has recently been applied to optimize  multi-input multi-output control laws
for the stabilization of the fluidic pinball and was found to be distinctly superior
to genetic programming \citep{Cornejo2020jfm}.

The key idea of EGM is 
to balance exploration and exploitation
by aiming to optimizing each for one step in an alternating fashion.
If the dimension of the search space is too large,
the downhill simplex algorithm may be replaced by a subplex \citep{King2020r}
or a stochastic gradient method.
Similarly, for large dimensions, 
the LHS may need to be replaced by Monte Carlo sampling
or a genetic algorithm.
Note that LHS will first explore the edge of the search space before it explores the center.

Summarizing, EGM is a versatile optimizer framework with numerous future applications.
The very algorithms of EGM cannot only be applied to parameter optimization
but also to model-free control law optimization,
hitherto performed by genetic programming \citep{Gautier2015jfm,Ren2020jh} 
and deep reinforcement learning \citep{Rabault2019jfm,Bucci2019prsa}.
Preliminary results for the fluidic pinball indicate 
that the learning rate increases by one order of magnitude
as compared to linear genetic programming control \citep{Li2018am}.
Future versions of EGM will also incorporate the learning of response models 
with methods of machine learning \citep[see, e.g.,][]{Brunton2019book} for accelerate learning.

%% file: Acknowledgements.tex
\section*{Acknowledgements}

This work is supported by Shanghai Key Lab of Vehicle Aerodynamics and Vehicle Thermal Management Systems (Grant No.18DZ2273300), by public grants overseen by the French National Research Agency (ANR-17-ASTR-0022, FlowCon),
by the German Science Foundation (SE 2504/2-1, SE 2504/3-1)
and by Polish Ministry of Science and Higher Education
(MNiSW) under the Grant No.: 05/54/DSPB/6492.

We have profited from stimulating discussions  
with Steven Brunton, Valery Chernoray, Guy Yoslan Cornejo Maceda, 
Nan Gao, Bingxi Huang, 
Sini\v{s}a Krajnovi\'c, Hao Li, Francois Lusseyran, Navid Nayeri, Oliver Paschereit, Luc Pastur, Richard Semaan, Wolfgang Schr\"oder, Bingfu Zhang and Yu Zhou.

\section*{Declaration of interests}
The authors report no conflict of interest.

%% file: SA.tex
\section{Genetic algorithm}
\label{ToC:GA:Matlab}
This section provides further details about the chosen matlab realization of the genetic algorithm.

\begin{description}
\item[1) First generation. ]
The algorithm begins by creating an initial population with random inviduals.
Each parameter of each individual is taken with uniform probability from a given interval.
\item[2) Next generations. ] 
The algorithm uses the individuals in the current generation, called parents, 
to create individuals of the next population, called children:
\begin{description}
\item[a) Cost evaluation. ] Score each member of the current population by computing its cost function. 
The cost function is assumed to be sorted,
$J_1  \le J_2 \le \ldots \le J_r \le \ldots J_I$.
The index is called the rank.
\item[b) Scaled fitness. ] Scale the cost function based on relative ordering.
An individual with rank $r$
has fitness score of $1/\sqrt{r}$  (higher fitness, smaller rank).
\item[c) Parents. ] Select members, called parents, based on their expectation value.
The selection function chooses parents for the next generation based on their expectation values. 
An individual can be selected more than once as a parent, 
in which case it contributes its genes to more than one child. 
\item[d) Elitism. ] The best $N_e$ individuals are copied directly into the new generation.
This number corresponds to the probability $P_e=0.05$, i.e., $N_e = I * P_e$ 
\item[e) Mutation and crossover. ] Produce children from the parents with crossover or mutation. 
By combining parts of genes from a pair of parents, 
crossover children are produced with probability $P_c=0.8$.
The remaining individuals, other than elite children, 
are mutation children by making random changes to a single parent.
Scattered, the default crossover function, creates a random binary vector 
and selects the genes where the vector is a 1 from the first parent, 
and the genes where the vector is a 0 from the second parent, 
and combines the genes to form the child.
The default mutation function---Gaussian adds a random number 
taken from a Gaussian distribution with mean 0 to each entry of the parent vector. 
The standard deviation of this distribution is determined by the parameters 
Scale 1 and Shrink 1.(Matlab 2018b)
\item[f) Next generation. ]  This generation comprises all children as created above.
\end{description}
\item[3) Termination. ]
The algorithm stops when a  stopping criter is met.
Here, the stopping criteria is the maximum generation number $L$. 
\end{description}

%% file: Main.bbl
\begin{thebibliography}{51}
\expandafter\ifx\csname natexlab\endcsname\relax\def\natexlab#1{#1}\fi
\def\au#1{#1} \def\ed#1{#1} \def\yr#1{#1}\def\at#1{#1}\def\jt#1{\textit{#1}}
  \def\bt#1{#1}\def\bvol#1{\textbf{#1}} \def\vol#1{#1} \def\pg#1{#1}
  \def\publ#1{#1}\def\arxiv#1{#1}\def\org#1{#1}\def\st#1{\textit{#1}}

\bibitem[Aider {\em et~al.\/}(2010)Aider, Beaudoin \& Wesfreid]{Aider2010ef}
{\sc \au{Aider, J.-L.}, \au{Beaudoin, J.-F.} \& \au{Wesfreid, J.~E.}} \yr{2010}
   \at{Drag and lift reduction of a 3{D} bluff-body using active vortex
  generators}.  \jt{Exp.\ Fluids}  \bvol{48},  \pg{771--789}.

\bibitem[Barros(2015)]{Barros2015phd}
{\sc \au{Barros, D.}} \yr{2015}  \at{Wake and drag manipulation of a bluff body
  using fluidic forcing}. PhD thesis, \'Ecole Nationale Sup\'erieure de
  M\'ecanique et d'A\'erotechnique, Poitiers, France.

\bibitem[Barros {\em et~al.\/}(2016)Barros, Bor\'ee, Noack, Spohn \&
  Ruiz]{Barros2016jfm}
{\sc \au{Barros, D.}, \au{Bor\'ee, J.}, \au{Noack, B.~R.}, \au{Spohn, A.} \&
  \au{Ruiz, T.}} \yr{2016}  \at{Bluff body drag manipulation using pulsed jets
  and {C}oanda effect}.  \jt{J.\ Fluid Mech.}  \bvol{805},  \pg{442--459}.

\bibitem[Ben-Hamou {\em et~al.\/}(2007)Ben-Hamou, Arad \&
  Seifert]{ben2007generic}
{\sc \au{Ben-Hamou, E.}, \au{Arad, E.} \& \au{Seifert, A.}} \yr{2007}
  \at{Generic transport aft-body drag reduction using active flow control}.
  \jt{Flow Turbul. Combust.}  \bvol{78}~(3-4),  \pg{365}.

\bibitem[Bideaux {\em et~al.\/}(2011)Bideaux, Bobillier, Fournier,
  Gilli{\'e}ron, El~Hajem, Champagne, Gilotte \& Kourta]{Bideaux2011ija}
{\sc \au{Bideaux, E.}, \au{Bobillier, P.}, \au{Fournier, E.},
  \au{Gilli{\'e}ron, P.}, \au{El~Hajem, M.}, \au{Champagne, J.-Y.},
  \au{Gilotte, P.} \& \au{Kourta, A.}} \yr{2011}  \at{Drag reduction by pulsed
  jets on strongly unstructured wake: towards the square back control}.
  \jt{Int. J. Aerodyn.}  \bvol{1}~(3-4),  \pg{282--298}.

\bibitem[Bruneau {\em et~al.\/}(2010)Bruneau, Creus{\'e}, Depeyras,
  Gilli{\'e}ron \& Mortazavi]{Bruneau2010cf}
{\sc \au{Bruneau, C.-H.}, \au{Creus{\'e}, E.}, \au{Depeyras, D.},
  \au{Gilli{\'e}ron, P.} \& \au{Mortazavi, I.}} \yr{2010}  \at{Coupling active
  and passive techniques to control the flow past the square back {A}hmed
  body}.  \jt{Comput. \& Fluids}  \bvol{39}~(10),  \pg{1875--1892}.

\bibitem[Brunn \& Nitsche(2006)]{Brunn2006afc}
{\sc \au{Brunn, A} \& \au{Nitsche, W}} \yr{2006}  \at{Drag reduction of an
  ahmed car model by means of active separation control at the rear vehicle
  slant}.  \bt{In {\em New Results in Numerical and Experimental Fluid
  Mechanics V\/}},  \pg{pp. 249--256}.  \publ{Springer}.

\bibitem[Brunton \& Kutz(2019)]{Brunton2019book}
{\sc \au{Brunton, S.~L.} \& \au{Kutz, N.}} \yr{2019} {\em Data-Driven Science
  and Engineering\/}.  \publ{Cambridge University Press}.

\bibitem[Brunton \& Noack(2015)]{Brunton2015amr}
{\sc \au{Brunton, S.~L.} \& \au{Noack, B.~R.}} \yr{2015}  \at{Closed-loop
  turbulence control: {P}rogress and challenges}.  \jt{Appl.\ Mech.\ Rev.}
  \bvol{67}~(5),  \pg{050801:01--48}.

\bibitem[Bucci {\em et~al.\/}(2019)Bucci, Semeraro, Allauzen, Wisniewski,
  Cordier \& Mathelin]{Bucci2019prsa}
{\sc \au{Bucci, M.~A.}, \au{Semeraro, O.}, \au{Allauzen, A.}, \au{Wisniewski,
  G.}, \au{Cordier, L.} \& \au{Mathelin, L.}} \yr{2019}  \at{Control of chaotic
  systems by deep reinforcement learning}.  \jt{Proc.\ Roy.\ Soc.\ London A}
  \bvol{475},  \pg{20190351}.

\bibitem[Chen {\em et~al.\/}(2020)Chen, Ji, Alam, Williams \&
  Xu]{ChenAlam2020jfm}
{\sc \au{Chen, W.}, \au{Ji, C.}, \au{Alam, Md~M.}, \au{Williams, J.} \& \au{Xu,
  D.}} \yr{2020}  \at{Numerical simulations of flow past three circular
  cylinders in equilateral-triangular arrangements}.  \jt{Journal of Fluid
  Mechanics}  \bvol{891},  \pg{1--44}.

\bibitem[{Cornejo Maceda} {\em et~al.\/}(2020){Cornejo Maceda}, Li, Lusseyran,
  Morzy\'nski \& Noack]{Cornejo2020jfm}
{\sc \au{{Cornejo Maceda}, G.~Y.}, \au{Li, Y.}, \au{Lusseyran, F.},
  \au{Morzy\'nski, M.} \& \au{Noack, B.~R.}} \yr{2020}  \at{Stabilization of
  the fluidic pinball with gradient-based machine learning control}.  \jt{J.
  Fluid Mech.}  \bvol{(in preparation)},  \pg{1--44}.

\bibitem[{Cornejo Maceda} {\em et~al.\/}(2019){Cornejo Maceda}, R., Lusseyran,
  Deng, Pastur \& Morzy\'nski]{Cornejo2019pamm}
{\sc \au{{Cornejo Maceda}, G.~Y.}, \au{R., Noack~B.}, \au{Lusseyran, F.},
  \au{Deng, N.}, \au{Pastur, L.} \& \au{Morzy\'nski, M.}} \yr{2019}
  \at{Artificial intelligence control applied to drag reduction of the fluidic
  pinball}.  \jt{Proc.\ Appl.\ Math.\ Mech.}  \bvol{19}~(1),
  \pg{e201900268:1--2}.

\bibitem[Cox \& Cox(2000)]{Cox2000book}
{\sc \au{Cox, T.~F.} \& \au{Cox, M.~A.~A.}} \yr{2000} {\em {M}ultidimensional
  {S}caling\/}, 2nd edn.,  \st{Monographs on Statistics and Applied
  Probability},  \vol{vol.~88}.  \publ{Chapman and Hall}.

\bibitem[Dejoan {\em et~al.\/}(2005)Dejoan, Jang \&
  Leschziner]{dejoan2005comparative}
{\sc \au{Dejoan, A}, \au{Jang, Y.~J.} \& \au{Leschziner, M.~A.}} \yr{2005}
  \at{Comparative {LES} and unsteady {RANS} computations for a
  periodically-perturbed separated flow over a backward-facing step}.  \jt{J.
  Fluids Eng.}  \bvol{127}~(5),  \pg{872--878}.

\bibitem[Deng {\em et~al.\/}(2020)Deng, Noack, Morzyński \&
  Pastur]{Deng2020jfm}
{\sc \au{Deng, N.}, \au{Noack, B.~R.}, \au{Morzyński, M.} \& \au{Pastur,
  L.~R.}} \yr{2020}  \at{Low-order model for successive bifurcations of the
  fluidic pinball}.  \jt{J. Fluid Mech.}  \bvol{884},  \pg{A37}.

\bibitem[Fernex {\em et~al.\/}(2020)Fernex, Semann, Albers, Meysonnat,
  Schr\"oder \& Noack]{Fernex2020prf}
{\sc \au{Fernex, D.}, \au{Semann, R.}, \au{Albers, M.}, \au{Meysonnat, P.~S},
  \au{Schr\"oder, W.} \& \au{Noack, B.~R.}} \yr{2020}  \at{Self-similar drag
  reduction formula from sparse data---{O}ptimization of turbulent
  skin-friction via spanwise travelling surface waves}.  \jt{Phys. Rev. Fluids}
   \bvol{5}~(7),  \pg{073901:1--18}.

\bibitem[Gautier {\em et~al.\/}(2015)Gautier, Aider, Duriez, Noack, Segond \&
  Abel]{Gautier2015jfm}
{\sc \au{Gautier, N.}, \au{Aider, J.-L.}, \au{Duriez, T.}, \au{Noack, B.~R.},
  \au{Segond, M.} \& \au{Abel, M.~W.}} \yr{2015}  \at{Closed-loop separation
  control using machine learning}.  \jt{J.\ Fluid Mech.}  \bvol{770},
  \pg{424--441}.

\bibitem[Geropp(1995)]{Geropp1995patent}
{\sc \au{Geropp, D.}} \yr{1995} Process and device for reducing the drag in the
  rear region of a vehicle, for example, a road or rail vehicle or the like.
  United States Patent {\bfseries US$\>$5407245$\>$A}.

\bibitem[Geropp \& Odenthal(2000)]{Geropp2000ef}
{\sc \au{Geropp, D.} \& \au{Odenthal, H.-J.}} \yr{2000}  \at{Drag reduction of
  motor vehicles by active flow control using the {C}oanda effect}.
  \jt{Exp.~Fluids}  \bvol{28}~(1),  \pg{74--85}.

\bibitem[Gilli{\'e}ron \& Kourta(2013)]{Gillieron2013ef}
{\sc \au{Gilli{\'e}ron, P.} \& \au{Kourta, A.}} \yr{2013}  \at{Aerodynamic drag
  control by pulsed jets on simplified car geometry}.  \jt{Exp. Fluids}
  \bvol{54}~(2),  \pg{1457}.

\bibitem[Glezer {\em et~al.\/}(2005)Glezer, Amitay \& Honohan]{Glezer2005aiaaj}
{\sc \au{Glezer, A.}, \au{Amitay, M.} \& \au{Honohan, A.M.}} \yr{2005}
  \at{Aspects of low- and high-frequency actuation for aerodynamic flow
  control}.  \jt{AIAA Journal}  \bvol{43}~(7),  \pg{1501--1511}.

\bibitem[Gad-el Hak(2006)]{Gadelhak2006book}
{\sc \au{Gad-el Hak, M.}} \yr{2006} {\em Flow Control: Passive, Active, and
  Reactive Flow Management\/}.  \publ{Cambridge university press}.

\bibitem[Han {\em et~al.\/}(2013)Han, Krajnovi{\'c} \& Basara]{Han2013ijhff}
{\sc \au{Han, X.}, \au{Krajnovi{\'c}, S.} \& \au{Basara, B.}} \yr{2013}
  \at{Study of active flow control for a simplified vehicle model using the
  {PANS} method}.  \jt{Int. J. Heat Fluid Flow}  \bvol{42},  \pg{139--150}.

\bibitem[Hucho(2002)]{Hucho2011book}
{\sc \au{Hucho, W.-H.}} \yr{2002} {\em Aerodynamik der stumpfen K\"orper.
  Physikalische Grundlagen und Anwendungen in der Praxis\/}, 2nd edn.
  \publ{Wiesbaden: Vieweg Verlag}.

\bibitem[Humphrey \& Wilson(2000)]{Humphrey2000jc}
{\sc \au{Humphrey, D.~G.} \& \au{Wilson, J.~R.}} \yr{2000}  \at{A revised
  simplex search procedure for stochastic simulation response surface
  optimization}.  \jt{INFORMS Journal on Computing}  \bvol{12}~(4),
  \pg{272--283}.

\bibitem[Ishar {\em et~al.\/}(2019)Ishar, Kaiser, Morzynski, Albers, Meysonnat,
  Schr\"oder \& Noack]{Ishar2019jfm}
{\sc \au{Ishar, R.}, \au{Kaiser, E.}, \au{Morzynski, M.}, \au{Albers, M.},
  \au{Meysonnat, P.}, \au{Schr\"oder, W.} \& \au{Noack, B.~R.}} \yr{2019}
  \at{Metric for attractor overlap}.  \jt{J.~Fluid Mech.}  \bvol{874},
  \pg{720--752}.

\bibitem[Kim(2011)]{Kim2011ptrsa}
{\sc \au{Kim, J.}} \yr{2011}  \at{Physics and control of wall turbulence for
  drag reduction}.  \jt{Phil. Trans. Roy. Soc. A}  \bvol{369}~(1940),
  \pg{1396--1411}.

\bibitem[King \& Rowan(2020)]{King2020r}
{\sc \au{King, A.~A.} \& \au{Rowan, T.}} \yr{2020} Subplex: Unconstrained
  optimization using the subplex algorithm.

\bibitem[Krajnovi{\'c}(2009)]{Krajnovic2009prsa}
{\sc \au{Krajnovi{\'c}, S.}} \yr{2009}  \at{Large eddy simulation of flows
  around ground vehicles and other bluff bodies}.  \jt{Phil.\ Trans.\ R.\ Soc.\
  A}  \bvol{367}~(1899),  \pg{2917--2930}.

\bibitem[Li {\em et~al.\/}(2018{\natexlab{{\em a\/}}})Li, Noack, Cordier,
  Bor\'ee, Kaiser \& Harambat]{Li2018am}
{\sc \au{Li, R.}, \au{Noack, B.~R.}, \au{Cordier, L.}, \au{Bor\'ee, J.},
  \au{Kaiser, E.} \& \au{Harambat, F.}} \yr{2018{\natexlab{{\em a\/}}}}
  \at{Linear genetic programming control for strongly nonlinear dynamics with
  frequency crosstalk}.  \jt{Archives of Mechanics}  \bvol{70}~(6),
  \pg{505--534}.

\bibitem[Li {\em et~al.\/}(2018{\natexlab{{\em b\/}}})Li, Cui, Jia \&
  Yang]{Li2018icfm8}
{\sc \au{Li, Y.}, \au{Cui, W.}, \au{Jia, Q.} \& \au{Yang, Z.}}
  \yr{2018{\natexlab{{\em b\/}}}} Wake control on a simplified vehicle using
  steady blowing.  \bt{In {\em Proceedings of the 8th International Conference
  on Fluid Mechanics (ICFM8)\/}},  \pg{pp. 1--6}. Paper.

\bibitem[McKay {\em et~al.\/}(1979)McKay, Beckman \& Conover]{McKay1979t}
{\sc \au{McKay, M.~D.}, \au{Beckman, R.~J.} \& \au{Conover, W.~J.}} \yr{1979}
  \at{Comparison of three methods for selecting values of input variables in
  the analysis of output from a computer code}.  \jt{Technometrics}
  \bvol{21}~(2),  \pg{239--245}.

\bibitem[Minelli {\em et~al.\/}(2020)Minelli, Dong, Noack \&
  Krajnović]{Minelli2020jfm}
{\sc \au{Minelli, G.}, \au{Dong, T.}, \au{Noack, B.~R.} \& \au{Krajnović, S.}}
  \yr{2020}  \at{Upstream actuation for bluff-body wake control driven by a
  genetically inspired optimization}.  \jt{J. Fluid Mech.}  \bvol{893},
  \pg{A1}.

\bibitem[Muralidharan {\em et~al.\/}(2013)Muralidharan, Muddada \&
  Patnaik]{muralidharan2013numerical}
{\sc \au{Muralidharan, K}, \au{Muddada, S.} \& \au{Patnaik, B. S.~V.}}
  \yr{2013}  \at{Numerical simulation of vortex induced vibrations and its
  control by suction and blowing}.  \jt{Appl. Math. Model.}  \bvol{37}~(1-2),
  \pg{284--307}.

\bibitem[Nelder \& Mead(1965)]{Nelder1965jc}
{\sc \au{Nelder, J.~A.} \& \au{Mead, R.}} \yr{1965}  \at{A simplex method for
  function minimization}.  \jt{J.\ Comput.}  \bvol{7},  \pg{308--313}.

\bibitem[Park {\em et~al.\/}(2006)Park, Lee, Jeon, Hahn, Kim, Choi \&
  Choi]{Park2006jfm}
{\sc \au{Park, H.}, \au{Lee, D.}, \au{Jeon, W.-P.}, \au{Hahn, S.}, \au{Kim,
  J.}, \au{Choi, J.} \& \au{Choi, H.}} \yr{2006}  \at{Drag reduction in flow
  over a two-dimensional bluff body with a blunt trailing edge using a new
  passive device}.  \jt{J.\ Fluid Mech.}  \bvol{563},  \pg{389--414}.

\bibitem[Pastoor {\em et~al.\/}(2008)Pastoor, Henning, Noack, King \&
  Tadmor]{Pastoor2008jfm}
{\sc \au{Pastoor, M.}, \au{Henning, L.}, \au{Noack, B.~R.}, \au{King, R.} \&
  \au{Tadmor, G.}} \yr{2008}  \at{Feedback shear layer control for bluff body
  drag reduction}.  \jt{J.\ Fluid Mech.}  \bvol{608},  \pg{161--196}.

\bibitem[Pfeiffer \& King(2014)]{Pfeiffer2014aiaa}
{\sc \au{Pfeiffer, J.} \& \au{King, R.}} \yr{2014} Linear parameter varying
  active flow control for a 3d bluff body exposed to cross-wind gusts.  \bt{In
  {\em {AIAA} Paper, 2014-2406\/}}.

\bibitem[Press {\em et~al.\/}(2007)Press, Flamery, Teukolsky \&
  Vetterling]{Press2007book}
{\sc \au{Press, W.H.}, \au{Flamery, B.P.}, \au{Teukolsky, S.A.} \&
  \au{Vetterling, W.T.}} \yr{2007} {\em Numerical {R}ecipes, The {A}rt of
  {S}cientific {C}omputing\/}, 3rd edn.  \publ{Cambridge, UK, etc.: Cambridge
  University Press}.

\bibitem[Protas(2004)]{Protas2004pf}
{\sc \au{Protas, B.}} \yr{2004}  \at{Linear feedback stabilization of laminar
  vortex shedding based on a point vortex model}.  \jt{Phys.\ Fluids}
  \bvol{16}~(12),  \pg{4473--4488}.

\bibitem[Rabault {\em et~al.\/}(2019)Rabault, Kuchta, Jensen, R{\'e}glade \&
  Cerardi]{Rabault2019jfm}
{\sc \au{Rabault, J.}, \au{Kuchta, M.}, \au{Jensen, A.}, \au{R{\'e}glade, U.}
  \& \au{Cerardi, N.}} \yr{2019}  \at{Artificial neural networks trained
  through deep reinforcement learning discover control strategies for active
  flow control}.  \jt{J. Fluid Mech.}  \bvol{865},  \pg{281--302}.

\bibitem[Raibaudo {\em et~al.\/}(2019)Raibaudo, Zhong, Noack \&
  Martinuzzi]{Raibaudo2020pf}
{\sc \au{Raibaudo, C.}, \au{Zhong, P.}, \au{Noack, B.~R.} \& \au{Martinuzzi,
  R.~J.}} \yr{2019}  \at{Machine learning strategies applied to the control of
  a fluidic pinball}.  \jt{Phys. Fluids}  \bvol{32},  \pg{015108}.

\bibitem[Ren {\em et~al.\/}(2020)Ren, Hu \& Tang]{Ren2020jh}
{\sc \au{Ren, F.}, \au{Hu, H.-B.} \& \au{Tang, H.}} \yr{2020}  \at{Active flow
  control using machine learning: A brief review}.  \jt{J. Hydrodyn.}
  \bvol{32}~((2)),  \pg{247--253}.

\bibitem[Schmidt {\em et~al.\/}(2015)Schmidt, Woszidlo, Nayeri \&
  Paschereit]{Schmidt2015ef}
{\sc \au{Schmidt, H.-J.}, \au{Woszidlo, R.}, \au{Nayeri, C.~N.} \&
  \au{Paschereit, C.~O.}} \yr{2015}  \at{The effect of flow control on the wake
  dynamics of a rectangular bluff body in ground proximity}.  \jt{Exp. Fluids}
  \bvol{56},  \pg{151}.

\bibitem[Thiria {\em et~al.\/}(2006)Thiria, Goujon-Durand \&
  Wesfreid]{Thiria2006jfm}
{\sc \au{Thiria, B.}, \au{Goujon-Durand, S.} \& \au{Wesfreid, J.~E.}} \yr{2006}
   \at{The wake of a cylinder performing rotary oscillations}.  \jt{J.\ Fluid
  Mech.}  \bvol{560},  \pg{123--147}.

\bibitem[Viken {\em et~al.\/}(2003)Viken, Vatsa, Rumsey \&
  Carpenter]{viken2003flow}
{\sc \au{Viken, S.}, \au{Vatsa, V.}, \au{Rumsey, C.} \& \au{Carpenter, M.}}
  \yr{2003} Flow control analysis on the hump model with {RANS} tools.  \bt{In
  {\em 41st Aerospace Sciences Meeting and Exhibit\/}},  \pg{p. 218}.

\bibitem[Wahde(2008)]{Wahde2008book}
{\sc \au{Wahde, M.}} \yr{2008} {\em Biologically Inspired Optimization Methods:
  An Introduction\/}.  \publ{WIT Press}.

\bibitem[Wood(1964)]{Wood1964jras}
{\sc \au{Wood, C.~J.}} \yr{1964}  \at{The effect of base bleed on a periodic
  wake}.  \jt{Journal of the Royal Aeronautical Society}  \bvol{68}~(643),
  \pg{477--482}.

\bibitem[Wright(1991)]{wright1991genetic}
{\sc \au{Wright, Alden~H}} \yr{1991}  \at{Genetic algorithms for real parameter
  optimization}.  \bt{In {\em Foundations of genetic algorithms\/}}, ,
  \vol{vol.~1},  \pg{pp. 205--218}.  \publ{Elsevier}.

\bibitem[Zhang {\em et~al.\/}(2018)Zhang, Liu, Zhou, To \& Tu]{Zhang2018jfm}
{\sc \au{Zhang, B.~F.}, \au{Liu, K.}, \au{Zhou, Y.}, \au{To, S.} \& \au{Tu,
  J.~Y.}} \yr{2018}  \at{Active drag reduction of a high-drag {A}hmed body
  based on steady blowing}.  \jt{J. Fluid Mech.}  \bvol{856},
  \pg{351–--396}.

\end{thebibliography}
